\documentclass[epsfig,psfig,aps,twocolumn,prb,showpacs]{revtex4-1}
\usepackage{amsfonts,amsmath,bbm,wasysym}
\usepackage{amsfonts,dsfont}
\usepackage[english]{babel}
\usepackage{float}
\usepackage{lipsum}
\usepackage{graphicx}
\usepackage{epsfig}
\usepackage{relsize}
\usepackage[colorlinks=true,citecolor=red,urlcolor=blue]{hyperref}
\usepackage{soul}
 
\newcommand{\red}{\color{red}}

\newcommand{\magenta}{\color{\magenta}}
 \usepackage{bm}

\begin{document}

\title{Twisted bilayer graphene reveals its flat bands under spin pumping
}
\author{Sonia Haddad$^{1,2,3}$}
\email{sonia.haddad@fst.utm.tn}
\author{Takeo Kato$^{2}$}
\author{Jihang Zhu$^{3}$}
\author{Lassaad Mandhour$^{1}$}
\affiliation{
$^1$Laboratoire de Physique de la Mati\`ere Condens\'ee, Facult\'e des Sciences de Tunis, Universit\'e Tunis El Manar, Campus Universitaire 1060 Tunis, Tunisia\\
$^2$ Institute for Solid State Physics, University of Tokyo, Kashiwa, Chiba 277-8581, Japan\\
$^3$ Max Planck Institute for the Physics of Complex Systems, N\"{o}thnitzer Strasse 38, Dresden 01187, Germany}
\date{\today}

\begin{abstract}
The salient property of the electronic band structure of twisted bilayer graphene (TBG), at the so-called magic angle (MA), is the emergence of flat bands around the charge neutrality point. These bands are associated with the observed superconducting phases and the correlated insulating states.
Scanning tunneling microscopy combined with angle resolved photoemission spectroscopy are usually used to visualize the flatness of the band structure of TBG at the MA.
Here, we theoretically argue that spin pumping (SP) provides a direct probe of the flat bands of TBG and an accurate determination of the MA. 
We consider a junction separating a ferromagnetic insulator and a heterostructure of TBG adjacent to a monolayer of a transition metal dichalcogenide.
We show that the Gilbert damping of the ferromagnetic resonance experiment, through this junction, depends on the twist angle of TBG, and exhibits a sharp drop at the MA. We discuss the experimental realization of our results which open the way to a twist switchable spintronics in twisted van der Waals heterostructures.
\end{abstract}

\maketitle

{\it Introduction. -- } Stacking two graphene layers with a relative twist angle $\theta$ results in a moir\'e superstructure which is found to host, in the vicinity of the so-called magic angle (MA) $\theta_M\sim1.1^{\circ}$, unconventional superconductivity and strongly correlated insulating states~\cite{Herrero1,Herrero2,Yank}. There is a general consensus that such strong electronic correlations originate from the moir\'e flat bands emerging at the MA around the charge neutrality point~\cite{Volovik,Senthil,Wu,Roy,Bernevig,Efetov,Young,Herrero3}.
The tantalizing signature of the flat bands have been experimentally demonstrated by probing the corresponding peaks of the density of states using transport~\cite{Herrero2,Herrero1,Yank,Efetov19,Dean19}, electronic compressibility measurements~\cite{Pablo19,Pablo20}, scanning tunneling microscopy (STM) and spectroscopy (STS)~\cite{Eva,Kerelsky,Yazdani19,Nadj19,Yazdani20,Nadj21,Eva2,Yazdani22}. The direct evidence of these flat bands has been reported by angle resolved photoemission spectroscopy (ARPES) measurements combined to different imaging techniques~\cite{Utama,Efetov21,Sato}.
However, spectroscopic measurements on magic-angle TBG raise many technical challenges related to the need of an accurate control of the twist angle, and the necessity to have non-encapsulated samples which can degrade in air~\cite{Efetov21}.

Here we propose a noninvasive method to probe the flat bands of TBG and accurately determine the MA. 
This method is based on spin pumping (SP) induced by ferromagnetic resonance (FMR)~\cite{Bauer1,Bauer2,SP-book,Hellman}, where the increase in the FMR linewidth, given by the Gilbert damping (GD) coefficient, provides insight into the spin excitations of the nonmagnetic (NM) material adjacent to the ferromagnet \cite{Qiu2016,Yang2018,Han2020}.
SP is expected to be efficient if the NM has high spin-orbit coupling (SOC) strength~\cite{Hait}. \newline
In our work, we consider spin injection from a ferromagnetic insulator (FI) into a TBG aligned on a monolayer of transition metal dichalcogenides (TMD) which are considered as good substrate candidates to induce relatively strong SOC in graphene and TBG~\cite{Castro,Morpu15,Morpu16,Bock16,Casa,Shi,Wees,Eroms,Eroms2,Makk,BouchitaPRL,BLG,Omar,Valen,Roche,Zaletel,Wang,Bouchiat19,David,Zaletel20,Bouchiat21,Lin,Alex,Bhowmick}. 

\begin{figure}[hpbt] 
\begin{center}
\includegraphics[width=0.9\columnwidth]{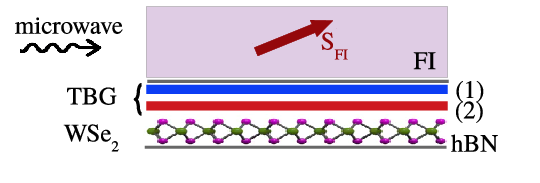}
\end{center}
\caption{Schematic representation of the junction between a ferromagnetic insulator (FI) and a heterostructure of TBG adjacent to a monolayer of $\mathrm{WSe}_2$. The labels (1) and (2) denote the graphene layers of TBG represented by the red and the blue lines. The red arrow indicates the spin orientation of the FI characterized by an average spin $\langle\mathbf{S}_{FI} \rangle=\left(S_0,0,0\right)$, written in the coordinate frame of the FI magnetization. The gray lines represent the boron-nitride (hBN) layers encapsulating the TBG/$\mathrm{WSe}_2$ heterostructure.}
\label{junction}
\end{figure}

% \textcolor{blue}{Recent experimental studies revealed that SOC, induced in TBG adjacent to $\mathrm{WSe}_2$, gives rise to ordered phases even at non-integer moir\'e band fillings~\cite{Bhowmick}. It was also found that it stabilizes superconductivity at a twist angle ($\theta\sim0.8^{\circ}$) smaller than the MA~\cite{Alex} and it transforms, at a given filling factors, the correlated insulating state of TBG into a ferromagnetic phase~\cite{Lin}}.\
We theoretically study a planar junction of a FI and a TBG adjacent to $\mathrm{WSe}_2$ (TBG/$\mathrm{WSe}_2$) as depicted in Fig.~\ref{junction}. 
We consider the case where a microwave of a frequency $\Omega$ is applied to this junction, and focus on the twist angle dependence of the FMR linewidth~\cite{temp}.

{\it Continuum model. -- } In TBG with a twist angle $\theta$, the Hamiltonian $h_{l}(\mathbf{k})$ of a graphene layer $l$ ($l=1,2$), rotated at an angle $\theta_l$, is 
$
h_{l}(\mathbf{k})=e^{i\frac{\theta_l}2\sigma_z} h_{l}^{(0)}(\mathbf{k})e^{-i\frac{\theta_l}2\sigma_z}
$,
where $\theta_2=-\theta_1=\frac{\theta}2$ and $h_{l}^{(0)}(\mathbf{k})$ is the unrotated monolayer Hamiltonian. In the continuum limit, $h_{1}^{(0)}(\mathbf{k})$ reduces to $h_{1}^{(0)}(\mathbf{k})=-\hbar v_F \mathbf{k}\cdot\mathbf{\sigma}^{\ast}$, where $v_F$ is the Fermi velocity, $\mathbf{\sigma}^{\ast}=\left(\xi \sigma_x,\sigma_y\right)$, and $\sigma_i$ ($i=x,y,z$) are the sublattice-Pauli matrices and $\xi$ is the valley index.\
We assume that the SOC is only induced in the graphene layer adjacent to the TMD layer, since the SOC arises from overlaps between atomic orbitals~\cite{Zaletel}. This assumption is consistent with recent studies on bilayer graphene and TBG aligned on TMD layers~\cite{Alex,Gmitra-BL,Zaletel,Alex2}.
Layer (2), in contact with the $\mathrm{WSe}_2$ monolayer, is then descried by the Hamiltonian
$
h_{2}^{(0)}(\mathbf{k})=h_{1}^{(0)}(\mathbf{k})+h_{\text{SOC}}+\frac{m}2\sigma_z
$~\cite{Alex}, where $h_{\text{SOC}}$ is given by
\begin{eqnarray}
h_{\text{SOC}}=\frac{\lambda_I}2\xi s_z+\frac{\lambda_R}2\left(\xi \sigma_x s_y-\sigma_y s_x\right)+\frac{\lambda_{\text{KM}}}2\xi \sigma_z s_z,
\label{hSOC}
\end{eqnarray}
$s_i$ ($i=x,y,z$) are the spin-Pauli matrices, $\lambda_I$, $\lambda_R$ and $\lambda_{\text{KM}}$ correspond, respectively, to the Ising, Rashba and Kane-Mele SOC parameters~\cite{Alex}. The variation ranges of these parameters are $\lambda_I\sim 1-5 \;\mathrm {meV}$, $\lambda_R\sim 1-15 \;\mathrm {meV}$, while $\lambda_{\text{KM}}$ is expected to be small~\cite{SOC1,SOC2,Makk,Zaletel,Wang,Morpu16,SOC7}. The last term in $h_{2}^{(0)}(\mathbf{k})$ is due to the inversion symmetry breaking induced by the TMD layer. Hereafter, we neglect this term regarding the small value of $m$ compared to the SOC parameters~\cite{Alex}.\newline
As in the case of TBG~\cite{Mc11}, the low-energy Hamiltonian of TBG/$\mathrm{WSe}_2$ reduces, at the valley $\xi$, to

\begin{eqnarray}
H_{\xi,SOC}(\mathbf{k})= 
\begin{pmatrix}
h_{1}(\mathbf{k}) & T_1 & T_2 & T_3\\
T^{\dagger}_1 & h_{2,1}(\mathbf{k}) & 0&0\\
T^{\dagger}_2 & 0& h_{2,2}(\mathbf{k}) & 0\\
T^{\dagger}_3 & 0& 0& h_{2,3}(\mathbf{k})\\
\end{pmatrix}.
\label{HTBG}
\end{eqnarray}
$H_{\xi,\text{SOC}}(\mathbf{k})$ is written in the basis 
$\Psi=\left(\psi_0(\mathbf{k}),\psi_1(\mathbf{k}),\psi_2(\mathbf{k}),\psi_3(\mathbf{k})\right)$ constructed on the four-component spin-sublattice spinor $\psi_0(\mathbf{k})$ and $\psi_j(\mathbf{k})$, ($j=1,2,3$) corresponding, respectively, to layer $(1)$ and layer $(2)$ (see Secs. I and II of the Supplemental Material~\cite{supp} and Refs.~\cite{Mc11,Koshino18,Bernevig,Falko19,Bi,Alex,Marwa}). The momentum $\mathbf{k}$ is measured relatively to the Dirac point $\mathbf{K}_{1\xi}$ of layer (1). In Eq.~(\ref{HTBG}), $T_j$ are the spin-independent interlayer coupling matrices, $h_{2,j}\left(\mathbf{k}\right)=h_2\left(\mathbf{k}+\mathbf{q}_{j\xi}\right)$, ($j=1,2,3$) where $\mathbf{q}_{j\xi}$ are the vectors connecting $\mathbf{K}_{1\xi}$ to its three neighboring Dirac points $\mathbf{K}_{2\xi}$ of layer (2) in the moir\'e Brillouin zone (mBZ)~\cite{Mc11}, and are given by
$\mathbf{q}_{1\xi}=\mathbf{K}_{1\xi}-\mathbf{K}_{2\xi}$, 
 $\mathbf{q}_{2\xi}=\mathbf{q}_{1\xi}+\xi\mathbf{G}^M_1$, $\mathbf{q}_{3\xi}=\mathbf{q}_{1\xi}+\xi\left(\mathbf{G}^M_1+\mathbf{G}^M_2\right)$,
where $\left(\mathbf{G}^M_1,\mathbf{G}^M_2\right)$ is the mBZ basis (see Sec. I of the Supplemental Material~\cite{supp}).\newline
In the unrelaxed TBG, and choosing sublattice A as the origin of the unit cell in each layer, the $T_j$ matrices take the form \cite{Bernevig}
$T_1= w\left(\mathbb{I}_{\sigma}+\sigma_x\right)$, 
$T_2=w \left(\mathbb{I}_{\sigma}-\frac 12\sigma_x+\xi \frac {\sqrt{3}}2\sigma_y\right)$ and $T_3=w\left( \mathbb{I}_{\sigma}-\frac 12\sigma_x-\xi\frac {\sqrt{3}}2\sigma_y\right)$~\cite{supp}, where $w\sim 110\, \mathrm{meV}$~\cite{w} is the interlayer tunneling amplitude and $\mathbb{I}_{\sigma}$ is the identity matrix acting on the sublattice indices.\

Using the perturbative approach of Ref.~\cite{Mc11}, we derive, from Eq.~(\ref{HTBG}), the effective low-energy Hamiltonian $H_{\xi,\text{SOC}}^{(1)}(\mathbf{k})$ of TBG/$\mathrm{WSe_2}$ (see Sec. II of the Supplemental Material~\cite{supp}). To the leading order in $\mathbf{k}$, $H_{\xi,\text{SOC}}^{(1)}(\mathbf{k})$ reads as~\cite{supp}
\begin{widetext}
\begin{eqnarray}
&& H_{\xi,\text{SOC}}^{(1)}(\mathbf{k})=\frac{\langle\Psi|H_{\xi,\text{SOC}}|\Psi\rangle}{\langle\Psi|\Psi\rangle}=
 \psi^{\dagger}_0 \left[h_{\text{eff}}\left(\mathbf{k}\right) + h_{\text{eff}}^{\text{SOC}} \right]\psi_0,
 \label{H1}\\
 &&h_{\text{eff}}\left(\mathbf{k}\right)=-\frac{\hbar v_F}{\langle\Psi|\Psi\rangle}
 \left\{ k_x\left[ \left(1-3\alpha^2\right) \xi \sigma_x \mathbb{I}_{s}
 -\frac{3\alpha^2}{\hbar v_F q_0}\left( \xi \lambda_I \sigma_y s_z+\lambda_R\left(\xi \sigma_ys_y-\sigma_x s_x\right)
 \right)\right]\right.\nonumber\\
 &&\hspace{2.8cm}\left.+k_y \left[ \left(1-3\alpha^2\right) \sigma_y \mathbb{I}_{s}
 -\frac{3\alpha^2}{\hbar v_F q_0}\left( - \lambda_I \sigma_x s_z+\lambda_R\left(\sigma_xs_y+\xi \sigma_y s_x\right)
 \right)\right]
 \right\},
 \label{heff}\\
 &&h_{\text{eff}}^{\text{SOC}}=\frac{3\alpha^2}{\langle\Psi|\Psi\rangle}\left[ \xi \lambda_Is_z\mathbb{I}_{\sigma}+\frac {\lambda_R}2\left( s_x\sigma_y-\xi s_y\sigma_x\right) \right],
 \label{heff_SOC}
\end{eqnarray}
\end{widetext}
where $\langle\Psi|\Psi\rangle\sim 1+6\alpha^2$, $\alpha=\frac{w}{\hbar v_F q_0}$, $q_0=|\mathbf{q}_{j\xi}|=\frac{4\pi}{3a}\theta$, $a$ is the graphene lattice constant and
 $\sigma_i$, ($i=x,y,z$) act now on the band indices $\sigma=\pm$ of the eigenenergies of $H_{\xi,\text{SOC}}^{(1)}$, denoted $E_{\sigma,\pm}$, and given to the leading orders in $\mathbf{k}$ and $\frac{\lambda_{I,R}}{\hbar v_Fq_0}$ by
\begin{eqnarray}
E(\mathbf{k})_{\sigma,\pm}= \frac {\sigma}{\langle \Psi |\Psi\rangle }
\sqrt{f_1(\mathbf{k})\pm 6 \alpha^2\sqrt{f_2(\mathbf{k})}}
\label{energy_H1}
\end{eqnarray}
\begin{eqnarray}
&&f_1(\mathbf{k})=(\hbar v_F)^2\left(1-3\alpha^2\right)^2||{\mathbf{k}}||^2+\frac 92\alpha^4\left(2\lambda_I^2+\lambda_R^2\right)\nonumber\\
&&f_2(\mathbf{k})= (\hbar v_F)^2\left(1-3\alpha^2\right)^2||{\mathbf{k}}||^2\left(\lambda_I^2+\frac 14\lambda_R^2\right)+ \frac9{16} \alpha^4 \lambda_R^4.\nonumber
\end{eqnarray}
Equation~\ref{heff_SOC} shows that the SOC parameters $\lambda_I$ and $\lambda_R$ are renormalized by the moir\'e structure of TBG to 
\begin{eqnarray}
 \tilde{\lambda}_I\sim \frac{6\alpha^2}{1+6\alpha^2}\lambda_I,\;
 \tilde{\lambda}_R\sim\frac{3\alpha^2}{1+6\alpha^2}\lambda_R,
 \label{SOCeff}
\end{eqnarray}
which increase by decreasing the twist angle.\newline
The expression of $H_{\xi,\text{SOC}}^{(1)}$ [Eq.~(\ref{H1})] can be taken as a starting point to unveil the role of SOC in the emergence of the stable superconducting phase observed, at $\theta\sim 0.8^{\circ}$, in TBG adjacent to $\mathrm{WSe}_2$~\cite{Alex}.\newline
To probe the validity of the effective Hamiltonian $H_{\xi,\text{SOC}}^{(1)}$ [Eq.~(\ref{H1})], we compared the corresponding eigenenergies with the numerical band structure obtained within the continuum model and taking into account 148 bands per valley and spin projection (see Sec. II of the Supplemental Material~\cite{supp}). The results show that $H_{\xi,\text{SOC}}^{(1)}$ describes correctly the band structure of TBG/$\mathrm{WSe}_2$ down to a twist angle $\theta\sim 0.7^{\circ}$. At smaller angles, the effective Fermi velocities of $H_{\xi,\text{SOC}}^{(1)}$ are overestimated. Such a discrepancy is expected since the lattice relaxation effect is important at small angles~\cite{Alex}. It is worth noting that, for the sake of simplicity, we did not consider a relaxed TBG, since we are interested in the SP around the MA.\\

{\it Gilbert damping. -- } In the absence of a junction, the magnon Green function of the FI is defined as~\cite{Ohnuma,Matsuo18,Kato19,Kato20,Matsuo-JP,Matsuo20,Yama}
$G_0\left(\mathbf{q}_m,i\omega_n\right)=\frac{2S_0/\hbar}{i\omega_n-\omega_{\mathbf{q}_m}-\alpha_G |\omega_n|}
$, where $\omega_n=2\pi n/\hbar \beta$ are the Matsubara frequencies for bosons, $S_0$ is the amplitude of the average spin per site, and $\alpha_G$ is the GD strength. The term $-\alpha_G |\omega_n|$ describes the spin relaxation within the FI.
In FMR experiments, the microwave excitation induces a uniform spin precession, which limits the magnon self-energy to the processes with $\mathbf{q}_m=0$~\cite{Funato}.\

In the presence of the interfacial coupling, a correction, $\delta \alpha_{G}(\omega)$, to the GD term is induced by the adjacent heterostructure TBG/$\mathrm{WSe_2}$. $\delta \alpha_{ G}(\omega)$ can be expressed in terms of the the self-energy $\Sigma^{R}_{\mathbf{0}}\left(\omega\right)\equiv\Sigma_{\mathbf{q}_m=\mathbf{0}}\left(i\omega_n\rightarrow \omega + i\delta \right)$, resulting from the interfacial exchange interactions, as~\cite{Funato}
\begin{eqnarray}
\delta\alpha_G\left(\omega\right)\equiv -\frac{2S_0}{\hbar \omega}\mathrm{Im}\, \Sigma_{\mathbf{0}}^R \left(\omega\right).
\label{GD-self}
\end{eqnarray}
For simplicity, we neglect the real part of $ \Sigma^R_{\mathbf{0}}(\omega)$ which simply shifts the FMR line and did not affect the linewidth, in which we are interested.
%%%%
The self-energy, in Eq.~(\ref{GD-self}), includes the contributions of all the interfacial spin transfer processes and can be written as $\Sigma_{\mathbf{0}}\left(i\omega_n\right)=\sum_{\mathbf{q}}\Sigma_{\mathbf{0}} (\mathbf{q},i\omega_n)$. Each process, described by the self-energy $\Sigma_{\mathbf{0}} (\mathbf{q},i\omega_n)$, is characterized by a momentum transfer $\mathbf{q}$ and a matrix element $T_{\mathbf{q},\mathbf{q}_m=\mathbf{0}}\equiv T_{\mathbf{q},\mathbf{0}}$.\

In the second order perturbation, with respect to the interfacial exchange interaction $T_{\mathbf{q},\mathbf{0}}$, the self-energy $\Sigma_{\mathbf{0}} (\mathbf{q},i\omega_n)$, is written as~\cite{Yama}
%\begin{widetext}
\begin{eqnarray}
\Sigma_{\mathbf{0}} (\mathbf{q},i\omega_n)&=&\frac{|T_{\mathbf{q},\mathbf{0}}|^2}{4\beta}\sum_{\mathbf{k},i\omega_m} 
\mathrm{Tr}\left[ \sigma_s^{x^{\prime},-}\; \hat{g} (\mathbf{k},\omega_m)\right.\nonumber\\
&\times&\left.\sigma_s^{x^{\prime},+}\;\hat{g} (\mathbf{k}+\mathbf{q},i\omega_m+i\omega_n)
\right].
\label{self}
\end{eqnarray}
%\end{widetext}
$\sigma_s^{x^{\prime},\pm}$ are the electronic spin ladder operators written in the coordinate system $\left(x^{\prime},y^{\prime},z^{\prime}\right)$ of the FI magnetization characterized by an average spin $\langle\mathbf{S}_{FI}  \rangle=\left(S_0,0,0\right)$.
$\hat{g}(\mathbf{k},i\omega_m)$ is the electronic Matsubara Green function given by $
\hat{g}(\mathbf{k},i\omega_n)=\left[ i\omega_n \mathbb{I}- H_{\text{SOC}}^{(1)}(\mathbf{k})\right]^{-1}$,
where $\omega_n=(2n+1)\pi/\hbar \beta$ are the fermionic Matsubara frequencies.\
In the basis of the spin-band four-component spinor $\Psi=\left( \psi_{+,\uparrow},\psi_{+,\downarrow},\psi_{-,\uparrow},\psi_{-,\downarrow}\right)$~\cite{Alex}, $\hat{g}(\mathbf{k},i\omega_n)$ reads as
$
\hat{g}(\mathbf{k},i\omega_n)=\hat{g}_0(\mathbf{k},i\omega_n)\mathbb{I}_s+
\mathbf{\hat{g}}(\mathbf{k},i\omega_n)\cdot\mathbf{s}
$,
where $\mathbf{s}=(s_x,s_y,s_z) $ are the spin-Pauli matrices; $\mathbf{\hat{g}}=(\hat{g}_x,\hat{g}_y,\hat{g}_z)$, $\hat{g}_0$, and $\hat{g}_i$ ($i=x,y,z$) are expressed, to the leading order in the SOC, as a function of the band-Pauli matrices $\sigma_i$ (see Sec. III of the Supplemental Material~\cite{supp}).\

% Taking the analytical continuation $i\omega_n\rightarrow \omega+i\delta$, one can obtain, form Eqs.~\ref{GD-self} and~\ref{self}, the frequency depenence $\delta\alpha_G\left(\omega\right)$ of the GD correction.
Since the ferromagnetic peak, given by $\mathrm{Im}G_0^R$, is sharp enough, namely $\alpha_G+\delta\alpha_G\ll 1$, one can replace the resonance frequency $\omega_{\mathbf{q}_m=\mathbf{0}}$ by the FMR frequency $\Omega$. The GD correction can then be expressed as \cite{Yama,supp}
\begin{eqnarray}
\delta\alpha_G\left(\Omega\right)= -\frac{2S_0}{\hbar \Omega}\mathrm{Im}\Sigma_{\mathbf{0}}^R\left(\Omega\right).
\label{GD-exp}
\end{eqnarray}
%
%%%%
In general, the interfacial spin transfer includes clean and dirty processes. The former (latter) take place with conserved (non-conserved) electron momentum, which turns out to take $\mathbf{q}=\mathbf{0}$ ($\mathbf{q}\neq\mathbf{0}$) in Eq.~(\ref{self}~\cite{Funato}).\newline
We first consider a clean interface, for which an analytical expression of the GD correction [Eq.~\ref{GD-exp}] can be derived (see Sec. IV of the Supplemental Material~\cite{supp} and reference~\cite{Funato,Yama}). The case of a dirty junction is discussed in the next section.\

% This assumption is justified regarding the large unit cell of the TBG over which the impurity contribution can be averaged out.\
Carrying out the summation over $\omega_m$ in Eq.~(\ref{self}), we obtain the analytical expression of the interfacial self-energy (see Sec. IV of the Supplemental Material~\cite{supp}). The sum over the electronic states $\mathbf{k}=\left(k,\varphi_{\mathbf{k}}\right)$ runs over the states included within a cutoff, $k_c\sim q_0/2$, on the momentum amplitude $k$, where the low-energy Hamiltonian [Eq.~(\ref{H1})] is expected to hold (see Sec. IV of the Supplemental Material~\cite{supp}).\

In the following, we discuss the behavior of the normalized GD coefficient 
\begin{eqnarray}
 \delta\alpha_G/\alpha_G^0=\left(\frac{\lambda}{\hbar\Omega} \right)^2 \tilde{\Sigma}\left(\mathbf{q}=\mathbf{0},\Omega\right),
 \label{alpha_R}
\end{eqnarray}
where $\tilde{\Sigma}$ is a dimensionless function depending on the twist angle $\theta$, temperature $T$, the chemical potential $\mu$ and the orientation of the FI magnetization, $\alpha_G^0=2S_0\left(\frac{|T_\mathbf{0}|}{\lambda}\right)^2$ and $\lambda=\frac{\lambda_I+\lambda_R}2$ is the average SOC (for details, see Sec. IV of the Supplemental Material~\cite{supp} and reference~\cite{Guinea22}).\\

\begin{figure}[htbp] 
\begin{center}
\includegraphics[width=0.7\columnwidth]{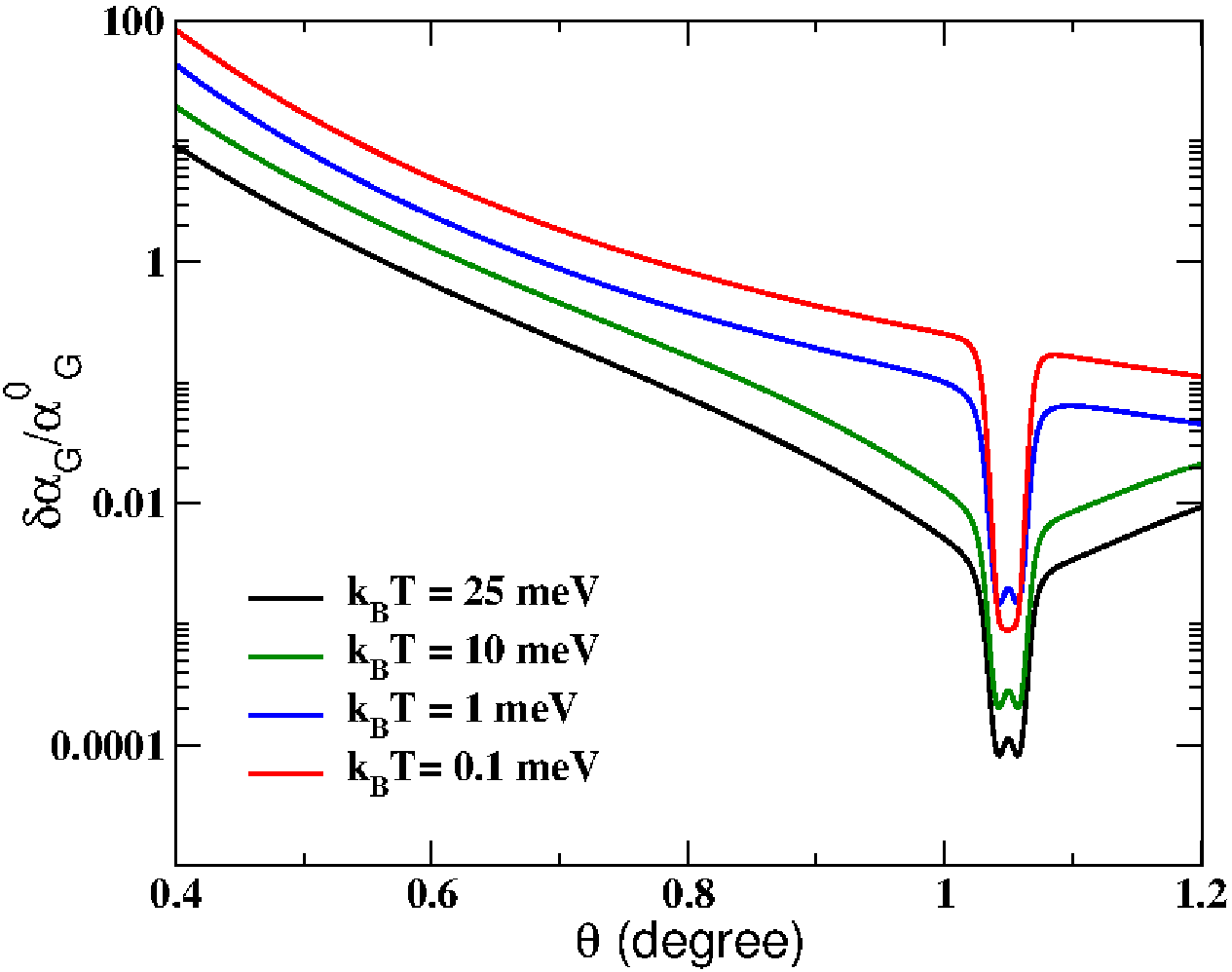}
\end{center}
\caption{Normalized GD, $\delta\alpha_G/\alpha_G^0$ [Eq.~(\ref{alpha_R})], as a function of the twist angle at different temperature ranges. Calculations are done for $\lambda_I=3\;\mathrm{meV}$, $\lambda_R=4\;\mathrm{meV}$, $\mu=0$, and for a FMR energy $\hbar\Omega=0.06\;\mathrm{meV}$. }
\label{GD2}
\end{figure}

%%%%%%%%%%%%%%%%%%%%%%%%
%%%%%%%%%%%%%%%%%%%%%%%%
%%%%%%%%%%%%%%% discussion%%%%%%%%%%%%%%%%%%%%
%
{\it Discussion. -- } In Fig.~\ref{GD2}, we plot $\delta\alpha_G/\alpha_G^0$ [Eq.~(\ref{alpha_R})], as a function of the twist angle $\theta$, for the undoped TBG, at different temperatures and for a fixed FMR energy $\hbar \Omega=0.06\,\mathrm{meV}$ which corresponds to the yttrium iron garnet. The SOC parameters are $\lambda_I=3 \,\mathrm{meV}$ and $\lambda_I=4\, \mathrm{meV}$ as in Ref.~[\onlinecite{Alex}].\\

Figure~\ref{GD2} shows that regardless of the temperature range, $\delta\alpha_G$ increases by decreasing $\theta$ but drops sharply at the MA, where it exhibits a relatively small peak which is smeared out at low temperature.\

\begin{figure*}[htbp] 
\centering
$
\begin{array}{cccc}
\includegraphics[width=0.5\columnwidth]{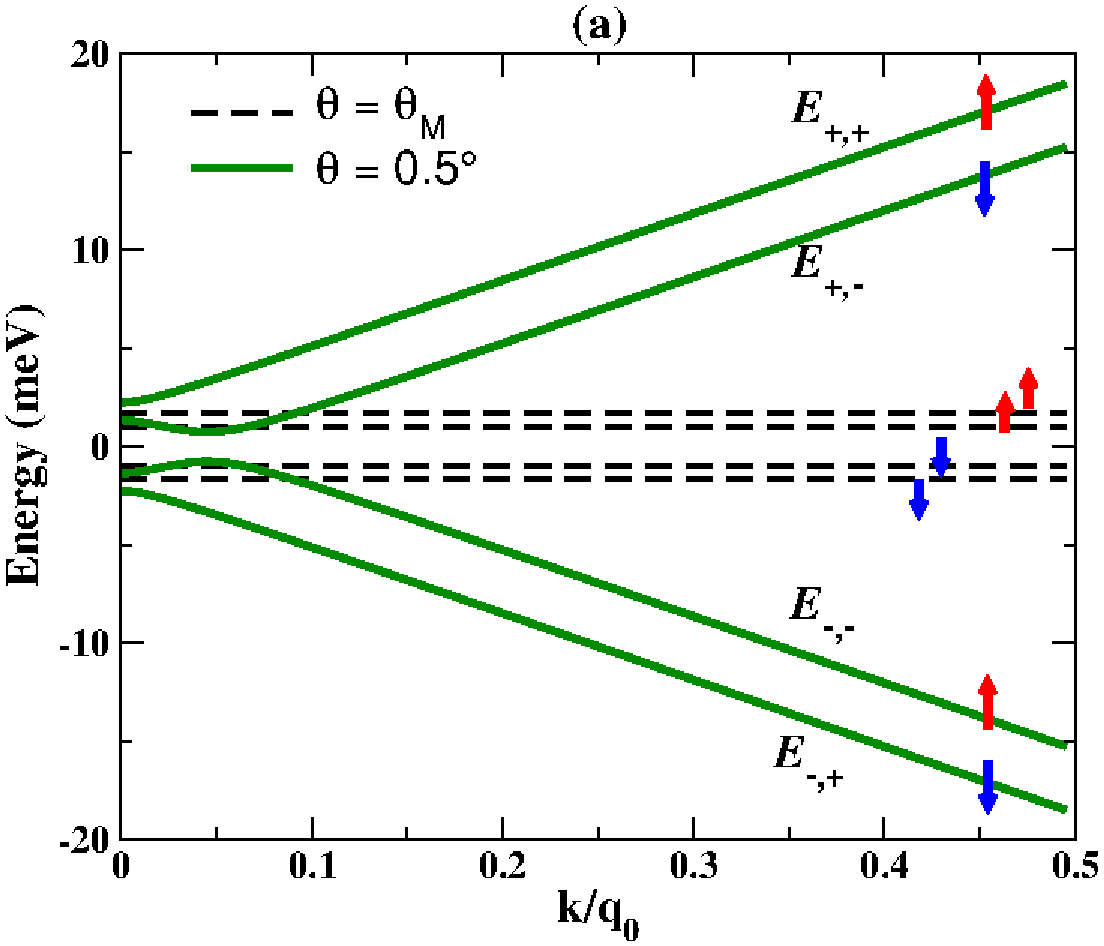}
\includegraphics[width=0.5\columnwidth]{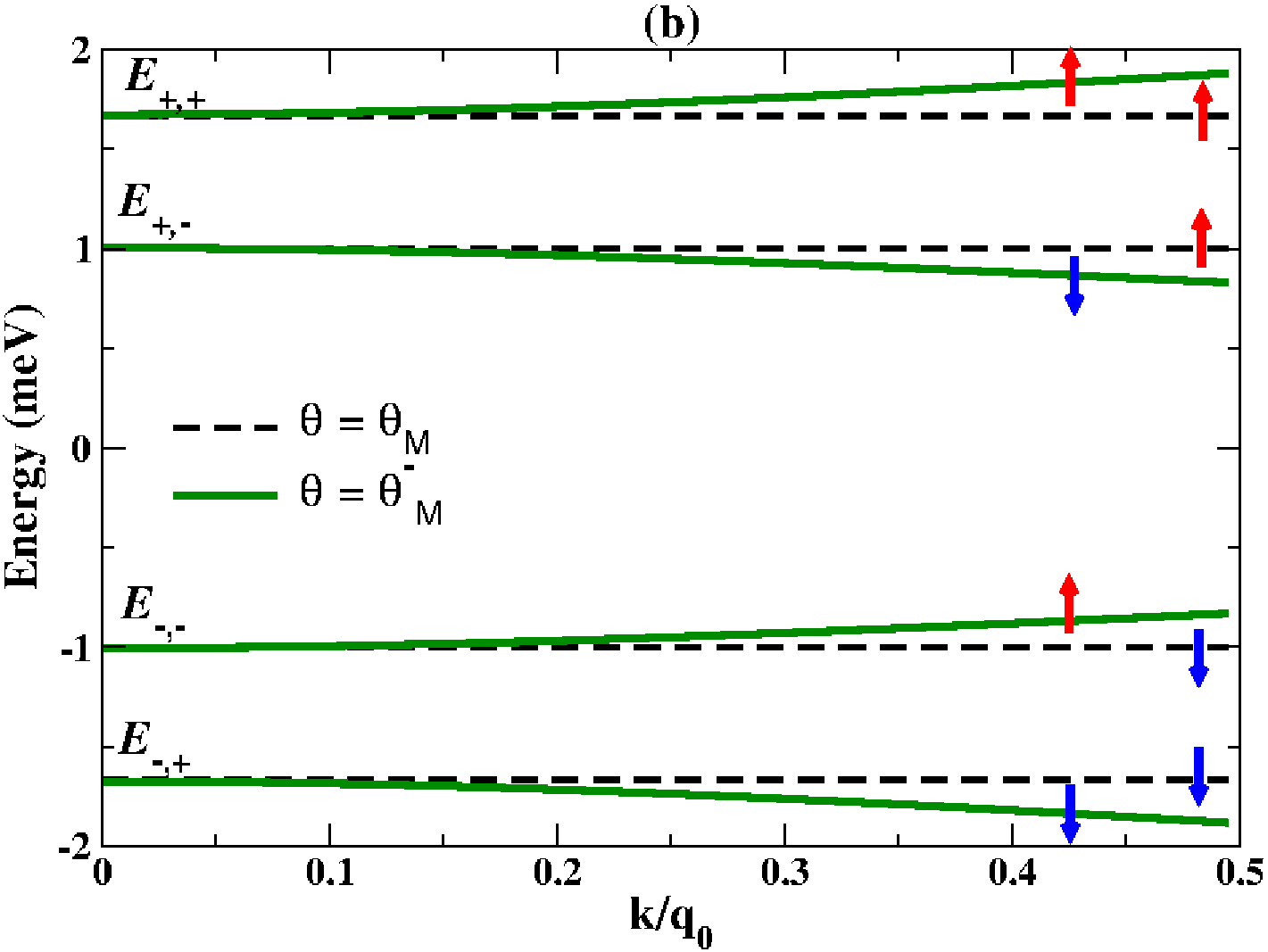}
\includegraphics[width=0.5\columnwidth]{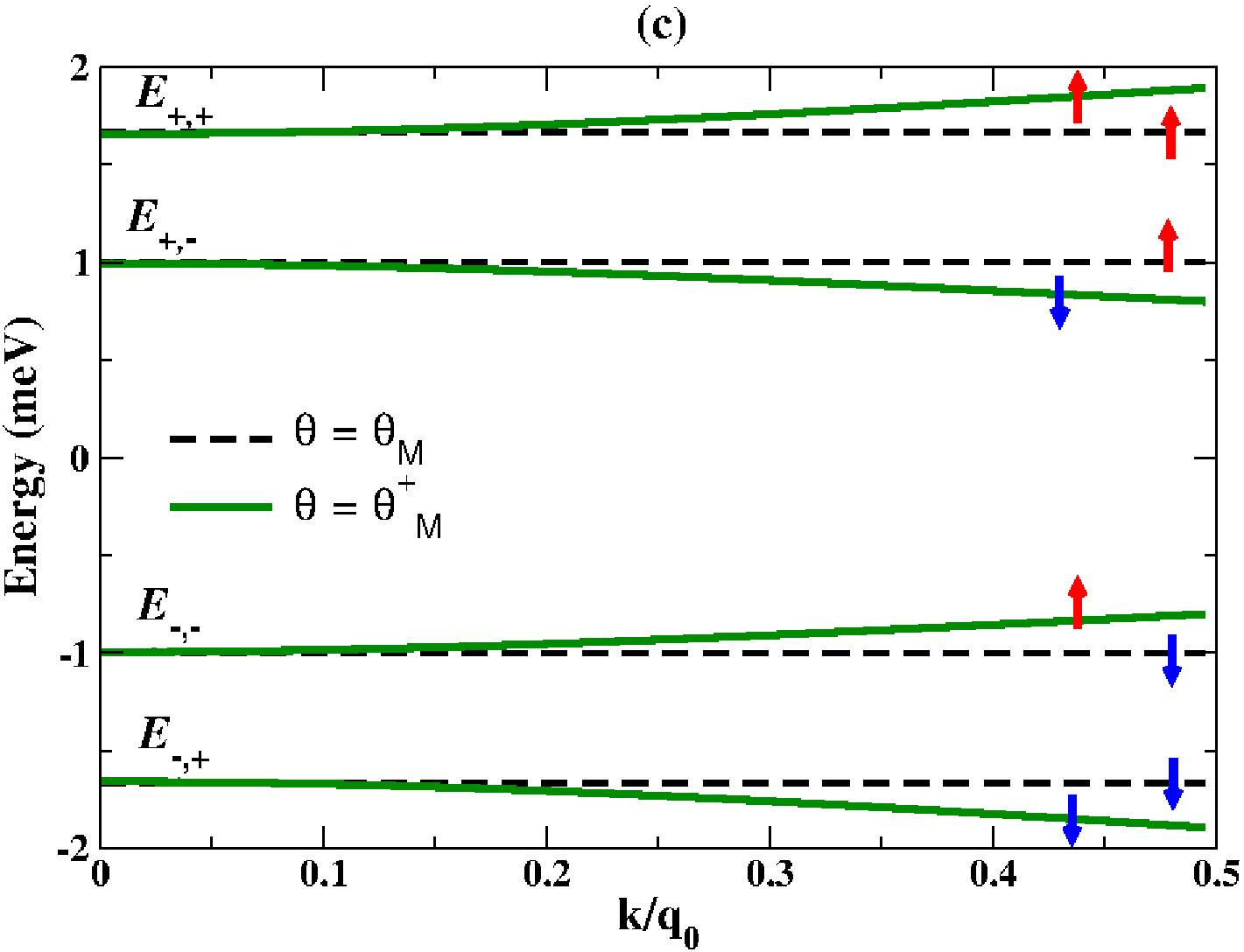}
\includegraphics[width=0.5\columnwidth]{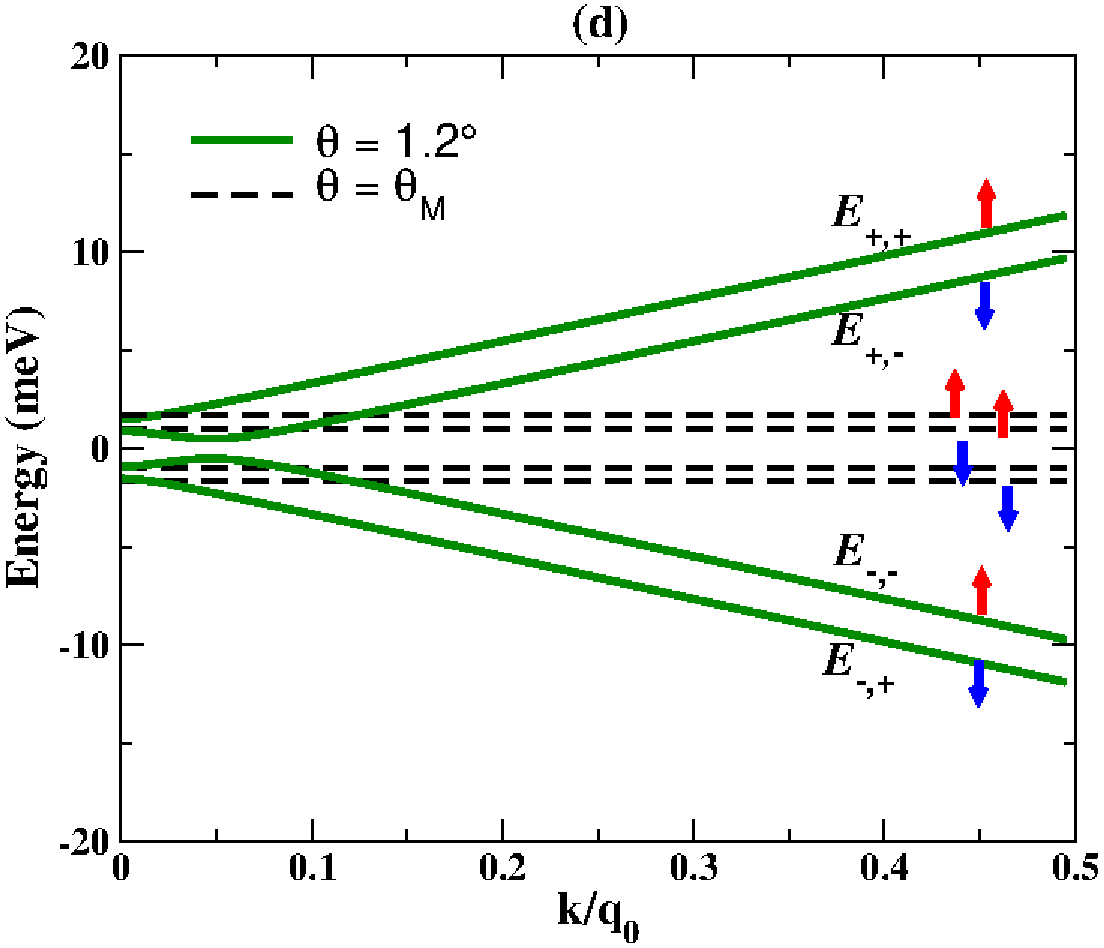}
\end{array}
$
\caption{Band structure of TBG/$\mathrm{WSe_2}$ in the continuum limit [Eq.~(\ref{energy_H1})] at  $\theta=0.5^{\circ}$ (a), $\theta=\theta^-_M=1.043^{\circ}$ (b), $\theta=\theta^+_M=1.058^{\circ}$ (c) and $\theta=1.2^{\circ}$ (d). The dashed lines represent the bands at the MA ($\theta_M=1.05^{\circ}$). The red (blue) arrows correspond to the out-of-plane electronic spin projection $\langle s_z\rangle=+1$ ($\langle s_z\rangle=-1$)~\cite{Alex}. Calculations are done for $\lambda_I=3\;\mathrm{meV}$ and $\lambda_R=4\;\mathrm{meV}$.}
\label{band}
\end{figure*}

Putting aside its drop at the MA, the enhancement of $\delta\alpha_G$, by decreasing $\theta$, can be, in a first step, ascribed to the dependence of the self-energy [Eq.~(\ref{self})] on the effective SOC, given by Eq.~(\ref{SOCeff}), which increase by decreasing $\theta$ . However, to understand the behavior of $\delta\alpha_G$ at the MA one needs to go back to the band structure, $E_{\sigma,\pm}(\mathbf{k})$ [Eq.~(\ref{energy_H1})], of the continuum Hamiltonian of TBG/$\mathrm{WSe_2}$, which is depicted in Fig.~\ref{band} at different twist angles. The arrows indicate the out-of-plane electronic spin projection $\langle s_z\rangle$ which we have numerically calculated for different twist angles in Sec. II of the Supplemental Material~\cite{supp}.\\

Away from the MA, the band dispersion gets larger as $\theta$ decreases and, in particular, the separation between bands with opposite $\langle s_z\rangle$, involved in the SP process, increases. This behavior is due to the angle dependence of the effective Fermi velocity $v^{\ast}$ of TBG/$\mathrm{WSe_2}$, which reduces, in the first order in the SOC, to that of TBG, namely (see Secs. I and II of the supplemental Material~\cite{supp})
\begin{eqnarray}
 v^{\ast}\sim v_F\frac{1-3\alpha^2}{1+6\alpha^2}
 \label{v*}
\end{eqnarray} 
The expression of the GD [Eq.~(\ref{alpha_R})] includes transitions between bands with opposite $\langle s_z\rangle$ (see Sec. IV of the Supplemental Material~\cite{supp}). These transitions depend on the statistical weight $\Delta f(E)=f(E_{\langle s_z\rangle})-f(E_{-\langle s_z\rangle})$ where $f(x)$ is the Fermi-Dirac function and $E_{\langle s_z\rangle}$ is the energy band with a spin orientation $\langle s_z\rangle$. \
%\begin{widetext}
%\end{widetext}

\begin{figure}[htbp] 
\begin{center}
\includegraphics[width=0.8\columnwidth]{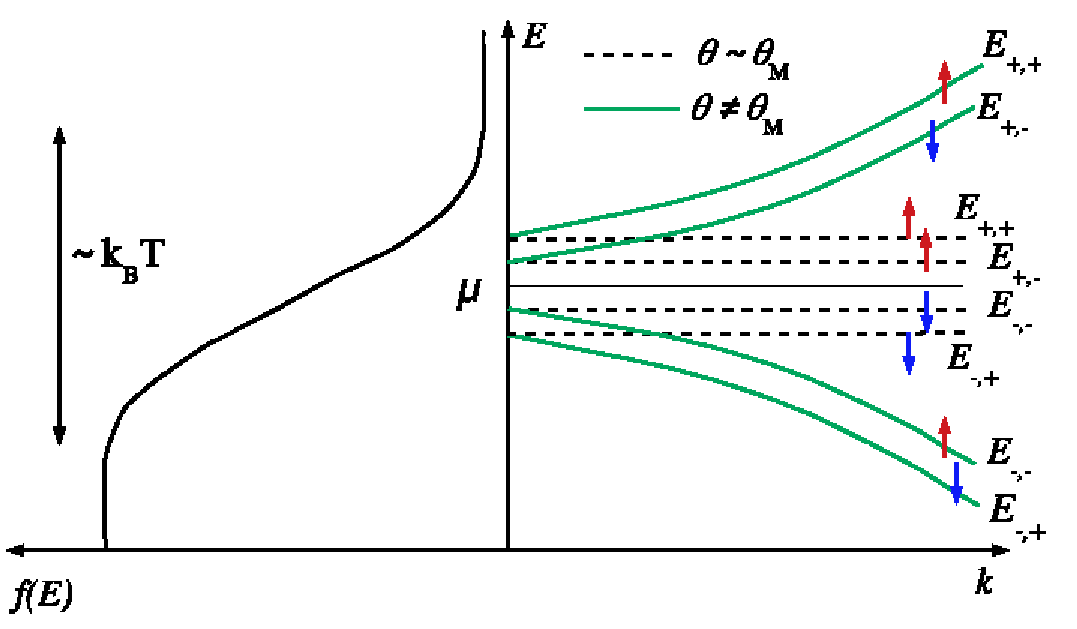}
\end{center}
\caption{Schematic representation of the band structure $E_{\sigma,\pm}$ (Eq.~\ref{energy_H1}) and the Fermi-Dirac distribution $f(E)$. The bands in dashed and green lines correspond, respectively, to the MA and to a twist angle $\theta$ far from the MA. The red (blue) arrows represent the projection of the out-of-plane spin projection $\langle S_z\rangle=+1$ ($\langle S_z\rangle=-1$).
Around the MA, the bands are almost flat and the statistical weights $\Delta f(E)$, corresponding to the transitions between 
$E_{-,+}\rightarrow E_{+,+}$ and $E_{-,-}\rightarrow E_{+,-}$, are small compared to the case of a twist angle away from the MA, where the band dispersion is larger. 
% As a consequence, the correction to the Gilbert damping $\delta\alpha_G$, which depends on $\Delta f(E)$, is reduced at the MA .
}
\label{FD}
\end{figure}

In Fig.~\ref{FD}, we plot a pictorial representation of the band structure of the continuum model [Eq.~(\ref{energy_H1})] and the Fermi-Dirac distribution $f(E)$ at a given temperature $T$. 
The band dispersion gets larger as $\theta$ moves away from the MA (Fig.~\ref{band}) and
the separation between the bands with opposite $\langle S_z\rangle$ increases.
As a consequence, the corresponding statistical weight $\Delta f(E)$ is enhanced compared to the case around the MA. 
This behavior explains the drop of the GD at the MA.\

Around the MA ($\theta^+_M$ and $\theta^-_M$), the statistical weight $\Delta f(E)$ is reduced compared to that at the MA since the bands $E_{+,-}$ and $E_{-,-}$ get closer (Fig.~\ref{band}).\newline
This behavior gives rise to the small peak at the MA (Fig.~\ref{GD2}), which disappears at low temperature ($k_BT<\lambda$) where bands around the MA have the same statistical weight $\Delta f(E)= 1$ (see Sec. IV of the Supplemental Material~\cite{supp}). In this case, the GD is basically dependent on the effective Fermi velocity $v^{\ast}$ [Eq.~(\ref{v*})] which vanishes at exactly the MA. Such dependence is responsible for the cancellation of several terms contributing to the self-energy [Eq.~(\ref{self})], as they are proportional to $v^{\ast}$ [Eq.~(\ref{v*})] (see Sec. IV of the Supplemental Material~\cite{supp}).\newline
Let us now turn to the case of a dirty interface where the spin transfer should now also include the non-conserved momentum processes. The corresponding self-energy [Eq.~(\ref{self})] can also be expressed in terms of the thermal weight $\Delta f(E)$ governing the interband transitions (see Sec. IV of the Supplemental Material~\cite{supp}).\newline
Regarding the flatness of the bands, the dirty processes at the MA acquire, as in the clean limit, small thermal weights compared to the twist angles away from the MA, where the band are dispersive. 
In the dirty limit, the Gilbert damping correction is, then, expected to drop at the MA as found in the case of a clean interface.\

It comes out that the twist angle dependence of $\delta\alpha_G$ is a direct probe of the emergence of the flat bands in TBG. On the other hand, the temperature dependence of the fine structure around the MA provides an accurate measurement of the MA, with a precision below $0.005^{\circ}$ (see Fig.~{\red{S.4}} of the Supplemental Material~\cite{supp}). It also gives an estimation of the SOC induced in TBG adjacent to a monolayer of TMD. \\

%%%%%%%%%%%%%%%%%%%%%%%%%%%%%%%%%%%%%%%%%%%%%
It is worth stressing that in our model we did not take into account the electron-electron interactions which significantly distort the electronic band structure of TBG~\cite{Guinea18,Guinea20,Guinea21,Guinea22}. Near the MA, the dominant electron-electron interaction is found to be the Coulomb interaction with an amplitude estimated to be 10-15 meV~\cite{Guinea20}, which is larger than the width of the flat bands $\sim 2-5$ meV and the SOC considered in the present work. How are the results of Fig.2 modified in the presence of Coulomb interaction?
Treating this interaction within the Hartree-Fock approximation revealed that the Hartree term considerably widens the bands while the exchange term leads, basically, to broken-symmetry phases.\
At the charge neutrality, the Hartree term vanishes and the exchange potential, which concerns bands with identical spins, opens a gap of 4 meV~\cite{Guinea20,Guinea21,Guinea22}, which is of the order of the SOC amplitudes. As a consequence, the statistical weight $\Delta f(E)$ of the bands with opposite $\langle S_z\rangle$ is expected to increase, but keeping larger values at small angles compared to the MA. Moreover, the bandwidth, around the MA, is found to relatively increase under the exchange term~\cite{Guinea20,Guinea21,Guinea22}, but remains smaller than 3 meV, which preserve the flatness of the bands. It comes out, that our results hold in undoped TBG under Coulomb interaction, and can be used to extract the value of the MA at which the Gilbert damping correction drops.
Away from the neutrality, the bands are substantially distorted by the Coulomb interaction~\cite{Guinea20,Guinea21,Guinea22} and our results should be taken with a grain of salt since they account for filling $\nu$ factors away from $-0.5<\nu<0.5$, where the bandwidth, at the MA, is less than 4 meV.\\

Besides interactions, strain is found to be a key parameter in the emergence of flat bands in TBG~\cite{Bi,Marwa}. The effect of strain can be included in our model by deriving the strain induced correction to the Hamiltonian given by Eq.~(\ref{H1}), taking into account the strain dependence of the vectors $\mathbf{q}_j$ connecting the Dirac points~\cite{Marwa}. The twist angle, at which $\delta \alpha_G$ drops, can then provide a way to measure the strain in TBG.

%%%%%%%%%%%%%%%%%%%%%%%%%%

{\it Experimental realization. -- } 
Our proposed setup consists of an interface between a FI and a fully hBN encapsulated TBG/$\mathrm{WSe_2}$ heterostructure (Fig.~\ref{junction}).
The hBN layer acts as a tunnel barrier which prevents the diffusion of the FI atoms into the graphene layer~\cite{SP-Gr}.
On the other hand, the encapsulation provides a clean interface and prevents the graphene degradation~\cite{SP-Gr} which is a challenging issue in the STM and ARPES experiments~\cite{Utama,Efetov21,Sato}, carried out on non-encapsulated TBG samples.\newline
It should be stressed that the hBN encapsulated TBG/$\mathrm{WSe}_2$ heterostructure has been already realized in Refs.~[\onlinecite{Lin,Alex}].
Furthermore, the spin transport through a clean interface between a FI and 2D material has been experimentally achieved~\cite{SP-Gr,SP-TMD}. The 2D materials were fully encapsulated by hBN~\cite{SP-Gr} or covered by a thin layer of an oxide insulator (as MgO)~\cite{SP-TMD} to avoid the  interdiffusion with the FI.\newline
Our proposed technique to measure the MA can, then, be implemented experimentally with a clean interface and at room temperature.
Moreover, an {\it insitu} manipulation of the twist angle can be realized as in Refs.~[\onlinecite{Rebeca,Hu,Geim,Inbar}].\\

%%%%%%%%%%%%%
{\it Conclusion. -- } To conclude, we have proposed an experiment to probe the flat bands of TBG and to measure its MA accurately. The experiment is based on a spin pumping measurement through a junction separating a FI and a TBG adjacent to a monolayer of $\mathrm{WSe_2}$. 
We first derived the continuum model of TBG with SOC, which constitutes a first step to develop an analytical understanding of the emergence of a stable superconducting state at small twist angles observed in TBG in proximity to $\mathrm{WSe_2}$~\cite{Alex}.
We then determined analytically the Gilbert damping correction $\delta\alpha_G$ induced by the presence of the TBG/$\mathrm{WSe_2}$ heterostructure.
Our results show that the twist angle dependence of $\delta\alpha_G$ exhibits a drop at the MA with a temperature-dependent fine structure. This feature provides an accurate determination of the MA and an estimation of the SOC induced in TBG by its proximity to the TMD layer. Our proposed set-up can be readily implemented regarding the state-of-the art of the experimental realizations of SP in 2D materials and TBG-based heterostructure.
Our work opens the gate to a twist tunable spintronics in twisted layered heterostructures.
\\

{\it Acknowledgments. -- } 
 We thank Mamoru Matsuo and Shu Zhang for stimulating discussions. We are indebted to Jean-No\"{e}l Fuchs and  Daniel Varjas for a critical reading of the manuscript.
S. H. acknowledges the kind hospitality of the Institute for Solid State Physics (ISSP) where this work was carried out.  S. H. also thanks the hospitality of the Max Planck Institute for the Physics of Complex Systems (MPI-PKS). S. H. acknowledges financial support from the ISSP International visiting professors program and the MPI-PKS visitors program.

%%%%%%%%%%%%%%%%%%%%%%%%%%%%%%%%%%%%%%%%%%%%%%%%%%%%%%%%%%%%%%%%%%%

%%%%%%%%%%%%%%%%%%%%%%%%%%%%%%%%%%%%%%%
%%%%%     Supplemental Material %%%%%%%%%%%%%%%%%%%
\newpage

%%%%%%%%%%%%%%%%%%%%%%%%%%%%%%%%%%%%%%%%%%%%%%%%%%%%%%%%%%%%%%%%%%%
\widetext
\begin{center}
\textbf{\large \uppercase{Supplemental Material} \quote{Twisted bilayer graphene reveals its flat bands under spin pumping}}
\vspace{0.3cm}

Sonia Haddad$^{1,2,3}$, Takeo Kato$^{2}$, Jihang Zhu$^{3}$, and Lassaad Mandhour$^{1}$\\

$^1$Laboratoire de Physique de la Mati\`ere Condens\'ee, Facult\'e des Sciences de Tunis, Universit\'e Tunis El Manar, Campus Universitaire 1060 Tunis, Tunisia\\
$^2$ Institute for Solid State Physics, University of Tokyo, Kashiwa, Chiba 277-8581, Japan\\
$^3$ Max Planck Institute for the Physics of Complex Systems, N\"{o}thnitzer Strasse 38, Dresden 01187, Germany
\end{center}
\renewcommand{\thefigure}{S.\arabic{figure}}
\renewcommand{\thesection}{S.\arabic{section}}
\renewcommand{\theequation}{S.\arabic{equation}}
\setcounter{figure}{0}
\setcounter{equation}{0}
\setcounter{section}{0}
\section{I. Derivation of the low-energy Hamiltonian of TBG without SOC}
\label{TBG}

We start by a brief overview of the perturbative approach proposed by Bistritzer and MacDonald~\cite{Mc11} to derive the continuum model of TBG. We consider a TBG where the two layers $l=1,2$ are rotated oppositely $\theta_2=-\theta_1=\frac{\theta}2$. The Hamiltonian of a graphene layer $l$ rotated at an angle $\theta_l$ is
\begin{eqnarray}
h_{l}(\mathbf{k})=e^{i\frac{\theta_l}2\sigma_z} h_{l}^{(0)}(\mathbf{k})e^{-i\frac{\theta_l}2\sigma_z}
\label{hl-supp}
\end{eqnarray}
where $h_{l}^{(0)}(\mathbf{k})$ is the Hamiltonian of the unrotated layer $(l)$ given, in the continuum limit, by 
\begin{eqnarray}
h_{l}^{(0)}(\mathbf{k})=-\hbar v_F \mathbf{k}\cdot\mathbf{\sigma}^{\ast},
\label{hgr}
\end{eqnarray}
where the momentum $\mathbf{k}$ is written relatively to the Dirac point $\mathbf{K}_{l,\xi}$,
$v_F$ is the Fermi velocity, $\xi$ is the valley index, $\mathbf{\sigma}^{\ast}=\left(\xi \sigma_x,\sigma_y\right)$ and $\sigma_i$ ($i=x,y,z$) are the sublattice-Pauli matrices.\

The leading contributions of the interlayer tunneling can be limited to three nearest hopping processes in the momentum space connecting states $|\mathbf {k}\rangle_1$, around the Dirac point $\mathbf{K}_{1,\xi}$ of layer $(1)$, to the states $|\mathbf{k+q_{j\xi}}\rangle_2$ around $\mathbf{K}_{2,\xi}$, the Dirac point of layer $(2)$. The $\mathbf{q_{j\xi}}$ vectors are given by~\cite{Mc11}
\begin{eqnarray}
 &&\mathbf{q}_{1\xi}=\xi k_{\theta}\left(0,1\right),\; 
 \mathbf{q}_{2\xi}=\mathbf{q}_{1\xi}+\xi\mathbf{G}^M_1=
 \xi k_{\theta}\left(-\frac{\sqrt{3}}2,-\frac12\right),\nonumber\\
 &&\mathbf{q_3}=\mathbf{q}_{1\xi}+\xi\left(\mathbf{G}^M_1+\mathbf{G}^M_2\right)=
 \xi k_{\theta}\left(\frac{\sqrt{3}}2,-\frac12\right),
 \label{q0}
\end{eqnarray}
where $k_{\theta}=2k_D\sin\frac{\theta}2\sim \theta k_D$ and $k_D=|\mathbf{K}_{1,\xi}|=|\mathbf{K}_{2,\xi}|=\frac {4\pi}{3a}$, $a$ being the graphene lattice parameter. The $\left(\mathbf{G}^M_1,\mathbf{G}^M_2\right)$ is the moir\'e BZ basis given by $\mathbf{G}^M_i=\mathcal{R}_t^T\mathbf{G}_i$, $\mathbf{G}_i$ are the lattice basis vectors of the monolayer reciprocal lattice $\mathbf{G}_1=\frac{2\pi}a \left(1,-1/\sqrt{3}\right)$ and $\mathbf{G}_2=\frac{2\pi}a \left(0,2/\sqrt{3}\right)$.
$\mathcal{R}_t$ is the rotation tensor written, in the sublattice basis, at a small twist angle as
\begin{eqnarray}
R(\theta)= 
\begin{pmatrix}
0 & -\theta \\
\theta & 0 
\end{pmatrix}.
\end{eqnarray}
In the basis $\left\{|\mathbf {k}\rangle_1,|\mathbf{k+q_{j,\xi}}\rangle_2 \right\}$, the Hamiltonian, at the valley $\xi$, reads as~\cite{Mc11}
\begin{eqnarray}
H(\mathbf{k})= 
\begin{pmatrix}
h_1(\mathbf{k}) & T_1 & T_2 & T_3\\
T^{\dagger}_1 & h_{2,1}(\mathbf{k}) & 0&0\\
T^{\dagger}_2 & 0& h_{2,2}(\mathbf{k}) & 0\\
T^{\dagger}_3 & 0& 0& h_{2,3}(\mathbf{k})\\
\end{pmatrix},
\label{HBL0}
\end{eqnarray}
For the relaxed TBG the $T_j$ matrices are given by~\cite{Koshino18}
\begin{eqnarray}
T_1= 
\begin{pmatrix}
w & w^{\prime}\\
w^{\prime} & w^{\prime\prime}\
\end{pmatrix},
T_2= e^{i\xi\mathbf{G}^M_1\cdot\mathbf{r}}
\begin{pmatrix}
w & w^{\prime}e^{-i\xi\Phi}\\
w^{\prime}e^{i\xi\Phi}& w^{\prime\prime}\\
\end{pmatrix},
T_3= e^{i\xi\left(\mathbf{G}^M_1+\mathbf{G}^M_2\right)\cdot\mathbf{r}}
\begin{pmatrix}
w & w^{\prime}e^{i\xi\Phi}\\
w^{\prime}e^{-i\xi\Phi}& w^{\prime\prime}\
\end{pmatrix},
\label{T}
\end{eqnarray}
Here $h_1(\mathbf{k})$ is given by Eq.~\ref{hl-supp} and $h_{2,j}(\mathbf{k})\equiv h_2(\mathbf{k}+\mathbf{q}_{j\xi})=h_1(\mathbf{k}+\mathbf{q}_{j\xi})$, where the momentum $\mathbf{k}$ is written relatively to $\mathbf{K}_{1,\xi}$.
$\Phi=\frac{2\pi}3$ and $\mathbf{r}$ is the shortest inplane shifts between carbon atoms of the two layers~\cite{Bernevig,Falko19}.
Hereafter, we neglect the relative sliding between the layers which is not relevant in the physics of TBG~\cite{Mc11,Bernevig}.
Choosing the A sublattice in both layers ($i=1,2$) as the origin of the unit cell, turns out to take $\mathbf{r}=\mathbf{0}$ in Eq.~\ref{T}~\cite{Bernevig}.
The parameters $w$, $w^{\prime}$ and $w^{\prime\prime}$ are the tunneling amplitudes which take the same value $w=w^{\prime}=w^{\prime\prime}\sim 110\; \mathrm{meV}$ in the rigid TBG~\cite{Bi}.\newline
In the relaxed lattice, these amplitudes are no more equal $w\sim w^{\prime\prime} \sim 90$ meV and $w^{\prime}=117$ meV~\cite{Koshino18}.
In the present work,  we do not consider the lattice relaxation effect, since the SOC parameters $\lambda_I,\; \lambda_R\sim 4\, \mathrm{meV} $ are small compared to the difference between the interlayer amplitudes $\Delta w=w-w^{\prime}\sim 20\, \mathrm{meV} $.
In the unrelaxed lattice, the $T_j$ matrices can be written as
\begin{eqnarray}
 T_1= w\left(\mathbb{I}_{\sigma}+\sigma_x\right),
 T_2=w \left(\mathbb{I}_{\sigma}-\frac 12\sigma_x+\xi \frac {\sqrt{3}}2\sigma_y\right),
 T_3=w\left( \mathbb{I}_{\sigma}-\frac 12\sigma_x-\xi\frac {\sqrt{3}}2\sigma_y\right).
 \label{Tj}
\end{eqnarray}
Here, $\mathbb{I}_{\sigma}$ is the identity matrix acting on the sublattice indices.\

Considering, in Eq.~\ref{HBL0}, the $\mathbf{k}$ dependent term  as a perturbation, the effective Hamiltonian can be written, to the leading order in $\mathbf{k}$, as
\begin{eqnarray}
 H^{(1)}\left(\mathbf{k}\right)=\frac{\langle\Psi|H(\mathbf{k})|\Psi\rangle}{\langle\Psi|\Psi\rangle},
 \label{H10}
\end{eqnarray}
where $\Psi=\left(\psi_0(\mathbf{k}),\psi_1(\mathbf{k}),\psi_2(\mathbf{k}),\psi_3(\mathbf{k})\right)$ is the zero energy eigenstate of $H\left(\mathbf{k}=\mathbf{0}\right)$.  $\Psi$ is constructed on the two-component sublattice spinor $\psi_0(\mathbf{k})$ ($\psi_j(\mathbf{k})$) of layer $1$ (layer $2$) taken at the momentum $\mathbf{k}$ ($\mathbf{k}+\mathbf{q}_{j\xi}$) around the Dirac point $\mathbf{K}_{1,\xi}$ ($\mathbf{K}_{2,\xi}$) at the valley $\xi$. $\psi_0$ is the zero energy eigenstate of $h_1$.
The $\Psi$ components satisfy
\begin{eqnarray}
 h_1\psi_0+\sum_j T_j\psi_j=0,\,\mathrm{and}\; T^{\dagger}_j\psi_0+h_j\psi_j=0, \;\mathrm{with}\; h_1\psi_0=0,
\end{eqnarray}
where $h_j\equiv h_2(\mathbf{q}_{j,\xi})$. Then
\begin{eqnarray}
 \psi_j=-h^{-1}_j T^{\dagger}_j\psi_0, \mathrm{and}\; \sum_jT_jh^{-1}_j T^{\dagger}_j=0.
 \label{psij}
\end{eqnarray}
To the leading order in $\mathbf{k}$, $H^{(1)}\left(\mathbf{k}\right)$, takes the following form
\begin{eqnarray}
 H^{(1)}\left(\mathbf{k}\right)=\frac{\langle\Psi|H(\mathbf{k})|\Psi\rangle}{\langle\Psi|\Psi\rangle}=\frac1{\langle\Psi|\Psi\rangle}
 \left[\psi^{\dagger}_0h_1\left(\mathbf{k}\right) \psi_0+ \psi^{\dagger}_0\sum_jT_j h^{-1}_j h_j\left(\mathbf{k}\right)h^{-1}_j T^{\dagger}_j \psi_0\right],
\label{psiHpsi}
\end{eqnarray}
\begin{eqnarray}
 H^{(1)}\left(\mathbf{k}\right)=-\hbar  v^{\ast} \psi^{\dagger}_0\, \mathbf{k}\cdot\mathbf{\sigma^{\ast}}\,\psi_0,
 \label{H1eff}
\end{eqnarray}
Eq.~\ref{H1eff} is obtained by neglecting $\theta$ in $h_j\left(\mathbf{k}\right)$, which turns out to take $\mathbf{q}_{j,\xi}=\mathbf{0}$ in $h_j\left(\mathbf{k}\right)$~\cite{Mc11}.\newline
$v^{\ast}$ is the effective velocity of the energy band of TBG around the zero energy which vanishes at the MA $\theta_m$, and is given by~\cite{Mc11}
\begin{eqnarray}
 v^{\ast}=v_F\frac{1-3\alpha^2}{1+6\alpha^2}
\end{eqnarray}
where $\alpha=\frac{w}{\hbar v_F q_{0}}$, $q_{0}=|\mathbf{q}_{j\xi}|\sim \frac{4\pi}{3a}\theta$. \newline
In our numerical calculations (Fig.~{\red3} of the main text), we take  $w=118\; \mathrm{meV}$ and $\hbar v_F/a\sim 2.68\; \mathrm{eV}$ which corresponds to $\theta_m=1.05^{\circ}$ for the first MA~\cite{Mc11}.

%%%%%%%%%%%%%%%%%%%%%%%%%%%%%%

\section{II. Derivation of the low-energy Hamiltonian of TBG with SOC}
\label{TBG-SOC}
We now consider the heterostructure consisting of TBG adjacent to a monolayer of $\mathrm{WSe_2}$ as shown in Fig.1 of the main text, where we denote the graphene layer in contact with the TMD by layer (2). This layer is subject to a SOC induced by proximity effect by the TMD, and the corresponding Hamiltonian can be written as~\cite{Alex} 
\begin{eqnarray}
h_{2}(\mathbf{k})=h_{1}(\mathbf{k})+h_{\text{SOC}}+\frac{m}2\sigma_z
\label{h2-supp}
\end{eqnarray}
where $h_{\text{SOC}}$ is given by Eq.~{\red1} of the main text.\\
 
To derive the continuum model of TBG/$\mathrm{WSe}_2$, we follow the perturbative approach of Ref.~\cite{Mc11} presented in the previous section. Now, the basis $\Psi=\left(\psi_0(\mathbf{k}),\psi_1(\mathbf{k}),\psi_2(\mathbf{k}),\psi_3(\mathbf{k})\right)$ is constructed on the four-component spin-sublattice spinor $\psi_0(\mathbf{k})$ and $\psi_j(\mathbf{k})$, ($j=1,2,3$) corresponding, respectively, to layer $(1)$ and layer $(2)$. 
$\psi_0(\mathbf{k})$ is written as 
$\psi_0(\mathbf{k})^T=\left(\psi_{0,A\uparrow},\psi_{0,A\downarrow},\psi_{0,B\uparrow},\psi_{0,B\downarrow}\right)$. In this basis, the Hamiltonian of TBG/$\mathrm{WSe}_2$ takes the form
\begin{eqnarray}
H_{\xi,\text{SOC}}(\mathbf{k})= 
\begin{pmatrix}
h_{1}(\mathbf{k}) & T_1 & T_2 & T_3\\
T^{\dagger}_1 & h_{2,1}(\mathbf{k}) & 0&0\\
T^{\dagger}_2 & 0& h_{2,2}(\mathbf{k}) & 0\\
T^{\dagger}_3 & 0& 0& h_{2,3}(\mathbf{k})\\
\end{pmatrix}.
\label{HTBG-supp}
\end{eqnarray}
The momentum $\mathbf{k}$ is measured relatively to the Dirac point $\mathbf{K}_{1\xi}$ of layer (1), $h_{2,j}\left(\mathbf{k}\right)=h_2\left(\mathbf{k}+\mathbf{q}_{j\xi}\right)$, ($j=1,2,3$) and $h_2\left(\mathbf{k}\right)$ includes now the SOC terms (Eq.~\ref{h2-supp}).\newline
We take the sublattice A as the origin of the unit cell in each layer. The $T_j$ matrices are written as the tensor product of those given by Eq.~\ref{Tj}, with the $2\times2$ identity spin-matrix $\mathbb{I}_s$.\\

Regarding the small values of the SOC, we assume that $H_{\xi,\text{SOC}}(\mathbf{k})$ has a zero eigenenergy and the corresponding eigenstate $\Psi$ satisfies the condition given by Eq.~\ref{psij}.\

Following the same procedure as in the previous section, we derive from Eq.~\ref{psiHpsi} the effective low energy Hamiltonian $H_{\xi,\text{SOC}}^{(1)}(\mathbf{k})$ of TBG/$\mathrm{WSe_2}$ by substituting $h_j\left(\mathbf{k}\right)$ by the Hamiltonian of layer (2), rotated at $\theta/2$ and including SOC as 
\begin{eqnarray}
 h_{\text{2,rot}}(\mathbf{k})=-\hbar \mathbf{k}\cdot\mathbf{\sigma}^{\ast}\mathbb{I}_s +\frac{\lambda_I}2\xi s_z+\frac{\lambda_R}2\left(\xi \sigma_x s_y-\sigma_y s_x\right)+\frac{\lambda_{\text{KM}}}2\xi \sigma_z s_z-\frac{\lambda_R}2\theta\left(\xi \sigma_y s_y+\sigma_x s_x\right),
\label{h2rot}
\end{eqnarray}
Hereafter, we neglect the Kane and Mele term whose contribution, to the leading order in $\mathbf{k}$, is found to vanish. We also disregard the last term in Eq.~\ref{h2rot}, which results into a higher order correction in $\theta$.\\

To the first order in the SOC coupling, we obtain the continuum model of TBG/$\mathrm{WSe_2}$ described by the Hamiltonian $H_{\xi,\text{SOC}}^{(1)}(\mathbf{k})$ given by Eq.~{\red3} in the main text.
This Hamiltonian contain a SOC term ($h_{\text{eff}}^{\text{SOC}}$) with renormalized Ising and Rashba interactions
\begin{eqnarray}
 \tilde{\lambda}_I=\frac{6\alpha^2}{\langle\Psi|\Psi\rangle}\lambda_I,\;
 \tilde{\lambda}_R=\frac{3\alpha^2}{\langle\Psi|\Psi\rangle}\lambda_R, \; \mathrm{and}\;
 \langle\Psi|\Psi\rangle\sim 1+ 6\alpha^2
 \label{SOCeff-supp}
\end{eqnarray}
$\tilde{\lambda}_I$ and $\tilde{\lambda}_R$ are enhanced by decreasing the twist angle from the MA.\\

\begin{figure}[hpbt] 
\begin{center}
\includegraphics[width=0.5\columnwidth]{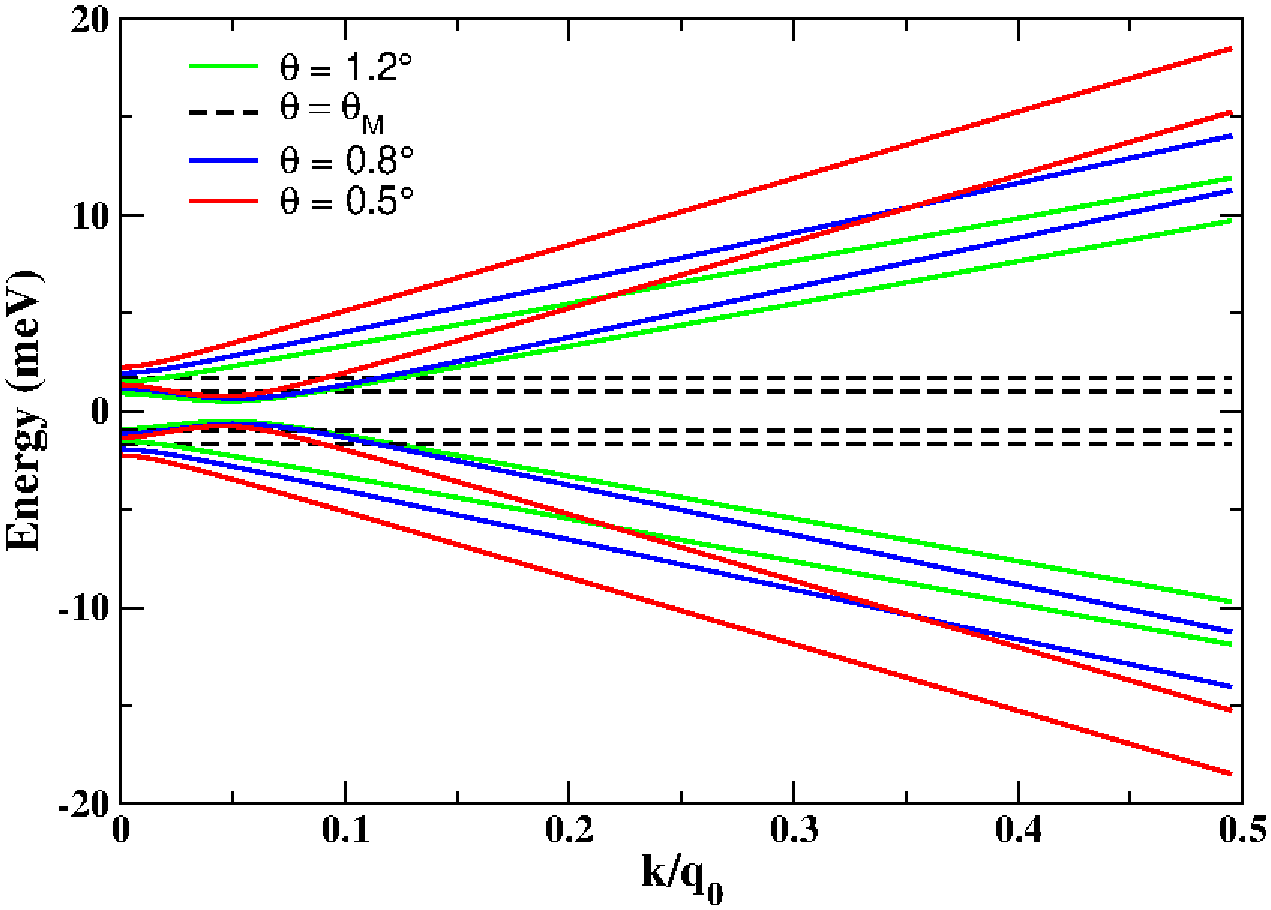}
\end{center}
\caption{Energy bands of TBG/$\mathrm{WSe}_2$ around zero energy as function of the dimensionless momentum amplitude $k/q_0$ at different twist angles. The bands are represented up to the cutoff $k_c=q_0/2$. Calculations are done for $\lambda_I=3\;\mathrm{meV}$ and $\lambda_R=4\;\mathrm{meV}$~\cite{Alex}. The MA is $\theta_M=1.05^{\circ}$. }
\label{energy}
\end{figure}

To the leading order in $\mathbf{k}$, the four eigenergies of the Hamiltonian $H_{\xi,\text{SOC}}^{(1)}$ (Eq.~{\red3} of the main text), denoted $E(\mathbf{k})_{\sigma,\pm}$, are given by
\begin{eqnarray}
E(\mathbf{k})_{\sigma,\pm}= \frac {\sigma}{\langle \Psi |\Psi\rangle }
\sqrt{f_1(\mathbf{k})\pm 6 \alpha^2\sqrt{f_2(\mathbf{k})}},
\label{band-supp}
\end{eqnarray}
\begin{eqnarray}
&&f_1(\mathbf{k})=(\hbar v_F)^2\left(1-3\alpha^2\right)^2||{\mathbf{k}}||^2+\frac 92\alpha^4\left(2\lambda_I^2+\lambda_R^2\right),\nonumber\\
&&f_2(\mathbf{k})= (\hbar v_F)^2\left(1-3\alpha^2\right)^2||{\mathbf{k}}||^2\left(\lambda_I^2+\frac 14\lambda_R^2\right)+ \frac9{16} \alpha^4 \lambda_R^4,
\end{eqnarray}
where $\sigma=\pm$ is the band index. $E(\mathbf{k})_{\sigma,\pm}$ are depicted in Fig.~\ref{energy} at different twist angles.\newline

At the Dirac point, the eigenergies reduce to
\begin{eqnarray}
E_{\pm,+}=\frac  {\pm 3\alpha^2}{\langle \Psi |\Psi\rangle } \sqrt{\lambda_I^2+\lambda_R^2}\;,
E_{\pm,-}=\frac  {\pm 3\alpha^2}{\langle \Psi |\Psi\rangle } \lambda_I.
\end{eqnarray}
It is worth to note that in TBG, the flatness of the bands around the charge neutrality point strongly depends on the heterostrain which may emerge in the graphene layers during the fabrication procedure~\cite{Marwa}. The interplay between strain and SOC in TBG/$\mathrm{WSe}_2$ goes beyond the scope of the present work.\\

\begin{figure*}[hpbt] 
\centering
$
\begin{array}{ccc}
\mathbf{(a)}&\mathbf{(b)}&\mathbf{(c)}\\
\includegraphics[width=0.33\columnwidth]{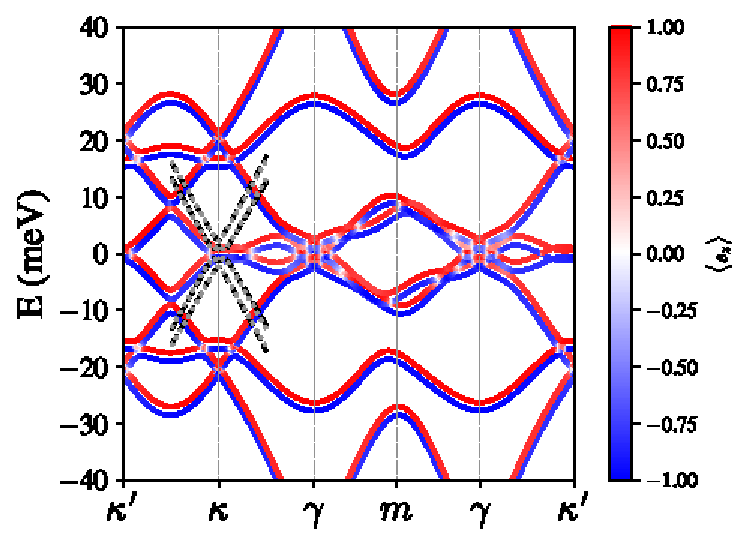}&
\includegraphics[width=0.33\columnwidth]{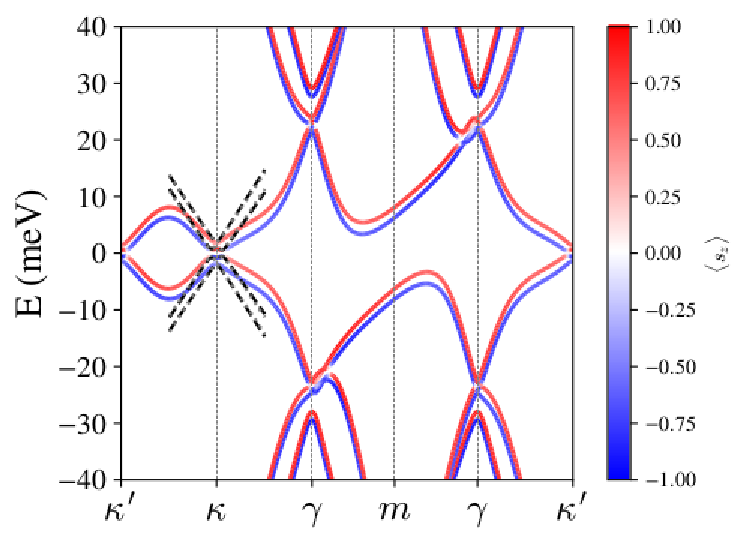}&
\includegraphics[width=0.33\columnwidth]{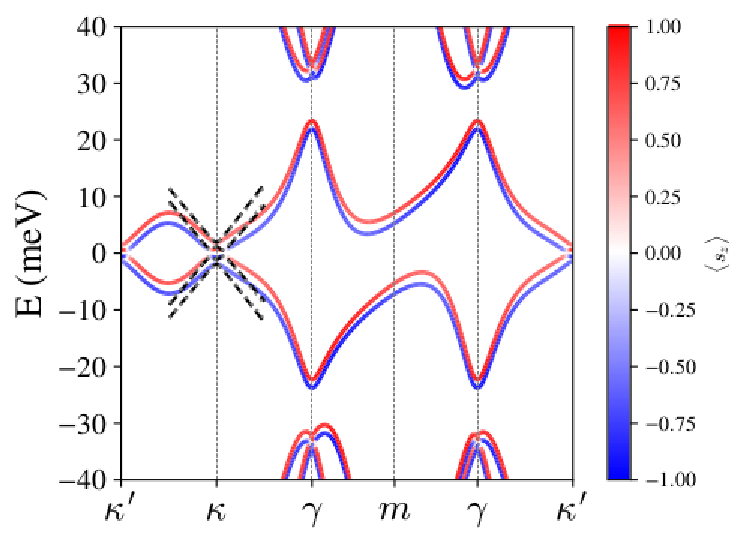}\\
\includegraphics[width=0.33\columnwidth]{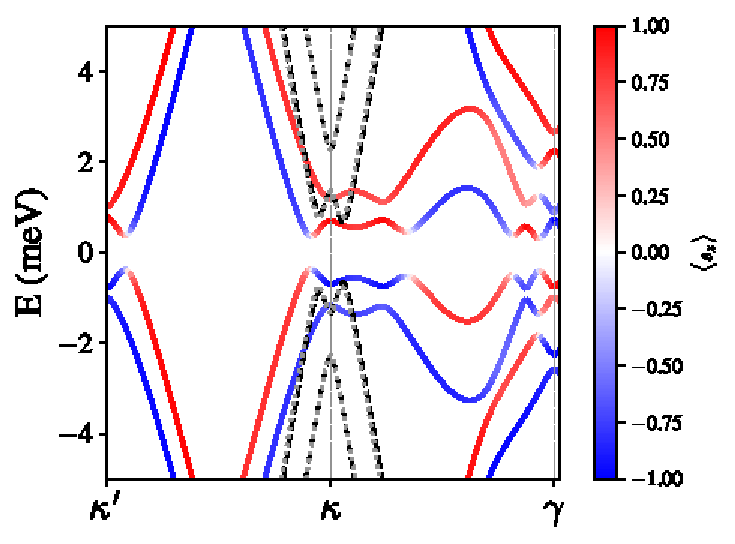}&
\includegraphics[width=0.33\columnwidth]{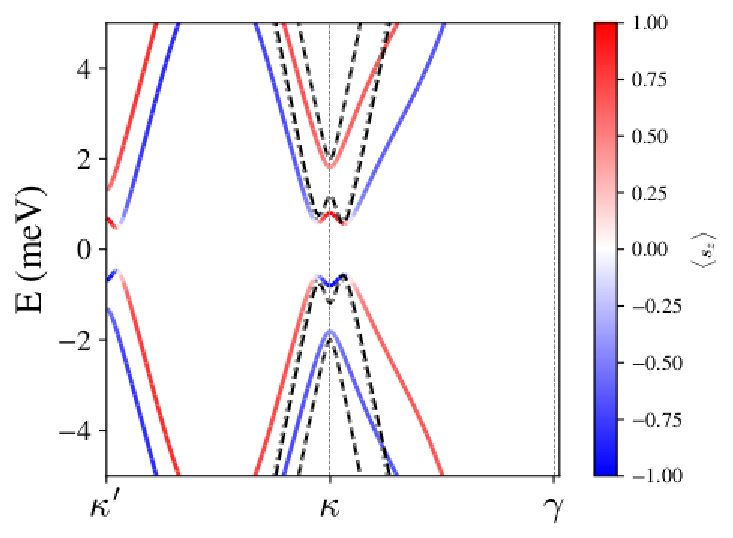}&
\includegraphics[width=0.33\columnwidth]{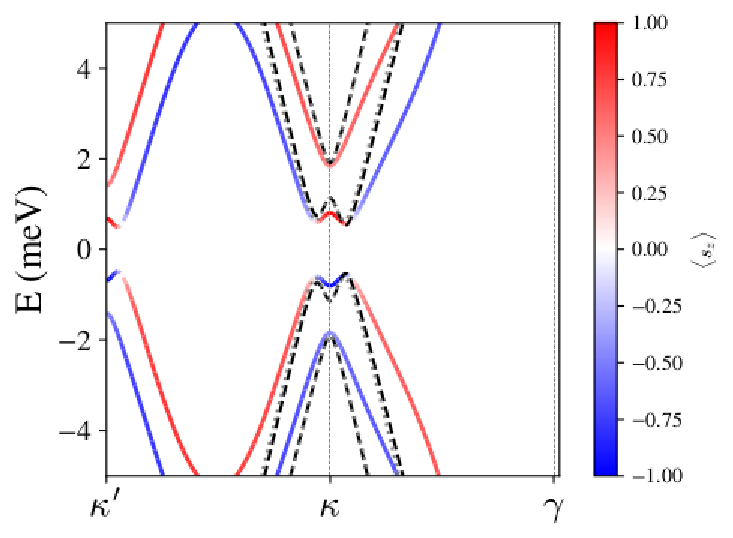}
\end{array}
$
\caption{Electronic band structure of TBG/$\mathrm{WSe_2}$ calculated, at a twist angle (a) $\theta=0.5^{\circ}$, (b) $\theta=0.79^{\circ}$, and (c) $\theta=0.87^{\circ}$. Calculations are based on the continuum model and including 148 bands for each moir\'e valley and spin. The line color denotes the value of the out-of-plane spin projection $\langle S_z\rangle$.
The dashed black lines represent the eigenergies of the four-band effective Hamiltonian given by Eq.~{\red 3} of the main text, and the gray dotted lines denote the approximated eigenergies given by Eq.~\ref{band-supp}. Calculations are done for $\lambda_R=4\, \mathrm{meV}$ and $\lambda_I=3\,\mathrm{meV}$. The band structure is represented in the moir\'e Brillouin zone where $\kappa$ and $\kappa^{\prime}$ correspond respectively to the Dirac point $\mathbf{K}_{1}$ of layer (1) and $\mathbf{K}_2$ layer (2) at the valley $\xi=+$. The bottom panel is a zoomed-in representation around the high symmetry point $\kappa$.}
\label{band-num}
\end{figure*}

\begin{figure*}[hpbt] 
\centering
$
\begin{array}{ccc}
\mathbf{(a)}&\mathbf{(b)}&\mathbf{(c)}\\
\includegraphics[width=0.33\columnwidth]{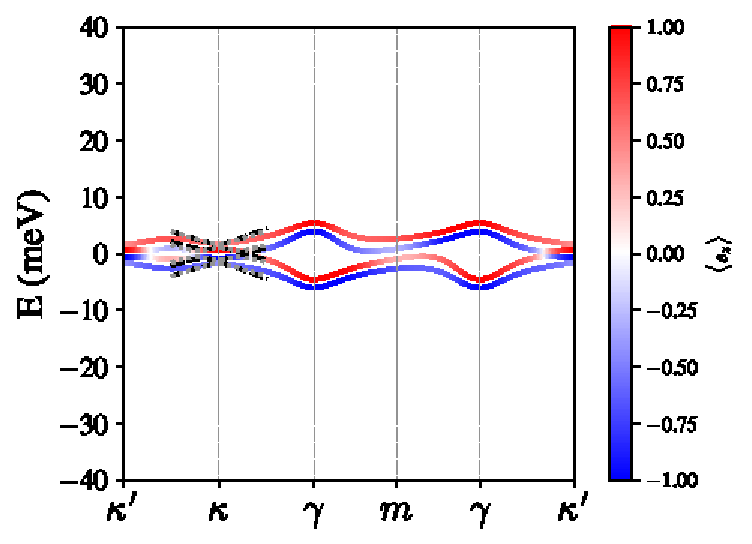}&
\includegraphics[width=0.33\columnwidth]{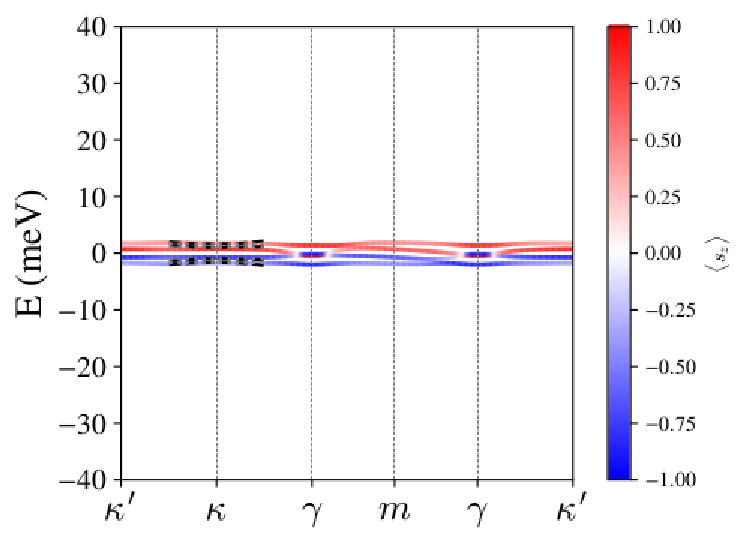}&
\includegraphics[width=0.33\columnwidth]{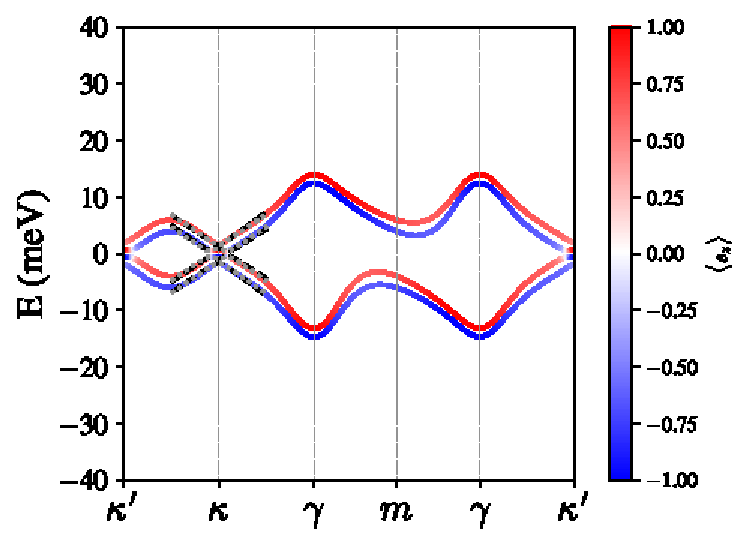}\\
\includegraphics[width=0.33\columnwidth]{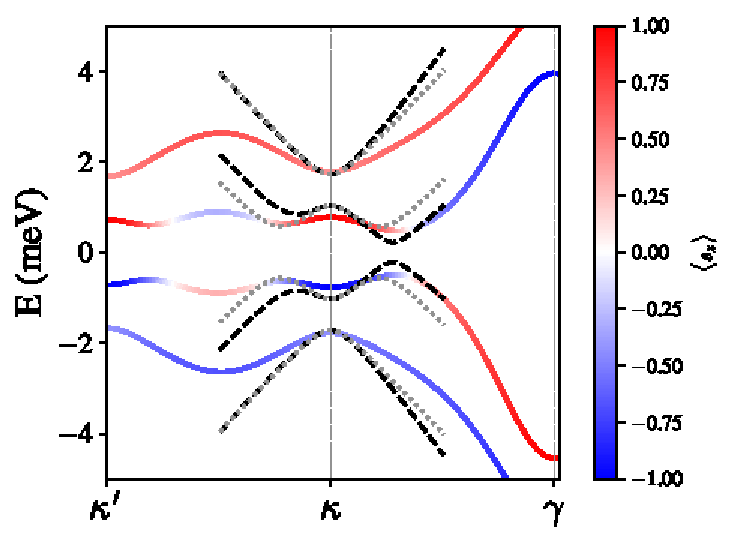}&
\includegraphics[width=0.33\columnwidth]{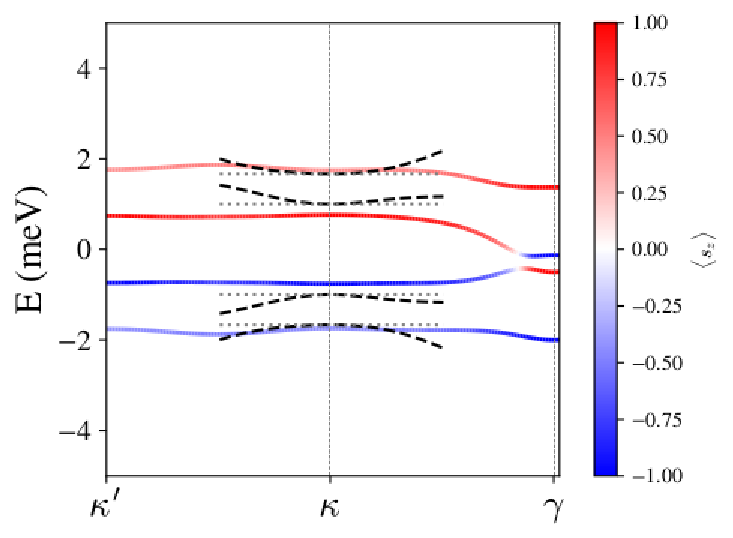}&
\includegraphics[width=0.33\columnwidth]{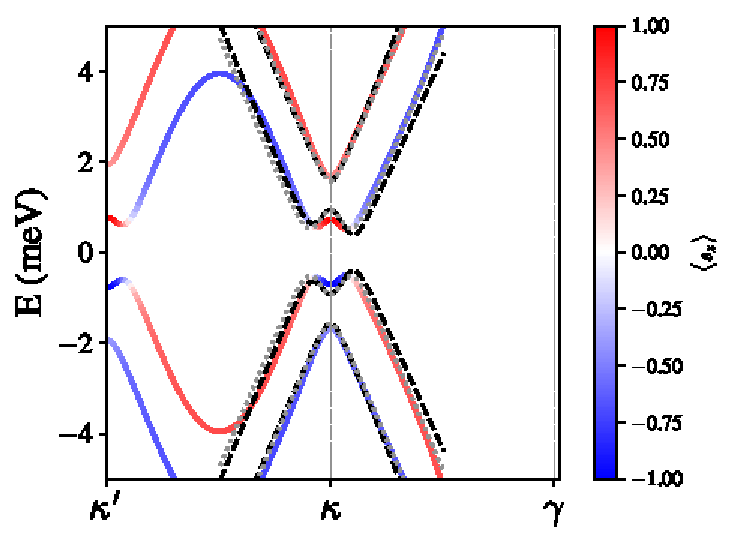}
\end{array}
$
\caption{Electronic band structure of TBG/$\mathrm{WSe_2}$ calculated around the MA $\theta_M$ at (a) $\theta=1.05^{\circ}$, (b) $\theta_M=1.1^{\circ}$, (c) $\theta=1.2^{\circ}$. The bottom panel is a zoomed-in representation around the high symmetry point $\kappa$. 
The data are the same as in Fig.~\ref{band-num}.}
\label{band-num2}
\end{figure*}

In figures~\ref{band-num} and ~\ref{band-num2}, we plot the electronic band structure of TBG/$\mathrm{WSe_2}$ at different twist angles. The solid lines correspond to the numerical results obtained within the continuum model taking into account 148 bands for each spin and valley. The calculations are done for the relaxed TBG with interlayer momentum hopping amplitudes $w=55\,\mathrm{meV}$ and $w=105\,\mathrm{meV}$ (Eq.~\ref{T}). We considered these values to reproduce the numerical band structure obtained in Ref.[~\onlinecite{Alex}] using the continuum model. We have taken into account the lattice relaxation, as in Ref.~\onlinecite{Alex}, since it is expected to be important at small angles.
The numerical results are compared to the eigenergies of the effective Hamiltonian $H_{\xi,\text{SOC}}^{(1)}$, given by Eq.~{\red 3} of the main text (dashed line), and to the approximated expressions (Eq.~\ref{band-supp}) represented by dotted gray lines. It should be stressed that $H_{\xi,\text{SOC}}^{(1)}$ is derived for a rigid TBG, for which we have taken an interlayer hopping amplitude $w=105\,\mathrm{meV}$, which gives rise to a MA $\theta_M\sim 1.1^{\circ}$ as in Ref~[\onlinecite{Alex}]. \newline
For the sake of simplicity, we did not consider the relaxation effect, which is not significant around the MA~\cite{Alex}, where we consider the spin pumping effect. The derivation of the effective Hamiltonian of the relaxed TBG adjacent to $\mathrm{WSe_2}$ is left to a future work.\newline

As shown by Fig.~\ref{band-num} and ~\ref{band-num2}, the effective Hamiltonian $H_{\xi,\text{SOC}}^{(1)}$ (Eq.~{\red 3} of the main text) provides a good description of the band structure of TBG/$\mathrm{WSe_2}$. It can be taken as a framework to unveil the origin of the observed stable superconducting state in this heterostructure~\cite{Alex}.
However, at relatively small angles ($\sim 0.5^{\circ}$), the Fermi velocities of $H_{\xi,\text{SOC}}^{(1)}$ are overestimated (Fig.~\ref{band-num}). This discrepancy is due to the assumption of a rigid TBG lattice which is not justified at small angles~\cite{Alex}.    

\section{III. Electronic Green function}\label{Green-electronic}

The Matsubara Green function associated to the effective Hamiltonian $H_{\xi,\text{SOC}}^{(1)}$ (Eq.~{\red5}) is

\begin{eqnarray}
\hat{g}(\mathbf{k},i\omega_n)=\left[ i\omega_n \mathbb{I}_{S}\mathbb{I}_{\sigma} - H_{\xi,\text{SOC}}(\mathbf{k})\right]^{-1},
\end{eqnarray}
where $\mathbb{I}_{S}$ and $\mathbb{I}_{\sigma}$ are the $2\times 2$ spin and band identity matrices, respectively.
$\hat{g}(\mathbf{k},i\omega_n)$ can be expressed as
\begin{eqnarray}
\hat{g}(\mathbf{k},i\omega_n)= \hat{g}_0(\mathbf{k},i\omega_n)\mathbb{I}_{s}+ \mathbf{\hat{g}}\cdot\mathbf{s}.
\label{green-elec}
\end{eqnarray}
$\mathbf{s}=(s_x,s_y,s_z)$ are spin-Pauli matrices, $\hat{g}_0(\mathbf{k},i\omega_n)$ and the components $\hat{g}_i$ ($i=x,y,z$) of $\mathbf{\hat{g}}$ are written as
\begin{eqnarray}
&&\hat{g}_0(\mathbf{k},i\omega_n)=A_0(\mathbf{k},i\omega_n)+C_0(\mathbf{k},i\omega_n),\,
\hat{g}_x(\mathbf{k},i\omega_n)=B_x(\mathbf{k},i\omega_n)+D_x(\mathbf{k},i\omega_n),\,\nonumber\\
&&g_y(\mathbf{k},i\omega_n)=B_y(\mathbf{k},i\omega_n)+D_y(\mathbf{k},i\omega_n),\,
\hat{g}_z(\mathbf{k},i\omega_n)=A_z(\mathbf{k},i\omega_n)+C_z(\mathbf{k},i\omega_n).
\label{gg}
\end{eqnarray}

The $A,B,C$ and $D$ operators are written in terms of the band-Pauli matrices $\sigma_{x,y,z}$ and the corresponding identity matrix $\mathbb{I}_{\sigma}$
\begin{eqnarray}
A_i(\mathbf{k},i\omega_n)=A_{i1}(\mathbf{k},i\omega_n) \mathbb{I}_{\sigma}+A_{iz}(\mathbf{k},i\omega_n)\sigma_z  \nonumber\\
C_{i}(\mathbf{k},i\omega_n)=C_{ix}(\mathbf{k},i\omega_n) \sigma_x +C_{iy}(\mathbf{k},i\omega_n)\sigma_y  \nonumber\\
B_{j}(\mathbf{k},i\omega_n)=B_{j1}(\mathbf{k},i\omega_n) \mathbb{I}_{\sigma}+B_{jz}(\mathbf{k},i\omega_n)\sigma_z  \nonumber\\
D_j(\mathbf{k},i\omega_n)=D_{jx}(\mathbf{k},i\omega_n) \sigma_x +D_{jy}(\mathbf{k},i\omega_n)\sigma_y
\label{abcd}
\end{eqnarray}
here $j=x,y$ and $i=0,z$. \\

In the limit of small SOC couplings $\lambda_R,\lambda_I \ll \hbar v_F q_0$, A, B, C, and D become
%\textcolor{red} {in the A,B,C D expressions change the $\lambda$ to $\lambda/2$}
\begin{eqnarray}
A_{01}(\mathbf{k},i\omega_m)&&= \frac {Y_{00}-E_1^2}{2(E_3^2-E_1^2)} 
\left[\frac 1 {i\hbar \omega_m-E_1}+\frac 1 {i\hbar \omega_m-E_2}\right]
+\frac {E_3^2-Y_{00}}{2(E_3^2-E_1^2)} 
\left[\frac 1 {i\hbar \omega_m-E_3}+\frac 1 {i\hbar \omega_m-E_4}\right]
\label{A01}\\
&& Y_{00}= \frac{(\hbar v_F)^2 (1-3\alpha^2)^2}{\langle\Psi|\Psi\rangle^2}||\mathbf{k}||^2
+\frac 12\left(\frac{3\alpha^2\lambda_R} {\langle\Psi|\Psi\rangle}\right)^2+
\left(\frac {3\alpha^2 \lambda_I}{\langle\Psi|\Psi\rangle}\right)^2\nonumber\\
A_{0z}(\mathbf{k},i\omega_m)&&=\frac {Y_0}{2\left(E_1^2-E_3^2 \right)}
\left\{ 
\frac 1{E_1}\left(\frac 1 {i\hbar \omega_m-E_1}-\frac 1 {i\hbar \omega_m-E_2}\right)
-\frac 1{E_3}\left(\frac 1 {i\hbar \omega_m-E_3}-\frac 1 {i\hbar \omega_m-E_4}\right)\right\}\\
&&Y_0=-\frac 32\left( \frac{3\alpha^2\lambda_R} {\langle\Psi|\Psi\rangle}\right)^2 \frac {\alpha^2 \lambda_I}{\langle\Psi|\Psi\rangle}\nonumber
\end{eqnarray}
\begin{eqnarray}
C_{0x}(\mathbf{k},i\omega_m)&&=
\frac 1{2\left(E_1^2-E_3^2 \right)} 
\left\{
Y_2\left[\frac 1{E_1}\left(\frac 1 {i\hbar \omega_m-E_1}-\frac 1 {i\hbar \omega_m-E_2}\right)-\frac 1{E_3} \left(\frac 1 {i\hbar \omega_m-E_3}-\frac 1 {i\hbar \omega_m-E_4}\right)\right]\right.\nonumber\\
&&\left.
-Y_1\left[E_1\left(\frac 1 {i\hbar \omega_m-E_1}-\frac 1 {i\hbar \omega_m-E_2}\right)-E_3\left(\frac 1 {i\hbar \omega_m-E_3}-\frac 1 {i\hbar \omega_m-E_4}\right)\right]\right\} \xi k_x\\
&& Y_1= \frac{\hbar v_F (1-3\alpha^2)} {\langle\Psi|\Psi\rangle},\;
Y_2=- Y_1  \left(\frac {3\alpha^2 \lambda_I}{\langle\Psi|\Psi\rangle} \right)^2 \nonumber
\end{eqnarray}
\begin{eqnarray}
C_{0y}(\mathbf{k},i\omega_m)&&=
\frac 1{2\left(E_1^2-E_3^2 \right)} 
\left\{
Y_2\left[\frac 1{E_1}\left(\frac 1 {i\hbar \omega_m-E_1}-\frac 1 {i\hbar \omega_m-E_2}\right)-\frac 1{E_3} \left(\frac 1 {i\hbar \omega_m-E_3}-\frac 1 {i\hbar \omega_m-E_4}\right)\right]\right.\nonumber\\
&&\left.
-Y_1\left[\left(E_1\frac 1 {i\hbar \omega_m-E_1}-\frac 1 {i\hbar \omega_m-E_2}\right)-E_3\left(\frac 1 {i\hbar \omega_m-E_3}-\frac 1 {i\hbar \omega_m-E_4}\right)\right]\right\} k_y
\end{eqnarray}
\begin{eqnarray}
A_{z1}(\mathbf{k},i\omega_m)&&=\xi
\frac 1{2\left(E_1^2-E_3^2 \right)} 
\left\{
Y_4\left[\frac 1{E_1}\left(\frac 1 {i\hbar \omega_m-E_1}-\frac 1 {i\hbar \omega_m-E_2}\right)-\frac 1{E_3} \left(\frac 1 {i\hbar \omega_m-E_3}-\frac 1 {i\hbar \omega_m-E_4}\right)\right]\right.\nonumber\\
&&\left.
-Y_3\left[E_1\left(\frac 1 {i\hbar \omega_m-E_1}-\frac 1 {i\hbar \omega_m-E_2}\right)-E_3\left(\frac 1 {i\hbar \omega_m-E_3}-\frac 1 {i\hbar \omega_m-E_4}\right)\right]\right\} \\
&& Y_3=- \frac {3\alpha^2 \lambda_I}{\langle\Psi|\Psi\rangle},\;
Y_4=-Y_3 \left[\frac{(\hbar v_F)^2}{\langle\Psi|\Psi\rangle^2} (1-3\alpha^2)^2 ||\mathbf{k}||^2- \frac12 \left(\frac{3\alpha^2 \lambda_R}{\langle\Psi|\Psi\rangle}\right)^2 -\left(\frac {3\alpha^2 \lambda_I}{\langle\Psi|\Psi\rangle} \right)^2  \right]\nonumber
\end{eqnarray}
\begin{eqnarray}
A_{zz}(\mathbf{k},i\omega_m)&&=\xi\frac {Y_5}{2\left(E_3^2-E_1^2 \right)}\left\{
\frac 1 {i\hbar \omega_m-E_1}+\frac 1 {i\hbar \omega_m-E_2}
-\frac 1 {i\hbar \omega_m-E_3}-\frac 1 {i\hbar \omega_m-E_4}\right\}\\
&&Y_5= \frac 12\left( \frac{\lambda_R} {\langle\Psi|\Psi\rangle}\right)^2\nonumber
\end{eqnarray}
\begin{eqnarray}
C_{zx}(\mathbf{k},i\omega_m)&&=\frac {Y_6}{2\left(E_3^2-E_1^2 \right)}\left\{
\frac 1 {i\hbar \omega_m-E_1}+\frac 1 {i\hbar \omega_m-E_2}
-\frac 1 {i\hbar \omega_m-E_3}-\frac 1 {i\hbar \omega_m-E_4}\right\} k_x\\
&&Y_6=  \hbar v_F (1-3\alpha^2)
\frac {6\alpha^2 \lambda_I}{\langle\Psi|\Psi\rangle^2}\nonumber
\end{eqnarray}
\begin{eqnarray}
C_{zy}(\mathbf{k},i\omega_m)&&=\frac {Y_6}{2\left(E_3^2-E_1^2 \right)}\left\{
\frac 1 {i\hbar \omega_m-E_1}+\frac 1 {i\hbar \omega_m-E_2}
-\frac 1 {i\hbar \omega_m-E_3}-\frac 1 {i\hbar \omega_m-E_4}\right\}\xi k_y\nonumber\\
\end{eqnarray}
\begin{eqnarray}
B_{x1}(\mathbf{k},i\omega_m)&&=\frac {Y_7}{2\left(E_3^2-E_1^2 \right)}\left\{
\frac 1 {i\hbar \omega_m-E_1}+\frac 1 {i\hbar \omega_m-E_2}
-\frac 1 {i\hbar \omega_m-E_3}-\frac 1 {i\hbar \omega_m-E_4}\right\} k_y\\
&&Y_7=  \hbar v_F (1-3\alpha^2)
\frac {3\alpha^2 \lambda_R}{\langle\Psi|\Psi\rangle^2}\nonumber
\end{eqnarray}
\begin{eqnarray}
B_{xz}(\mathbf{k},i\omega_m)&&=
Y_8\frac 1 {2\left(E_1^2-E_3^2 \right)}\,k_y \times\nonumber\\
&&\left\{\frac 1{E_1}\left[\frac 1 {i\hbar \omega_m-E_1}-\frac 1 {i\hbar \omega_m-E_2}\right]
+\frac 1 {E_3}\left[\frac 1 {i\hbar \omega_m-E_3}-\frac 1 {i\hbar \omega_m-E_4}\right]\right\} \\
&&Y_8=- \hbar v_F (1-3\alpha^2)
\frac {3\alpha^2 \lambda_R}{\langle\Psi|\Psi\rangle^2} \frac {3\alpha^2 \lambda_I}{\langle\Psi|\Psi\rangle}\nonumber
\end{eqnarray}
\begin{eqnarray}
D_{xx}(\mathbf{k},i\omega_m)&&=Y_9\frac 1 {2\left(E_1^2-E_3^2 \right)}\,\xi k_x k_y\times\nonumber\\
&&\left\{\frac 1{E_1}\left[\frac 1 {i\hbar \omega_m-E_1}-\frac 1 {i\hbar \omega_m-E_2}\right]
-\frac 1 {E_3}\left[\frac 1 {i\hbar \omega_m-E_3}-\frac 1 {i\hbar \omega_m-E_4}\right]\right\} \\
&&Y_9= (\hbar v_F)^2 (1-3\alpha^2)^2
\frac {3\alpha^2 \lambda_R}{\langle\Psi|\Psi\rangle} \nonumber
\end{eqnarray}
\begin{eqnarray}
D_{xy}(\mathbf{k},i\omega_m)=&&
\frac 1{2\left(E_1^2-E_3^2 \right)} 
\left\{
Y_{11}\left[\frac 1{E_1}\left(\frac 1 {i\hbar \omega_m-E_1}-\frac 1 {i\hbar \omega_m-E_2}\right)-\frac 1{E_3} \left(\frac 1 {i\hbar \omega_m-E_3}-\frac 1 {i\hbar \omega_m-E_4}\right)\right]\right.\nonumber\\
&&\left.
-Y_{10}\left[E_1\left(\frac 1 {i\hbar \omega_m-E_1}-\frac 1 {i\hbar \omega_m-E_2}\right)-E_3\left(\frac 1 {i\hbar \omega_m-E_3}-\frac 1 {i\hbar \omega_m-E_4}\right)\right]\right\} \\
&& Y_{10}=-\frac {3\alpha^2 \lambda_R}{2\langle\Psi|\Psi\rangle},\;
Y_{11}=Y_{10} \left[(\hbar v_F)^2 (1-3\alpha^2)^2 (k_x^2-k_y^2)+ \left(\frac {3\alpha^2 \lambda_I}{\langle\Psi|\Psi\rangle} \right)^2  \right]\nonumber
\end{eqnarray}
\begin{eqnarray}
&&B_{y1}(\mathbf{k},i\omega_m)=-Y_{7} \frac 1 {2\left(E_3^3-E_1^2 \right)}\left\{ 
\frac 1 {i\hbar \omega_m-E_1}+\frac 1 {i\hbar \omega_m-E_2}
-\frac 1 {i\hbar \omega_m-E_3}-\frac 1 {i\hbar \omega_m-E_4}\right\} k_x
\end{eqnarray}

\begin{eqnarray}
&&B_{yz}(\mathbf{k},i\omega_m)=-Y_8\frac 1 {2\left(E_1^2-E_3^2 \right)}\,k_x
\left\{\frac 1{E_1}
\left[\frac 1 {i\hbar \omega_m-E_1}-\frac 1 {i\hbar \omega_m-E_2}\right]
-\frac 1{E_3}\left[\frac 1 {i\hbar \omega_m-E_3}-\frac 1 {i\hbar \omega_m-E_4}\right]\right\} 
\end{eqnarray}
\begin{eqnarray}
D_{yx}(\mathbf{k},i\omega_m)&=&\xi
\frac 1{2\left(E_1^2-E_3^2 \right)} 
\left\{
Y_{12}\left[\frac 1{E_1}\left(\frac 1 {i\hbar \omega_m-E_1}-\frac 1 {i\hbar \omega_m-E_2}\right)-\frac 1{E_3} \left(\frac 1 {i\hbar \omega_m-E_3}-\frac 1 {i\hbar \omega_m-E_4}\right)\right]\right.\nonumber\\
&&\left.
+Y_{10}\left[E_1\left(\frac 1 {i\hbar \omega_m-E_1}-\frac 1 {i\hbar \omega_m-E_2}\right)-E_3\left(\frac 1 {i\hbar \omega_m-E_3}-\frac 1 {i\hbar \omega_m-E_4}\right)\right]\right\} \\
&&Y_{12}= \frac{3\alpha^2 \lambda_R}{3\langle\Psi|\Psi\rangle} \left[-(\hbar v_F)^2 (1-3\alpha^2)^2 (k_x^2-k_y^2)+ \left(\frac {3\alpha^2 \lambda_I}{\langle\Psi|\Psi\rangle} \right)^2  \right]\nonumber
\end{eqnarray}
\begin{eqnarray}
D_{yy}(\mathbf{k},i\omega_m)&&=-Y_9\frac 1 {\left(E_1^2-E_3^2 \right)}k_x k_y
\left\{\frac1{E_1}
\left[\frac 1 {i\hbar \omega_m-E_1}-\frac 1 {i\hbar \omega_m+E_1}\right]
-\frac1{E_3}\left[\frac 1 {i\hbar \omega_m-E_3}-\frac 1 {i\hbar \omega_m-E_4}\right]\right\},
\label{Dyy}
\end{eqnarray}
\

where $E_1=E_{+,+}(\mathbf{k})$, $E_2=E_{-,+}(\mathbf{k})$, $E_3=E_{+,-}(\mathbf{k})$ and $E_4=E_{-,-}(\mathbf{k})$ (Eq.~\ref{band-supp}).

\section{IV. Magnon Green function and Gilbert damping}\label{Green-mag}

\subsection{Interfacial exchange coupling between a ferro-
magnetic insulator (FI) and a TBG}

We consider the Hamiltonian of the ferromagnetic insulator (FI) in the independent magnon approximation as
\begin{align}
H_{\rm FI} = \sum_{\bm k} \hbar \omega_{\bm k} b_{\bm k}^\dagger b_{\bm k},
\end{align}
where $\hbar \omega_{\bm k}\simeq {\cal D}|{\bm k}|^2 + \hbar \gamma h_{\rm dc}$ is a dispersion of magnons, ${\cal D}$ is a spin stiffness, $\gamma$ is the gyromagnetic ratio, $h_{\rm dc}$ is a static magnetic field.
In the spin pumping setup, only the static part associated to $\mathbf{k}=\mathbf{0}$ is relevant.
Considering only the uniform spin precession, the Hamiltonian of the FI can be simply written as
\begin{align}
H_{\rm FI} = \hbar \omega_{\bm 0} b_{\bm 0}^\dagger b_{\bm 0},  
\end{align}
where $b_{\bm q}$ is the Fourier transformation of the site representation $b_i$ defined as
\begin{align}
b_i &= \frac{1}{\sqrt{N_{\rm FI}}} \sum_{\bm q}
e^{i{\bm q}\cdot {\bm r}_i} b_{\bm q}
\simeq \frac{1}{\sqrt{N_{\rm FI}}} b_{\bm 0},\label{bapp1} \\
b_i^\dagger &= \frac{1}{\sqrt{N_{\rm FI}}} \sum_{\bm q}
e^{-i{\bm q}\cdot {\bm r}_i} b^\dagger_{\bm q} \simeq \frac{1}{\sqrt{N_{\rm FI}}} b^\dagger_{\bm 0},
\label{bapp2}
\end{align}
where $N_{\rm FI}$ is the number of unit cells in the FI.\newline

We consider the (retarded) magnon Green function as
\begin{align}
G^R({\bm q},\omega) &= \int dt \, G^R({\bm q},t) e^{i\omega t}, \\
G^R({\bm q},t) &= -\frac{i}{\hbar} \theta(t) \langle [S^+_{\bm q}(t),S^-_{\bm q}(0)] \rangle 
= -\frac{2iS_0}{\hbar} \theta(t) \langle [b_{\bm q}(t),b_{\bm q}^\dagger(0)] \rangle.
\end{align}
In the absence of the junction, the magnon Green function is
\begin{align}
G_0^R({\bm q},\omega) = \frac{2S_0/\hbar}{\omega - \omega_{\bm q}+i\delta} .
\end{align}
We introduce spin relaxation of the bulk FI phenomenologically as
\begin{align}
G_0^R({\bm q},\omega) = \frac{2S_0/\hbar}{\omega - \omega_{\bm q}+i\alpha_{\rm G}\omega} ,
\end{align}
where $\alpha_{\rm G}$ is a dimensionless strength of the Gilbert damping, which is of order of $10^{-4}$--$10^{-3}$.
We note that a line shape of the ferromagnetic resonance is proportional to ${\rm Im}\, G_0^R({\bm q}={\bm 0},\omega)$~\cite{Funato}.

In the presence of the interfacial coupling and for a uniform spin precession, the magnon Green function is given by the Dyson equation as
\begin{align}
G_0^R({\bm {q=0}},\omega) = \frac{2S_0/\hbar}{\omega - \omega_{\bm q=0}+i\alpha_{\rm G}\omega-(2S_0/\hbar) \Sigma^R(\bm {q=0},\omega)} ,
\end{align}
where $\Sigma^R(\omega)$ is the self-energy.
Although the real part of $\Sigma^R(\omega)$ is related to the shift of the ferromagnetic resonance, we neglect it for simplicity.
Then, the magnon Green function is rewritten as
\begin{align}
G_0^R({\bm 0},\omega) &= \frac{2S_0/\hbar}{\omega - \omega_{\bm q=0}+i(\alpha_{\rm G}+\delta \alpha_{\rm G})\omega}, \\
\delta \alpha_{\rm G}(\omega) &= - \frac{2S_0}{\hbar \omega} \rm{Im}\Sigma^R(\mathbf {q}=\mathbf{0},\omega).
\end{align}
We note that $\delta \alpha_{\rm G}(\omega)$ depends on $\omega$ in general.
However, since the ferromagnetic resonance peak is sharp enough ($\alpha_{\rm G}+\delta \alpha_{\rm G}\ll 1$), we can replace $\omega$ with $\omega_{\rm 0}=\Omega$ (the peak position of the ferromagnetic resonance):
\begin{align}
\delta \alpha_{\rm G} &\simeq - \frac{2S_0}{\hbar \Omega} \Sigma^R(\bm {q=0},\Omega).
\end{align}

The Hamiltonian of the interfacial coupling is given as
\begin{align}
H_{\rm int} = \sum_{\langle i,j \rangle} T_{ij} (S^+_i s^-_j + {\rm h.c.}).
\end{align}
Here, $S^\pm_i$ is a spin ladder operator of the FI and is described by magnon annihilation/creation operators ($b_i$ and $b_i^\dagger$) as
\begin{align}
S^+_i = \sqrt{2S_0} b_i, \quad S^-_i = \sqrt{2S_0} b_i^\dagger,
\end{align}
where $S_0$ is an amplitude of the localized spin in the FI.
$s^\pm_j$ is a spin ladder operator of electrons in two-dimensional electron systems (twisted bilayer graphene) and is described by the electron annihilation/creation operators ($c_{j\sigma}$ and $c_{j\sigma}^\dagger$) as
\begin{align}
s^+_j = c_{j\uparrow}^\dagger c_{j\downarrow}, \quad
s^-_j = c_{j\downarrow}^\dagger c_{j\uparrow}.
\end{align}
We define the Fourier transformation as
\begin{align}
c_{j\sigma} &= \frac{1}{\sqrt{N}} \sum_{\bm k}
e^{i{\bm k}\cdot {\bm r}_j} c_{{\bm k}\sigma}, \label{fourier1} \\
c^\dagger_{j\sigma} &= \frac{1}{\sqrt{N} } \sum_{\bm k}
e^{-i{\bm k}\cdot {\bm r}_j} c^\dagger_{{\bm k}\sigma}, \label{fourier2}
\end{align}
where $N$ is the number of unit cells and ${\bm r}_j$ is the position of the site $j$ in TBG.
Then, we obtain
\begin{align}
s^+_j &= \frac{1}{N} \sum_{{\bm k},{\bm k}'} e^{-i{\bm k}\cdot {\bm r}_j+i{\bm k}'\cdot {\bm r}_j} c_{{\bm k}\uparrow}^\dagger c_{{\bm k}'\downarrow}, \\
s^-_j &= \frac{1}{N} \sum_{{\bm k},{\bm k}'} e^{-i{\bm k}\cdot {\bm r}_j+i{\bm k}'\cdot {\bm r}_j} c_{{\bm k}\downarrow}^\dagger c_{{\bm k}'\uparrow}.
\end{align}
We define the Fourier transformation of $s^\pm_j$ as
\begin{align}
s^+_j &= \frac{1}{N} \sum_{\bm q} e^{i{\bm q}\cdot {\bm r}_j} s^+_{\bm q},  \\
s^-_j &= \frac{1}{N} \sum_{\bm q} e^{i{\bm q}\cdot {\bm r}_j} s^-_{\bm q} .
\end{align}
From $s^-_j=(s^+_j)^\dagger$, we obtain the relation $s^-_{\bm q} = (s^+_{-{\bm q}})^\dagger$.
%Note that $(s^-_{\bm q})^\dagger \ne s^+_{{\bm q}}$.
The inverse Fourier transformation is given as
\begin{align}
s^+_{\bm q} &= \sum_{j} e^{-i{\bm q}\cdot {\bm r}_j} s^+_j , \\
s^-_{\bm q} &= (s^+_{-{\bm q}})
= \sum_{j} e^{-i{\bm q}\cdot {\bm r}_j} s^-_j . 
\end{align}
Especially for ${\bm q}={\bm 0}$, we obtain
\begin{align}
s^+_{\bm 0} = \sum_{j} s^+_j , 
\quad
s^-_{\bm 0} = \sum_{j} s^-_j , 
\end{align}
Using Eqs.~(\ref{fourier1}) and (\ref{fourier2}), we obtain
\begin{align}
s^+_{\bm q} &= \frac{1}{N}
\sum_{j} e^{-i{\bm q}\cdot {\bm r}_j} \sum_{{\bm k},{\bm k}'} e^{-i{\bm k}\cdot {\bm r}_j+i{\bm k}'\cdot {\bm r}_j} c_{{\bm k}\uparrow}^\dagger c_{{\bm k}'\downarrow} =\sum_{{\bm k}}c_{{\bm k}\uparrow}^\dagger c_{{\bm k}+{\bm q} \downarrow} , \\
s^-_{\bm q} &= \frac{1}{N}
\sum_{j} e^{-i{\bm q}\cdot {\bm r}_j} \sum_{{\bm k},{\bm k}'} e^{-i{\bm k}\cdot {\bm r}_j+i{\bm k}'\cdot {\bm r}_j} c_{{\bm k}\downarrow}^\dagger c_{{\bm k}'\uparrow} 
=\sum_{{\bm k}}c_{{\bm k}\downarrow}^\dagger c_{{\bm k}+{\bm q} \uparrow} .
\end{align}

For a clean interface, we can set $T_{ij}=T$.
Then, using Eqs.~(\ref{bapp1}) and (\ref{bapp2}), the Hamiltonian of the interface is written as
\begin{align}
H_{\rm int} &= \frac{T\sqrt{2S_0}}{\sqrt{N_{\rm FI}}} \sum_{\langle i,j\rangle} (b_{\bm 0} s_j^- +
b_{\bm 0}^\dagger s_j^+) 
\simeq \frac{T\sqrt{2S_0}}{\sqrt{N_{\rm FI}}} \left[ b_{\bm 0} \Bigl(\sum_j s_j^-\Bigr) +
b_{\bm 0}^\dagger \Bigl(\sum_j s_j^+\Bigr) \right] \nonumber \\
&= \sqrt{2S_0} b_{\bm 0} \tilde{s}^- + \sqrt{2S_0} b_{\bm 0} \tilde{s}^+,
\end{align}
where $\tilde{s}^{\pm}$ is defined as
$\tilde{s}^\pm = (T/\sqrt{N_{\rm FI}}) s_{\bm 0}^\pm$.

By the second-order perturbation, the self-energy of the magnon at ${\bm q}={\bm 0}$ is calculated as
\begin{align}
\Sigma^R(\omega) &= \int dt \, \Sigma^R(t) e^{i\omega t} , \\
\Sigma^R(t) &= -\frac{i}{\hbar}\theta(t) \langle [\tilde{s}^+(t),\tilde{s}^-(0)] \rangle.
\end{align}
The self-energy can be related to a retarded component of the dynamic spin susceptibility {\it per unit cell} as
\begin{align}
\Sigma^R(\omega) &= -\frac{T^2N}{N_{\rm FI}} \chi({\bm 0},\omega), \\
\chi({\bm q},\omega) &= \int dt \, \chi({\bm q},t) e^{i\omega t} , \\
\chi({\bm q},t) &= \frac{i}{N\hbar}\theta(t) \langle [s_{-{\bm q}}^+(t),s_{\bm q}^-(0)] \rangle.
\end{align}
$\chi({\bm q},t)$ is calculated for one-band of TBG without spin-orbit interaction as
\begin{align}
\chi({\bm q},t) = \frac{1}{N} \sum_{\bm q}
\frac{f(\xi_{
\bm k})-f(\xi_{{\bm k}+{\bm q}})}
{\hbar \omega + i\delta + \xi_{{\bm k}+{\bm q}}-\xi_{\bm k}},
\end{align}
where $\xi_{\bm k}=\epsilon_{\bm k}-\mu$, $\epsilon_{\bm k}$ is a dispersion of electrons, $\mu$ is a chemical potential.
This is just a Lindhard function.
We note that $\chi({\bm q},t)$ is independent of the system size (area).
For systems with spin-orbit interaction, we have to extend the Lindhard function into the spin-dependent one.

Then, the enhancement of the Gilbert damping is written as
\begin{align}
\delta \alpha_{\rm G} &= - \frac{2S_0}{\hbar \Omega} \, {\rm Im} \Sigma^R(\mathbf{q}=\mathbf{0},\Omega) \nonumber \\
&= \frac{2S_0T^2N}{N_{\rm FI}\hbar \omega_{\bm 0}} \chi(\mathbf{q}=\mathbf{0},\Omega).
\end{align}
We note that the number of the unit cell of twisted bilayer graphene is written as $N=S/A$ where $S$ is a area of the junction and $A$ is an area of a unit cell of twisted bilayer graphene.
We also note that the number of the unit cell of the FI is written as $N_{\rm FI}=Sd/a^3$ where $d$ is a thickness of the FI, $a$ is a lattice constant of the FI.
Using these parameters, we obtain
\begin{align}
\delta \alpha_{\rm G} &= \frac{2S_0T^2a^3}{Ad\hbar \omega_{\bm 0}} \chi(\mathbf{q}=\mathbf{0},\Omega).
\end{align}
We note that $\delta \alpha_{\rm G}$ is proportional to $1/d$ in consistent with experimental results.
If YIG (Yttrium Iron Garnet) is chosen as the ferromagnet insulator, the parameter is given in the Table.

\begin{table}[tp]
 \caption{Experimental parameters.}
 \label{table:Parameters}
 \centering
  \begin{tabular}{lll}
   \hline
   Microwave frequency & $\omega_{\bm 0}$ & $1\, {\rm GHz}$ \\
   Amplitude of spins of FI & $S_0$ & 10 \\
   Lattice constant of FI & $a$ & $12.376\,$\AA \\
   Thickness of FI & $d$ & $\ge 10\, {\rm nm}$ \\
   Interfacial exchange coupling & $J$ & $\sim 1\, {\rm K}$ (not known) \\
   \hline
 %  \label{Table1}
  \end{tabular}
\end{table}

\subsection{Electronic spins in the FI magnetization frame}

\begin{figure}[hpbt] 
\begin{center}
\includegraphics[width=0.2\columnwidth]{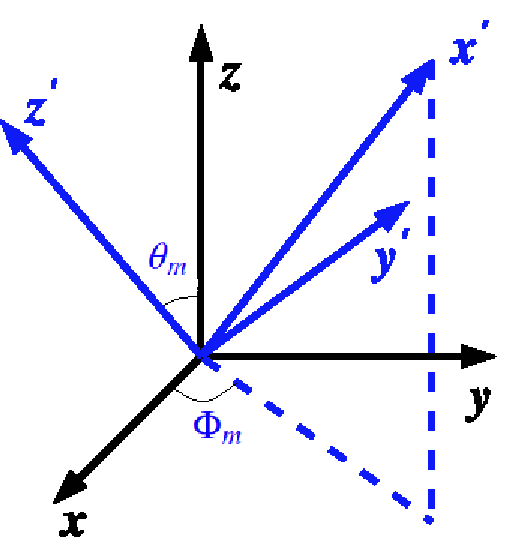}
\end{center}
\caption{Magnetization-fixed coordinate frame $\left(x^{\prime},y^{\prime},z^{\prime}\right)$ with respect to the Laboratory frame $\left(x,y,z\right)$.}
\label{axis}
\end{figure}

Regarding the dependence on $s_z$ of the electronic Hamiltonian (Eq.~{\red3} of the main text), one should consider a $3D$ FI magnetization as in Ref.~\cite{Funato}.  The average spin vector is along the orthoradial spherical vector $\langle \mathbf{S}_{FI}\rangle= \langle \mathbf{S}_{FI}\rangle \,\mathbf{u}_{x^{\prime}}$. The radial vector $\mathbf{u}_{z^{\prime}}$ forms an angle $\theta_m$ with the $z$ axis perpendicular to the interface. The third axis $y^{\prime}$ is in the $(xoy)$ plane and its unit vector is the orthoradial inplane vector $\mathbf{u}_{y^{\prime}}=-\sin\Phi_m \;\mathbf{u}_x+\cos\Phi_m\; \mathbf{u}_y $ as shown in Fig.\ref{axis}.\

In the FI spin frame ($x^{\prime},y^{\prime},z^{\prime}$), the components of the electronic spin operators are given by:
\begin{eqnarray}
&&s^{x^{\prime}}_{\mathbf{k}}=\mathbf{s}_{\mathbf{k}}\cdot \mathbf{u}_{x^{\prime}}=
\cos \theta_m\cos\Phi_m \,s^x_{\mathbf{k}}+\cos\theta_m\sin\Phi_m\, s^y_{\mathbf{k}}-\sin \theta_m \,s^z_{\mathbf{k}}\nonumber\\
&&s^{y^{\prime}}_{\mathbf{k}}=\mathbf{s}_{\mathbf{k}}\cdot \mathbf{u}_{y^{\prime}}=
-\sin \Phi_m\, s^x_{\mathbf{k}}+\cos\Phi_m\, s^y_{\mathbf{k}}\nonumber\\
&&s^{z^{\prime}}_{\mathbf{k}}=\mathbf{s}_{\mathbf{k}}\cdot \mathbf{u}_{z^{\prime}}=
\sin \theta_m\cos\Phi_m \,s^x_{\mathbf{k}}+\sin\theta_m\sin\Phi_m\, s^y_{\mathbf{k}}+\cos \theta_m \,s^z_{\mathbf{k}}\nonumber\\
\end{eqnarray}
We define the ladder electronic spin operators as
\begin{eqnarray}
s^{x^{\prime},\pm}_{\mathbf{k}}&=&s^{y^{\prime}}_{\pm\mathbf{k}}\pm i s^{z^{\prime}}_{\pm\mathbf{k}}
=\frac 12 \sum_{\sigma,\sigma^{\prime},\mathbf{k}^{\prime}} c^{\dagger}_{\mathbf{k}^{\prime},\sigma} \left(\sigma_s^{x^{\prime},\pm} \right)_{\sigma,\sigma'}
c^{\dagger}_{\mathbf{k}^{\prime}\pm \mathbf{k},\sigma'}
\end{eqnarray}
where $s^{x^{\prime},\pm}= s_x \left( -\sin\Phi_m\pm i \sin \theta_m\cos\Phi_m\right)
+s_y\left(\cos\Phi_m\pm i \sin \theta_m\sin\Phi_m \right)\pm i \cos \theta_m s_z$.

\subsection{Magnon self-energy}\label{mag}

In the second order perturbation with respect to the interfacial exchange interaction $T_\mathbf{q}$, the interfacial self-energy is given by~\cite{Yama}
\begin{eqnarray}
\Sigma (\mathbf{q},i\omega_n)=\frac{|T_\mathbf{q}|^2}{4\beta}\sum_{\mathbf{k},i\omega_m}
\text{Tr}\left[ \sigma_s^{x^{\prime},-} \;\hat{g} (\mathbf{k},i\omega_m)\sigma_s^{x^{\prime},+}\;\hat{g} (\mathbf{k}+\mathbf{q},i\omega_m+i\omega_n)
\right]
\label{self-supp}
\end{eqnarray}
where $\hat{g} (\mathbf{k},i\omega_m)$ is the electronic Green function given by Eq.~\ref{green-elec}.\

The trace term is of the form:
\begin{eqnarray}
\text{Tr}\left[ \mathbf{a}^{\ast}\cdot\mathbf{\sigma}_s \left(\hat{g}_0+\hat{\mathbf{g}}\cdot\mathbf{\sigma}_s\right)
\mathbf{a}\cdot\mathbf{\sigma}_s 
\left(\hat{g}^{\prime}_0+\hat{\mathbf{g}}^{\prime}\cdot\mathbf{\sigma}_s\right)
\right]
\label{Trace}
\end{eqnarray}
where the vector $\mathbf{a}=\left(-\sin \Phi_m+i\sin\theta_m\cos\Phi_m,\cos \Phi_m+i\sin\theta_m\sin\Phi_m,i\cos \theta_m \right)$ is written in the laboratory frame $(x,y,z)$.\

We set $\hat{g}=\hat{g} (\mathbf{k},i\omega_m)$ and $\hat{g}^{\prime}=\hat{g} (\mathbf{k},i\omega_m+i \omega_n)$.
Taking into account the operator character of $\hat{g}$ one could use the identity
\begin{eqnarray}
\left( \mathbf{a}\cdot \mathbf{\sigma}_s\right)\left( \mathbf{b}\cdot \mathbf{\sigma}_s\right)=
\left( \mathbf{a}\cdot \mathbf{b}\right) \mathbb{I} +i \left( \mathbf{a}\times \mathbf{b}\right)\cdot \mathbf{\sigma}_s
\end{eqnarray}
Given the expressions of $\hat{g}$ and $\hat{g}^{\prime}$ in Eq.~\ref{gg}, the trace term (Eq.~\ref{Trace}) reduces to:
\begin{eqnarray}
\text{Tr}\left[ \mathbf{a}^{\ast}\cdot\mathbf{\sigma}_s \left(\hat{g}_0+\hat{\mathbf{g}}\cdot\mathbf{\sigma}_s\right)
\mathbf{a}\cdot\mathbf{\sigma}_s 
\left(\hat{g}^{\prime}_0+\hat{\mathbf{g}}^{\prime}\cdot\mathbf{\sigma}_s\right)
\right]= \sum_{i=0,1,2} F_i(\mathbf{k},i\omega_m,i\omega_n)
\label{Trace2}
\end{eqnarray}
where
\begin{eqnarray}
&& F_0(\mathbf{k},i\omega_m,i\omega_n)=4\left(A_{01}A'_{01}+A_{0z}A'_{0z}+C_{0x}C'_{0x}+C_{0y}C'_{0y}\right)\nonumber\\
&&F_{1}(\mathbf{k},i\omega_m,i\omega_n)=-2\left\{
\cos \theta_m\cos \Phi_m\right.\nonumber\\ 
&&\left(A_{01}B'_{x1}-B_{x1}A'_{01}+A_{0z}B'_{xz}-A'_{0z}B_{xz}
+C_{0x}D'_{xx}-C'_{0x}D_{xx}+C_{0y}D'_{xy}-C'_{0y}D_{xy}
\right)\nonumber\\
&&+\cos \theta_m\sin \Phi_m \left(A_{01}B'_{y1}-B_{y1}A'_{01}+A_{0z}B'_{yz}-A'_{0z}B_{yz}
+C_{0x}D'_{yx}-C'_{0x}D_{yx}+C_{0y}D'_{yy}-C'_{0y}D_{yy}
\right)\nonumber\\
&&\left.-\sin \theta_m\left(A_{01}A'_{z1}-A'_{01}A_{z1}+A_{0z}A'_{zz}-A'_{0z}A_{zz}+
C_{0x}C'_{zx}-C'_{0x}C_{zx}+C_{0y}C'_{zy}-C'_{zy}C_{0y}
\right)
\right\}\nonumber\\
&&F_{2}(\mathbf{k},i\omega_m,i\omega_n)=
-2\cos^2 \theta_m\cos^2 \Phi_m
\left(B_{x1}B'_{x1}+B_{xz}B'_{xz}+D_{xx}D'_{xx}+D_{xy}D'_{xy}
\right)\nonumber\\
&&-2\cos^2 \theta_m\sin^2 \Phi_m
\left(B_{y1}B'_{y1}+B_{yz}B'_{yz}+D_{yx}D'_{yx}+D_{yy}D'_{yy}
\right)\nonumber\\
&&-2\sin^2\theta_m
\left(A_{z1}A'_{z1}+A_{zz}A'_{zz}+C_{zx}C'_{zx}+C_{zy}C'_{zy}
\right)\nonumber\\
&&-\cos^2 \theta_m\sin 2 \Phi_m
\left(B_{x1}B'_{y1}+B_{xz}B'_{yz}+D_{xx}D'_{yx}+D_{xy}D'_{yy}
+B'_{x1}B_{y1}+B'_{xz}B_{yz}+D'_{xx}D_{yx}+D'_{xy}D_{yy}
\right)\nonumber\\
&&+\cos\Phi_m \sin 2 \theta_m
\left(B_{x1}A'_{z1}+B_{xz}A'_{zz}+D_{xx}C'_{zx}+D_{xy}C'_{zy}
B'_{x1}A_{z1}+B'_{xz}A_{zz}+D'_{xx}C_{zx}+D'_{xy}C_{zy}
\right)\nonumber\\
&&+\sin\Phi_m \sin 2 \theta_m
\left(B_{y1}A'_{z1}+B_{yz}A'_{zz}+D_{yx}C'_{zx}+D_{yy}C'_{zy}
B'_{y1}A_{z1}+B'_{yz}A_{zz}+D'_{yx}C_{zx}+D'_{yy}C_{zy}
\right)
\end{eqnarray}
The terms with a prime are expressed in terms of $i\omega_n'=i\omega_n+i\omega_m$.\

Regarding the $\mathbf{k}$ dependence of the $A,B,C$ and $D$ operators (Eqs.~\ref{A01}-~\ref{Dyy}), only $F_0$, the last term in $F_1$ and the three first terms in $F_2$ give non-vanishing contributions after summing over $\mathbf{k}$ in Eq.~{\red12}.\

On the other hand, the terms between parentheses in the first and second line in $F_2$ expression give the same contribution.
As a result, the GD is found to be independent of the azimuthal angle $\Phi_m$, which expresses isotropy of the electronic band structure $E_{\sigma,\pm}$ (Eq.~\ref{band-supp}). However, the GD depends on the out-of-plane orientation of the FI magnetization via the angle $\theta_m$.\\

According to Eq.~\ref{self-supp}, the terms to calculate are of the form 

\begin{eqnarray}
\sum_{\mathbf{k}} F(\mathbf{k})\sum_{\omega_m}\frac 1 {i\hbar \omega_m-E_i} \frac 1 {i\hbar \omega_m^{\prime}-E_j},
\label{sum}
\end{eqnarray}
 where $\omega_m^{\prime}=\omega_m+\omega_n$ and $F(\mathbf{k})$ is a function of $\mathbf{k}=\left(k,\varphi_{\mathbf{k}}\right)$.\
 
The summation over $\omega_m$ in Eq.~\ref{sum} can be written as
\begin{eqnarray}
\sum_{\omega_m}\frac 1 {i\hbar \omega_m-E_i} \frac 1 {i\hbar \omega_m^{\prime}-E_j}
&&= \frac 1 {i\hbar\omega_n- (E_j-E_i)}\sum_{\omega_m}\left[\frac 1 {i\hbar \omega_m-E_i}- \frac 1 {i\hbar \omega_m^{\prime}-E_j} \right]\nonumber\\
&&=-\frac 1 {i\hbar\omega_n- (E_j-E_i)} \frac 1{k_BT}\int_c \frac{dz}{2\pi i} h(z) f(z)
\end{eqnarray}
where $h(z)=\frac 1 {z-E_i}- \frac 1 {z+i\hbar \omega_n-E_j}$, $f(z)$ is the Fermi-Dirac function, $C$ is clockwise contour around the poles $z=i\hbar \omega_m$.\

Equation~\ref{sum} reduces, then, to
\begin{eqnarray}
\sum_{\omega_m}\frac 1 {i\hbar \omega_m-E_i} \frac 1 {i\hbar \omega_m^{\prime}-E_j}
=\frac {f(E_j)-f(E_i)} {i\hbar\omega_n- (E_j-E_i)}  
\end{eqnarray}
Taking the analytic continuation $ i\hbar\omega_n=\hbar\omega +i\eta$, Eq.~\ref{sum} becomes
\begin{eqnarray}
\lim_{\eta \to 0}\sum_{\mathbf{k}} F(\mathbf{k})  \eta\; \frac {f(E_j)-f(E_i)} {(\hbar\omega - E_j+E_i)^2+\eta^2}=\lim_{\eta \to 0}\sum_{\mathbf{k}}F(\mathbf{k}) (f(E_j)-f(E_i))
L(\hbar \omega-(E_j-E_i)),
\label{lor}
\end{eqnarray}
$L(x)=\frac {\eta}{x^2+\eta^2}$ being the Lorentzian function.\
The sum over $\mathbf{k}=(k,\varphi_{\mathbf{k}})$ in Eq.~\ref{lor} reduces to $\frac A{(2\pi)^2}\int_0^{k_c} k dk\int_0^{2\pi} d\varphi_{\mathbf{k}}$, where $A$ is the moir\'e superlattice area, $k_c$ is a cutoff on the momentum amplitude $k$, below which the continuum model for the monolayer is justified. We take $k_c=q_0/2$, where $q_0=\frac{4\pi}{3 a}\theta$ is the separation between the Dirac points $\mathbf{K}_{1,\xi}$ and $\mathbf{K}_{2,\xi}$ of, respectively, layer (1) and layer (2) at a given monolayer valley $\xi$.

\subsection{Gilbert damping correction}

For a uniform spin precession, the Gilbert damping correction $\delta\alpha_G$, at the FMR frequency $\Omega$, can be expressed as~\cite{Yama}
\begin{eqnarray}
\delta\alpha_G=-\frac {2S_0}{\hbar \Omega} \mathrm{Im} \Sigma(\mathbf{q=0},\Omega)
\end{eqnarray}
The imaginary part of the self-energy is of the form $\mathrm{Im} \Sigma(\mathbf{q=0},\Omega)=\frac{T_0^2}{\hbar \Omega} \tilde{\Sigma}(\mathbf{q=0},\Omega)$, where $\tilde{\Sigma}(\mathbf{q=0},\Omega)$ is a dimensionless integral.
Introducing the average SOC $\lambda=\frac12(\lambda_I+\lambda_R)$, $\delta\alpha_G$ can be written as
\begin{eqnarray}
\delta\alpha_G=-\alpha_G^0 \left(\frac{\lambda}{\hbar \Omega}\right)^2\tilde{\Sigma}(\mathbf{q=0},\Omega)
\end{eqnarray}
where $\alpha_G^0=2S_0\frac{T_0^2}{\lambda^2}$.
 
\begin{figure*}[hpbt]
\begin{center}
\includegraphics[width=0.8\columnwidth]{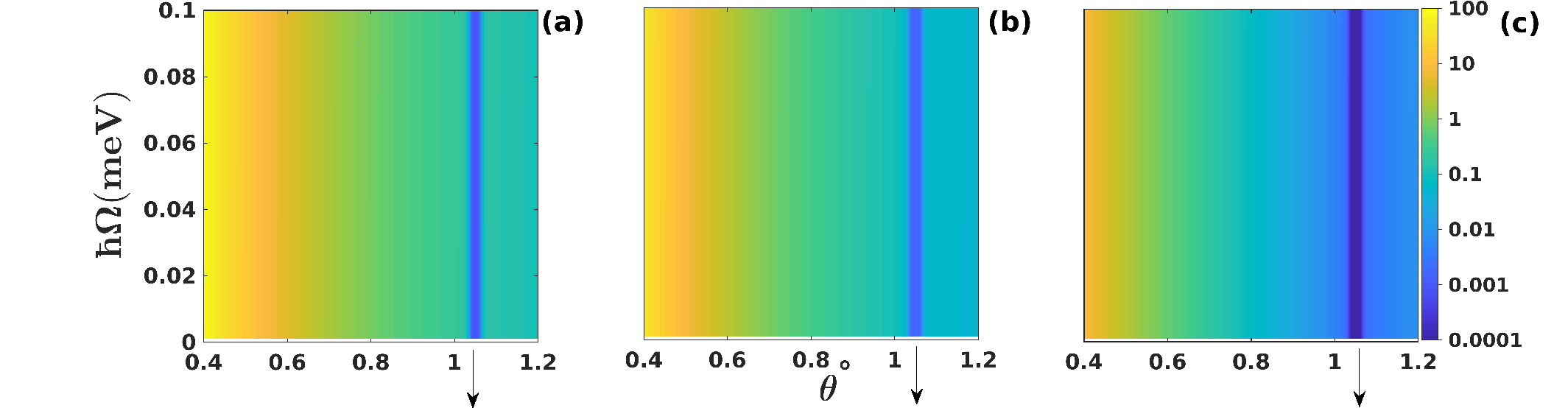}
\includegraphics[width=0.8\columnwidth]{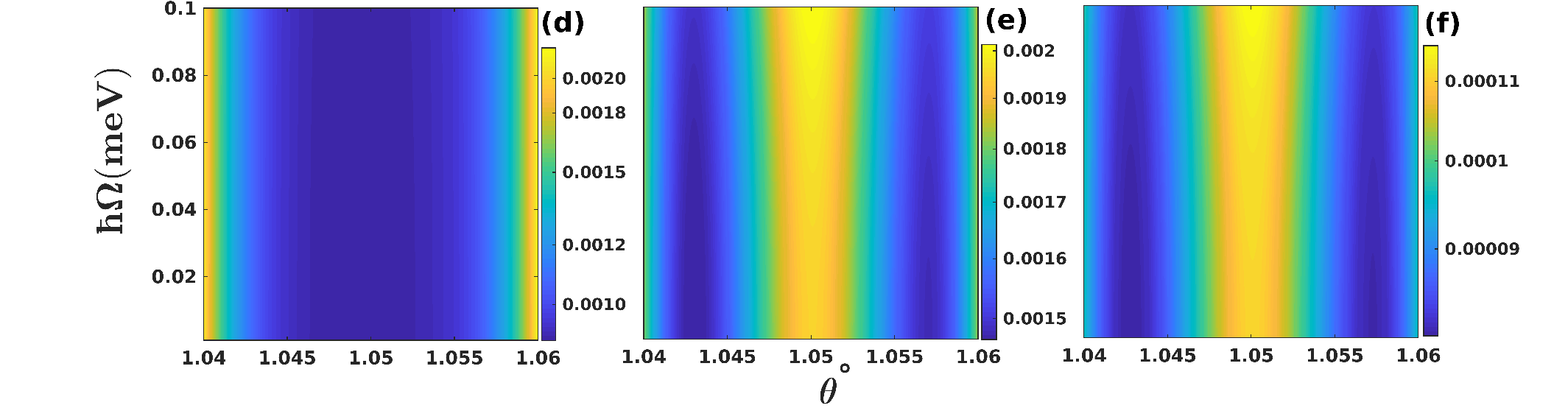}
\end{center}
\caption{Color plot of the normalized Gilbert damping correction $\delta\alpha_G/\alpha_G^0$ as a function of the twist angle $\theta$ and the FMR energy $\hbar \Omega$ at $k_BT=0.1\;\mathrm{meV} $ ((a) and (d)), $k_BT=1\;\mathrm{meV} $ ((b) and (e)) and $k_BT=25\;\mathrm{meV} $ ((c) and (f)). The bottom panels show the behavior of the GD around the MA.
Calculations are done for $\mu=0$, $\lambda_I=3\;\mathrm{meV}$ and $\lambda_R=4\,\mathrm{meV}$.}
\label{GD1-supp}
\end{figure*}

In Fig.~\ref{GD1-supp}, we plot the normalized Gilbert damping correction $\delta\alpha_G/\alpha_G^0$ as a function of the twist angle $\theta$ and the FMR energy $\hbar\Omega$ at different temperatures. The bottom panels are a zoomed representation around the MA.
Fig.~\ref{GD1-supp} shows that, the GD increases by decreasing the twist angle and sharply drops at the MA, regardless of the temperature range and the FMR frequency .\
 
At high temperature ($k_BT>\lambda$), the GD exhibits, around the MA, a fine structure characterized by a peak which disappears at low temperature.
The origin of this peak is, as discussed in the main text, due to the dispersion of the energy bands of TBG/$\mathrm{WSe}_2$ and their corresponding thermal weights $\Delta f(E)=f(E_{\langle S_z\rangle})-f(E_{-\langle S_z\rangle})$ (Eq.~\ref{lor}).

%%%%%%%%%%%%%%%%%%%%%%%%%%%%%%%%%%%%%%%%%%%%%%%%%%%%%%%%%%%

\begin{figure*}[hpbt] 
\begin{center}
$
\begin{array}{ccc}
\includegraphics[width=0.3\columnwidth]{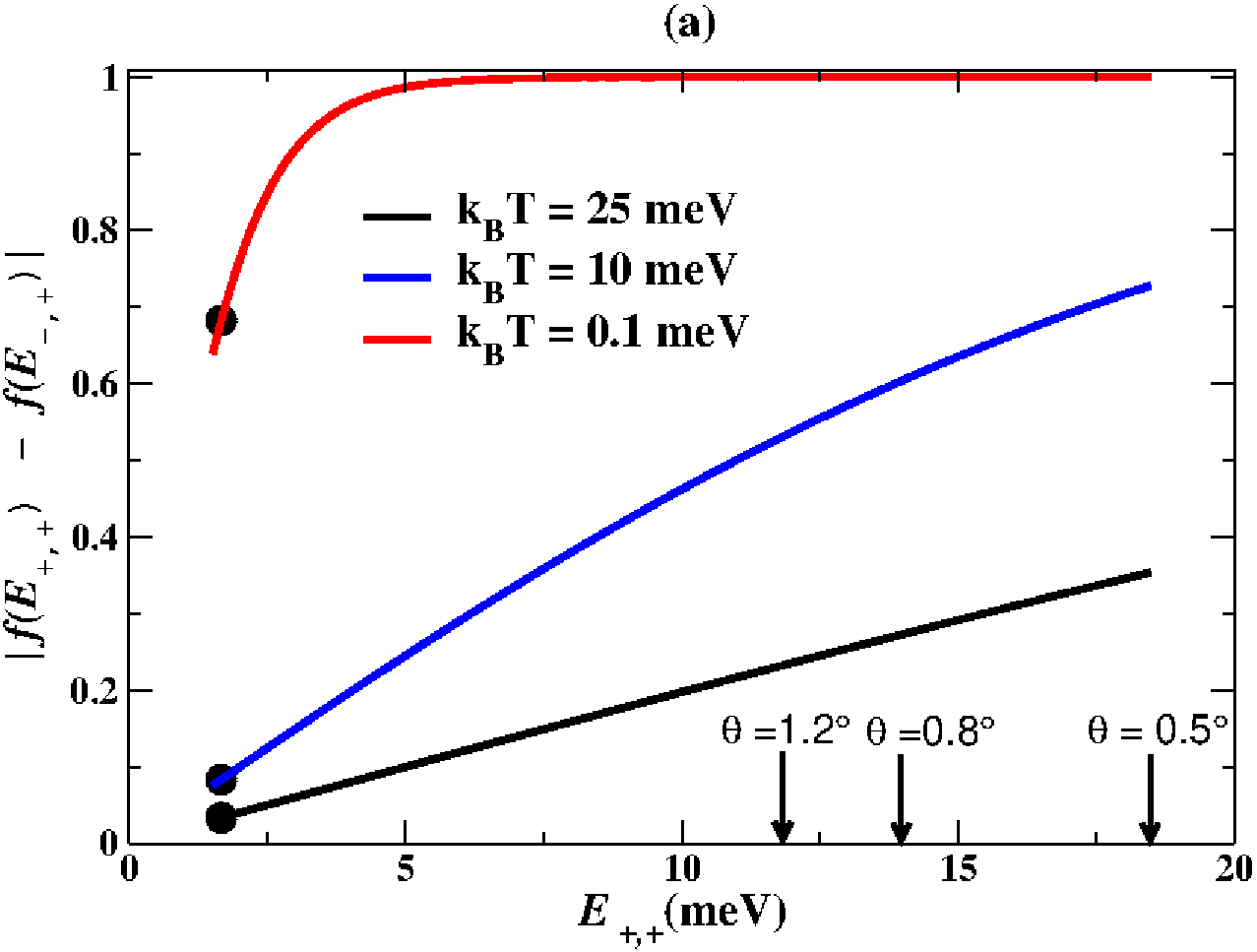}
\includegraphics[width=0.3\columnwidth]{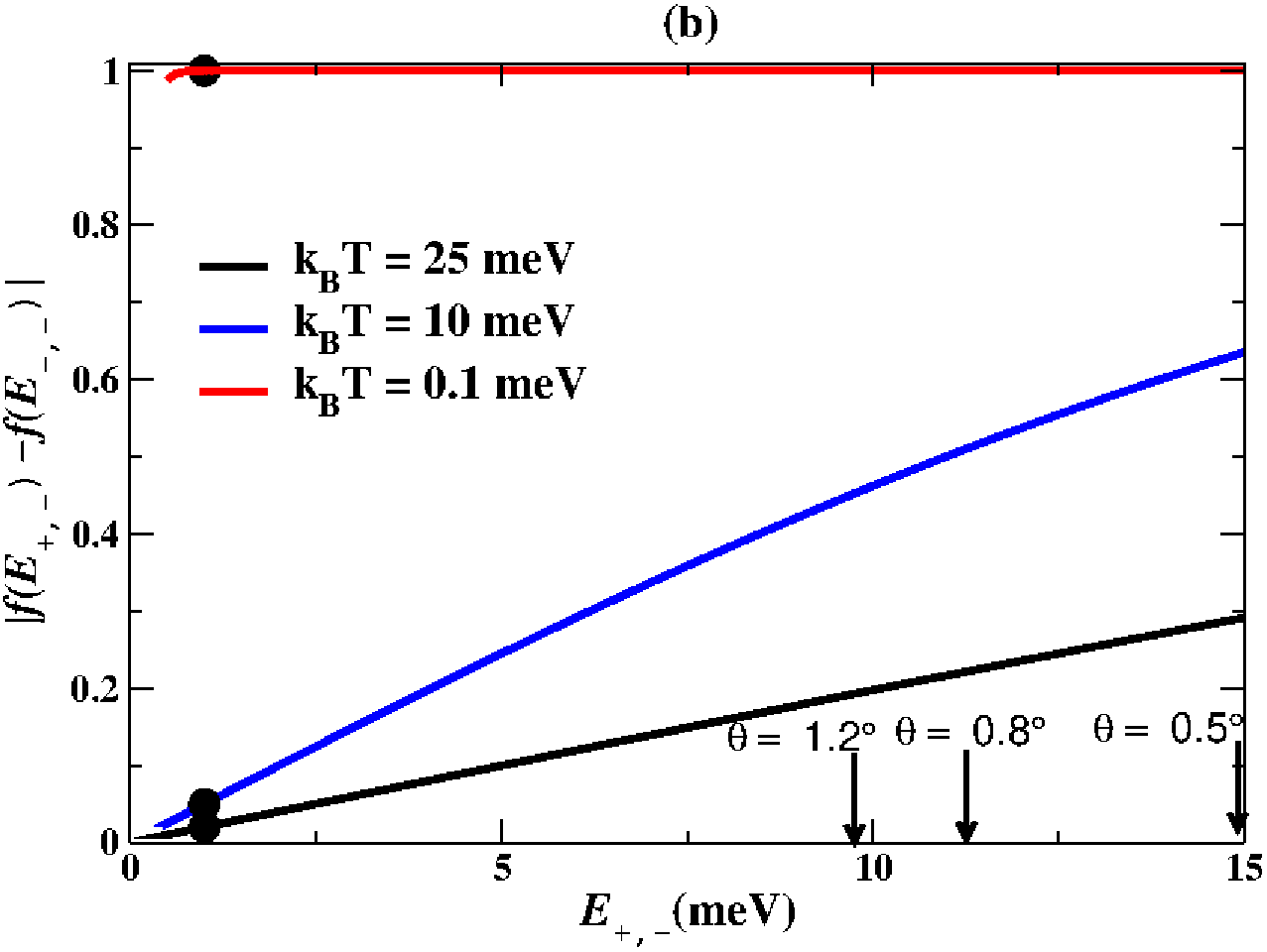}
\includegraphics[width=0.3\columnwidth]{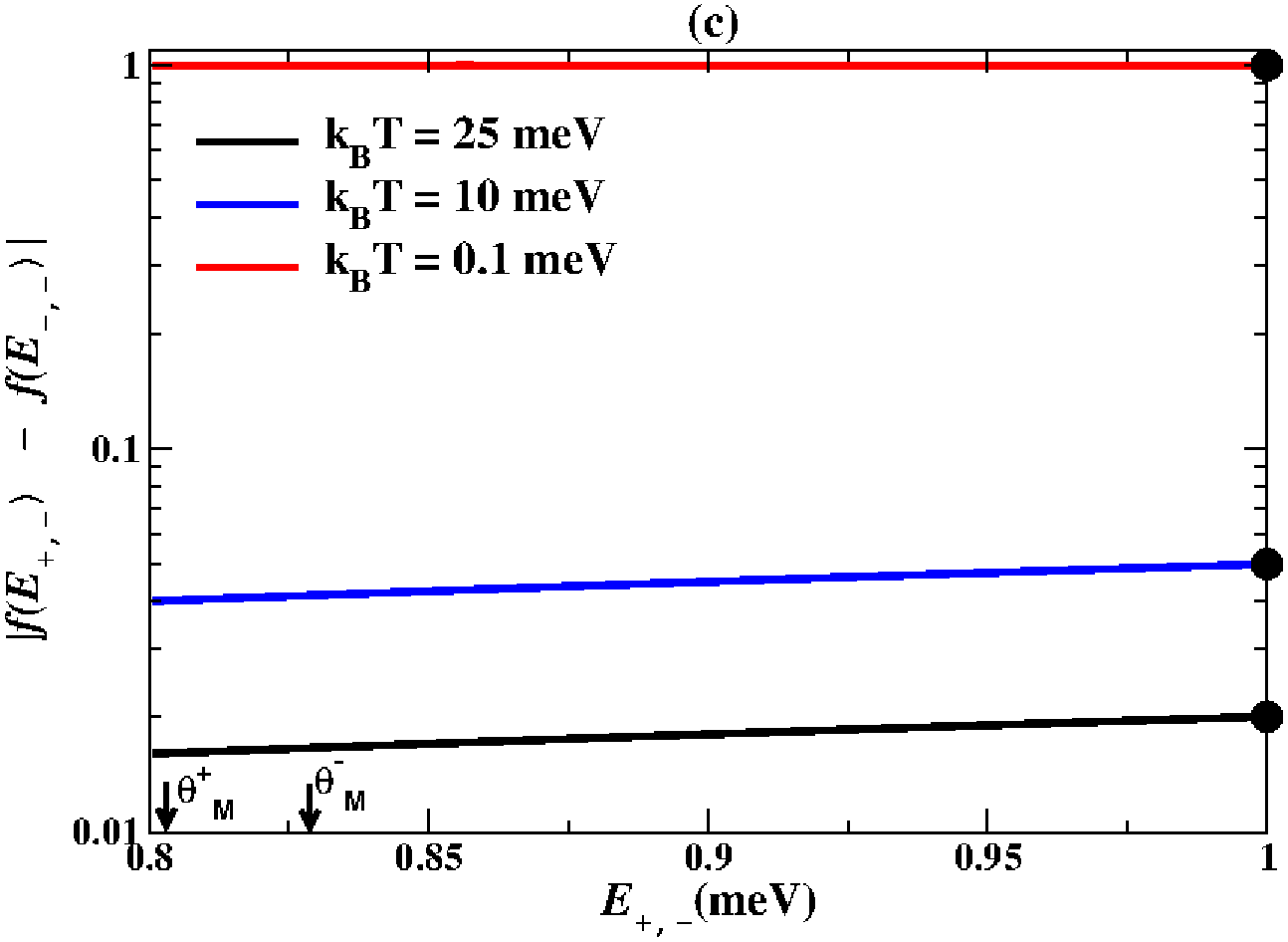}
\end{array}
$
\end{center}
\caption{Statistical weight $\Delta f(E)$ corresponding to the transitions between $E_{-,+}\rightarrow E_{+,+}$ (a)  and  $E_{-,-}\rightarrow E_{+,-}$ ((b) and (c)) at different temperatures. 
The dots represent the energy $E_{+,+}$ (a) and $E_{+,-}$ ((b) and (c)) at the MA and the arrows mark the limit of the band $E_{+,+}$ (a)  and $E_{+,-}$ ((b) and (c)) at the indicated twist angle. In (c), $\Delta f(E)$ is shown around the MA for the transition between $E_{-,-}\rightarrow E_{+,-}$. Calculations are done for the SOC $\lambda_I=3\;\mathrm{meV}$, $\lambda_R=4\;\mathrm{meV}$~\cite{Alex} and in the undoped TBG ($\mu=0$).}
\label{FD-supp}
\end{figure*}

In Fig.~\ref{FD-supp} we plot $\Delta f(E)$ corresponding to the transitions between $E_{-,+}\rightarrow E_{+,+}$ and $E_{-,-}\rightarrow E_{+,-}$ in the case of the undoped system.\

Figures~\ref{FD-supp} (a) and (b) show that, at high temperature ($k_BT>\lambda_I,\lambda_R$), $\Delta f(E)$ increases as the band dispersion gets larger and reaches its minimal value at the MA. This behavior explains the drop of the GD at the MA and its enhancement at small twist angles.\

In figure~\ref{FD-supp} (c), we plot $\Delta f(E)$ around the MA, at relatively high thermal energy compared to the SOC, where the GD exhibits a peak at the MA (Fig.~{\red2} of the main text).
In this case, $\Delta f(E)$ is maximal at the MA and decreases at the angles $\theta^+_M$ and $\theta^-_M$ close to the MA.
This feature results from the decrease of the energy separation between $E_{-,-}$ and $E_{+,-}$ at $\theta^+_M$ and $\theta^-_M$, compared to that at $\theta_M$ (Fig.~{\red3} of the main text). \
At low temperature, and around the MA, one gets $\Delta f(E)=1$ for the transitions  $E_{-,-}\rightarrow E_{+,-}$ and $E_{-,+}\rightarrow E_{+,+}$. As a consequence, the GD behavior is now only dependent on the effective Fermi velocity $v^{\ast}$ which vanishes at the MA. As a consequence, the small peak of the GD, emerging at the MA at relatively high temperature, disappears.\\

\begin{figure}[hpbt] 
\begin{center}
\includegraphics[width=0.35\columnwidth]{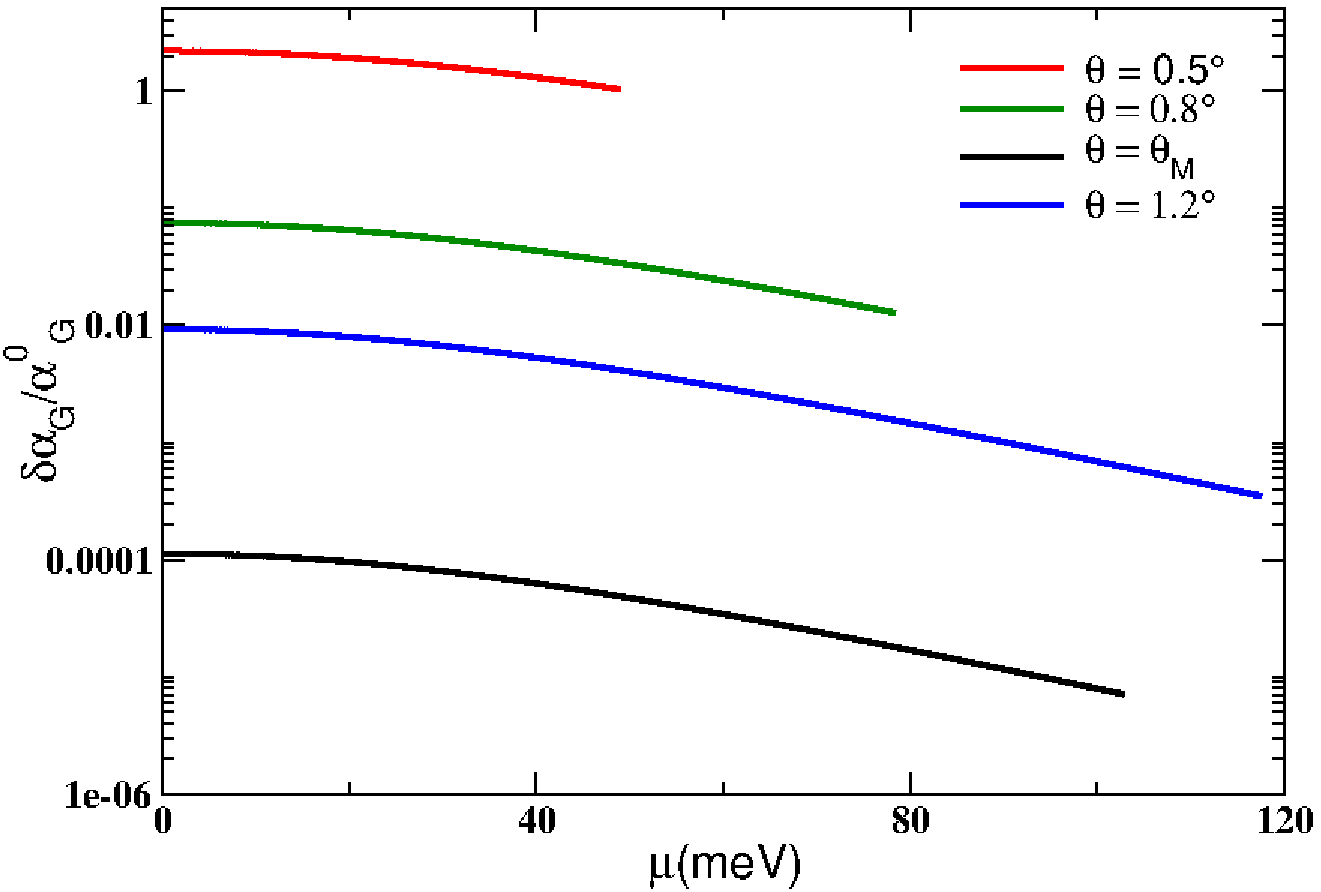}
\end{center}
\caption{Normalized GD correction $\delta\alpha_G/\alpha_G^0$ as function of the chemical potential $\mu$ at $k_BT=25\,\mathrm{meV}$ and for different twist angles. The upper limit of $\mu$ is $\mu_c=\hbar v_F k_c$ corresponding to the momentum cutoff $k_c=\frac{q_0}2$. Calculations are done for the SOC $\lambda_I=3\;\mathrm{meV}$, $\lambda_R=4\;\mathrm{meV}$~\cite{Alex}, $k_BT=25\,\mathrm{meV}$ and for the FMR energy $\hbar \Omega=0.06\,\mathrm{meV}$.}
\label{GD-mu}
\end{figure}

Figure~\ref{GD-mu} shows the behavior of the normalized GD correction $\delta\alpha_G/\alpha_G^0$ as function of the chemical potential $\mu$ at $k_BT=25\,\mathrm{meV}$ and for the FMR energy $\hbar \Omega=0.06\,\mathrm{meV}$. The decrease of $\delta\alpha_G$ is a consequence of the thermal weight. 
The results shown in Fig.~\ref{GD-mu} are expected to hold in the presence of Coulomb interaction if the width of the bands at the MA remains less than 4 meV, which is the case of the filling factor $\nu$ satisfying $-0.5<\nu<0.5$~\cite{Guinea22}.\\

\begin{figure*}[hpbt] 
\centering
$
\begin{array}{ccc}
\includegraphics[width=0.3\columnwidth]{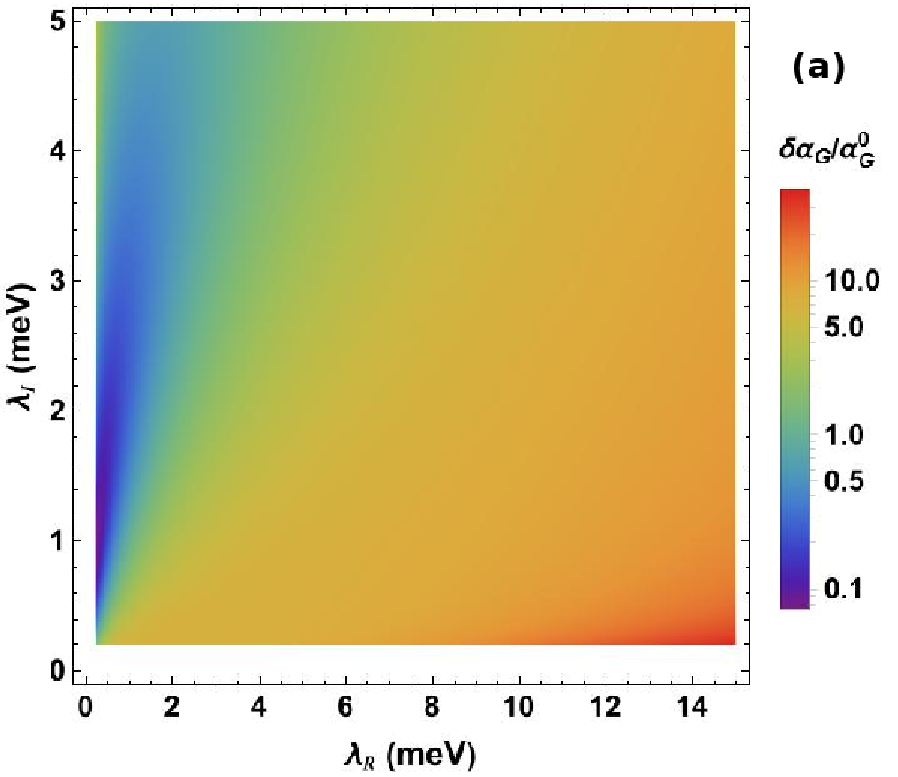}
\includegraphics[width=0.3\columnwidth]{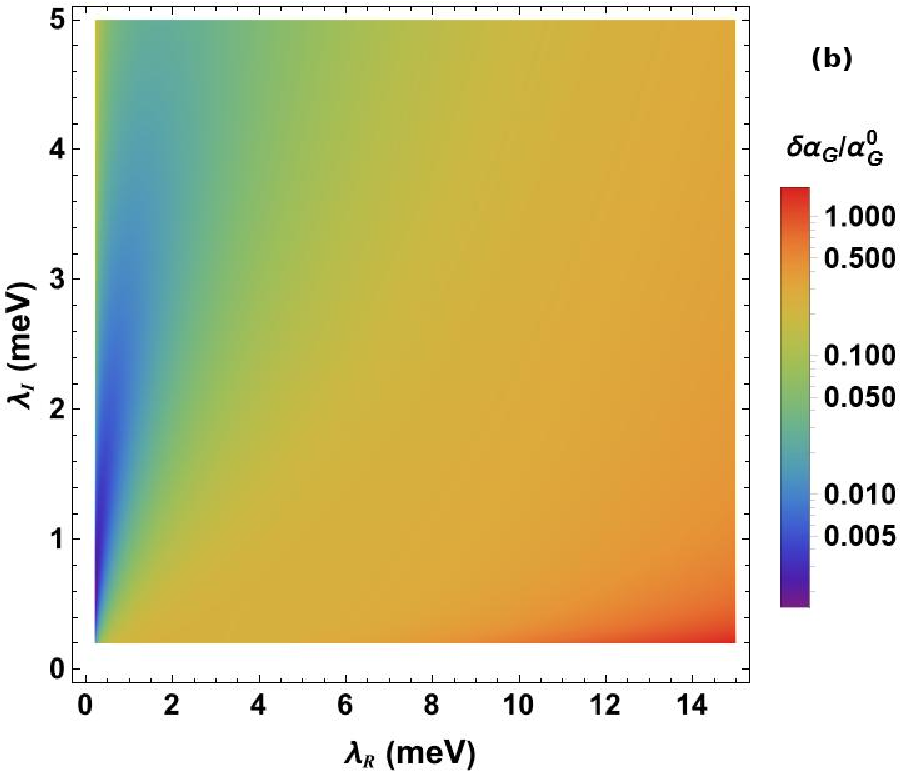}
\includegraphics[width=0.3\columnwidth]{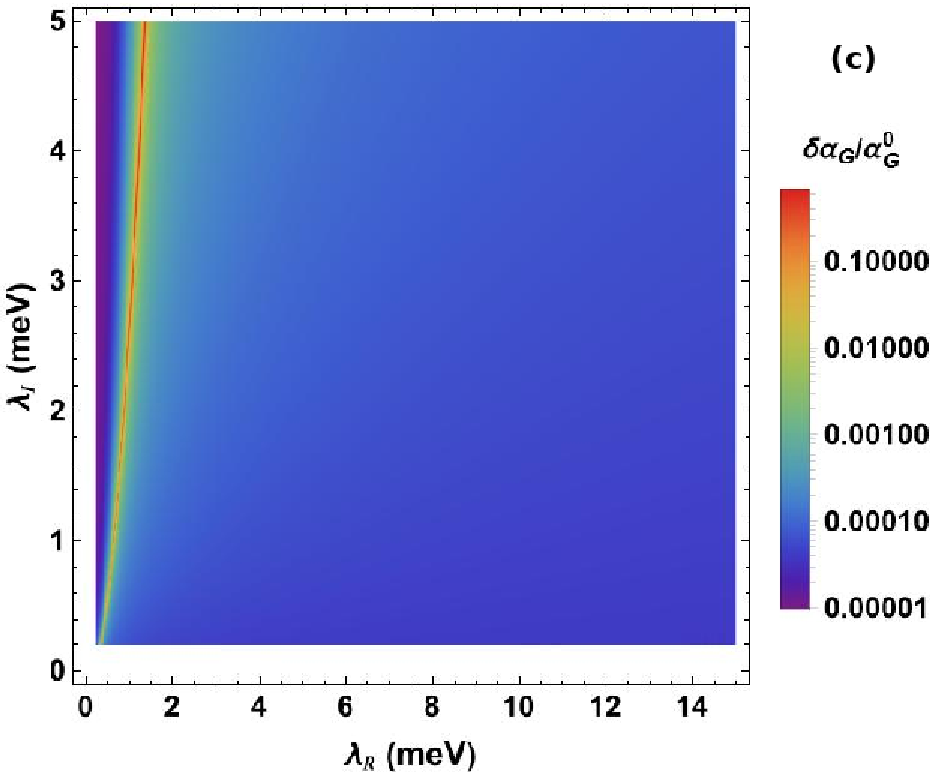}
\end{array}
$
\caption{Normalized GD correction $\delta\alpha_G/\alpha_G^0$ as function of the SOC $\lambda_R$ and $\lambda_I$ at a twist angle $\theta=0.5^{\circ}$ (a), $\theta=0.8^{\circ}$ (b) and at the MA $\theta=1.05^{\circ}$ (c). Calculations are done for $\mu=0$, $k_BT=25\,\mathrm{meV}$ and for the FMR energy $\hbar \Omega=0.06\,\mathrm{meV}$.}
\label{GD-lambda}
\end{figure*}

In Fig.~\ref{GD-lambda}, we plot the normalized GD correction $\delta\alpha_G/\alpha_G^0$ as function of the SOC parameters, $\lambda_I$ and $\lambda_R$, for different twist angles, at $k_BT=25\,\mathrm{meV}$, $\hbar \Omega=0.06\,\mathrm{meV}$ and in the case of the undoped system. The drop of $\delta\alpha_G$ at the MA is a robust feature regardless of the amplitude of the SOC. However, there is a relative increase of $\delta\alpha_G$, at the MA, if the bands $E_{-,+}$ and $E_{+,+}$ (or $E_{-,-}$ and $E_{+,-}$) are in resonance with the FMR energy, as shown in Fig.~\ref{GD-lambda}(c). This resonance can be only reached for relatively small values of $\lambda_R$. \\

\begin{figure*}[hpbt] 
\centering
\includegraphics[width=0.4\columnwidth]{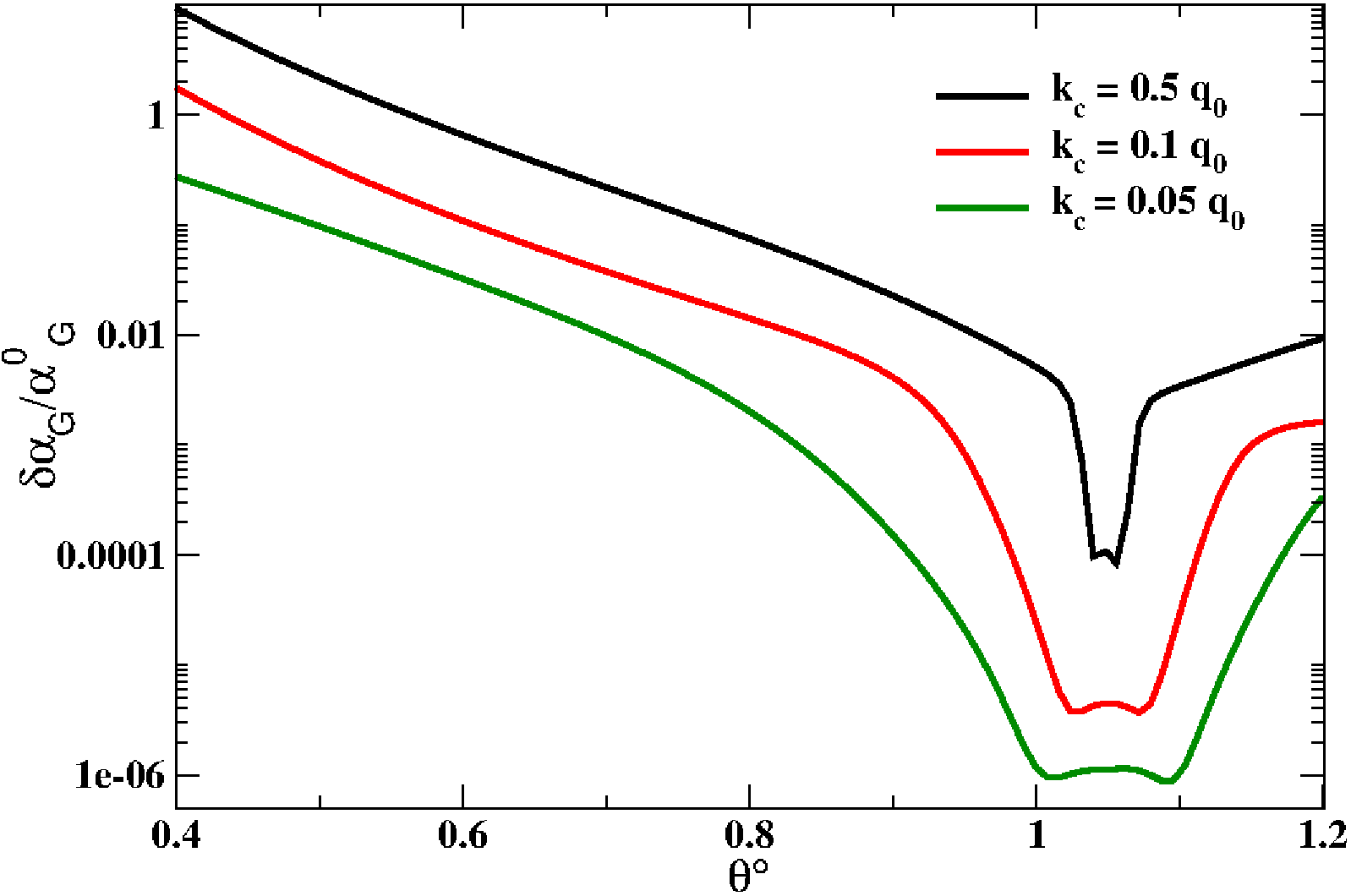}
\caption{Normalized GD correction $\delta\alpha_G/\alpha_G^0$ as function of the twist angle for different values of the cutoff parameter $k_c$. Calculations are done for $\mu=0$, $k_BT=25\,\mathrm{meV}$, $\lambda_I=3\,\mathrm{meV}$, $\lambda_R=4\,\mathrm{meV}$, and for the FMR energy $\hbar \Omega=0.06\,\mathrm{meV}$.}
\label{cutoff}
\end{figure*}

As shown in Fig.~\ref{band-num}, the energy spectrum of the effective model (dashed lines) are slightly more dispersive, at small twist angles ($\theta\sim 0.5^{\circ}$), than those obtained by including higher bands (solid lines). This discrepancy should be taken into account when fixing the value of the cutoff $k_c$ up to which the sum in Eq.~\ref{lor} is evaluated. To determine the role of the cutoff on the SP effect, we plot, in Fig.~\ref{cutoff}, the GD correction $\delta\alpha_G$ as a function of the twist angle at different cutoffs $k_c\le \frac{q_0}2$, where $q_0=|\mathbf{K}_{1,\xi}\mathbf{K}_{2,\xi}|$ is the momentum separation between the Dirac points $\mathbf{K}_{1,\xi}$ and $\mathbf{K}_{2,\xi}$ of respectively layer (1) and layer (2) at a given valley $\xi$.\newline
Fig.~\ref{cutoff} shows that the GD correction drops at the MA regardless of the cutoff values. The larger the cutoff, the sharper the drop. 

%%%%%%%%%%%%%%%%%%%%%%%%%%%%%%%%%%%%%%%%%%%%%%%%%%%%%%%%%%%%%

%%%%%%%%%%%%%%%%%%%%%%%%%%%%%%%%%%%%%%%%%%%%%%%%%%%%%%%%%%%%%

\begin{thebibliography} {200}

\bibitem{Herrero1} Y. Cao, V. Fatemi, S. Fang, K. Watanabe, T. Taniguchi, E. Kaxiras, and P. Jarillo-Herrero, Unconventional superconductivity in magic-angle graphene superlattices, Nature, {\bf 556}, 43 (2018).

\bibitem{Herrero2} Y. Cao, V. Fatemi, A. Demir, S. Fang, S. L. Tomarken, J. Y. Luo, J. D. Sanchez-Yamagishi, K. Watanabe, T. Taniguchi, E. Kaxiras, R. C. Ashoori, and P. Jarillo-Herrero, Correlated insulator behaviour at half-filling in magic-angle graphene superlattices, Nature, {\bf 556}, 80 (2018).

\bibitem{Yank} M. Yankowitz, S. Chen, H. Polshyn, Y. Zhang, K. Watanabe, T. Taniguchi, D. Graf, A. F. Young, C. R. Dean, Tuning superconductivity in twisted
bilayer graphene, Science {\bf 363}, 1059 (2019).

\bibitem{Volovik} N. B. Kopnin, T. T. Heikkila, and G. E. Volovik, High-temperature surface superconductivity in topological flat-band systems, Phys. Rev. B {\bf 83}, 220503(R) (2011).

\bibitem{Senthil} H. C. Po, L. Zou, A. Vishwanath, and T. Senthil, Origin of Mott Insulating Behavior and Superconductivity in Twisted Bilayer Graphene, Phys. Rev. X {\bf 8}, 031089 (2018).

\bibitem{Wu} F. Wu, A. H. MacDonald, and I. Martin, Theory of Phonon-Mediated Superconductivity in Twisted Bilayer Graphene, Phys. Rev. Lett. {\bf 121}, 257001 (2018).

\bibitem{Roy} B. Roy and V. Juricic, Unconventional superconductivity in nearly flat bands in twisted bilayer graphene, Phys. Rev. B {\bf 99}, 121407(R) (2019).

\bibitem{Bernevig} B. Lian, Z. Wang, and B. A. Bernevig, Twisted Bilayer Graphene: A Phonon-Driven Superconductor, Phys. Rev. Lett. {\bf 122}, 257002 (2019).

\bibitem{Efetov} P. Stepanov, I. Das, X. Lu, A. Fahimniya, K. Watanabe, T. Taniguchi, F. H. L. Koppens, J. Lischner, L. Levitov and D. K. Efetov, Untying the insulating and superconducting orders in magic-angle graphene, Nature {\bf 583}, 375 (2020).

\bibitem{Young} Y. Saito, J. Ge, K. Watanabe, T. Taniguchi, A. F. Young, Independent superconductors and correlated insulators in twisted bilayer graphene, Nature Physics {\bf 16}, 926 (2020).

\bibitem{Herrero3} Y. Cao, D. Rodan-Legrain, J. M. Park, F. N. Yuan, K. Watanabe, T. Taniguchi, R. M. Fernandes, L. Fu, P. Jarillo-Herrero, Nematicity and competing orders in superconducting magic-angle graphene, Science, {\bf 372}, 264 (2021).

\bibitem{Efetov19} X. Lu, P. Stepanov, W. Yang, M. Xie, M. A. Aamir, I. Das, C. Urgell, K. Watanabe, T. Taniguchi, G. Zhang, A.
Bachtold, A. H. MacDonald, and D. K. Efetov, Superconductors, orbital magnets and correlated states in magic-angle bilayer graphene, Nature
{\bf 574}, 653 (2019).

\bibitem{Dean19} H. Polshyn, M. Yankowitz, S. Chen, Y. Zhang, K.
Watanabe, T. Taniguchi, C. R. Dean, and A. F. Young, Large linear-in-temperature resistivity in twisted bilayer graphene,
Nat. Phys. {\bf 15}, 1011 (2019).

\bibitem{Pablo20} Y. Cao, D. Chowdhury, D. Rodan-Legrain, O.
Rubies-Bigorda, K. Watanabe, T. Taniguchi, T. Senthil,
and P. Jarillo-Herrero, Strange Metal in Magic-Angle Graphene with near Planckian Dissipation, Phys. Rev. Lett. {\bf 124}, 076801
(2020).

\bibitem{Pablo19} S. L. Tomarken, Y. Cao, A. Demir, K. Watanabe, T. Taniguchi, P. Jarillo-Herrero, and R. C. Ashoori, Electronic Compressibility of Magic-Angle Graphene Superlattices,
Phys. Rev. Lett. {\bf 123}, 046601 (2019).

\bibitem{Eva} G. Li, A. Luican, J. M. B. Lopes dos Santos, A. H. Castro Neto, A. Reina, J. Kong and E. Y. Andrei, Observation of Van Hove singularities in twisted graphene layers, Nature Physics, {\bf 6}, 109 (2010).

\bibitem{Kerelsky} A. Kerelsky, L. J. McGilly, D. M. Kennes, L. Xian, M. Yankowitz, S. Chen, K. Watanabe, T. Taniguchi, J. Hone, C. Dean, A. Rubio and A. N. Pasupathy, Maximized electron interactions at the magic angle in twisted bilayer graphene, Nature, {\bf 572}, 95 (2019).

\bibitem{Yazdani19} Y. Xie, B. Lian, B. J\"ack, X. Liu, C.-L. Chiu, K. Watanabe, T. Taniguchi, B. A. Bernevig, and A. Yazdani, Spectroscopic signatures of many-body correlations in magic-angle twisted bilayer graphene, Nature
{\bf 572}, 101 (2019).

\bibitem{Nadj19} Y. Choi, J. Kemmer, Y. Peng, A. Thomson, H. Arora, R. Polski, Y. Zhang, H. Ren, J. Alicea, G. Refael, F. von
Oppen, K. Watanabe, T. Taniguchi, and S. Nadj-Perge, Electronic correlations in twisted bilayer graphene near the magic angle, 
Nat. Phys. {\bf 15}, 1174 (2019).

\bibitem{Yazdani20} D. Wong, K. P. Nuckolls, M. Oh, B. Lian, Y. Xie, S. Jeon, K. Watanabe, T. Taniguchi, B. A. Bernevig, and A.
Yazdani, Cascade of electronic transitions in magic-angle twisted bilayer graphene, Nature {\bf 582}, 198 (2020).

\bibitem{Nadj21}Y. Choi, H. Kim, Y. Peng, A. Thomson, C. Lewandowski, R. Polski, Y. Zhang, H. S. Arora, K. Watanabe, T.
Taniguchi, J. Alicea, and S. Nadj-Perge, Correlation-driven topological phases in magic-angle twisted bilayer graphene, Nature
{\bf 589}, 536 (2021).

\bibitem{Eva2} N. Tilak, X. Lai, S. Wu, Z. Zhang, M. Xu, R. de Almeida Ribeiro, P. C. Canfield and E. Y. Andrei, Flat band carrier confinement in magic-angle twisted bilayer graphene, Nat Commun. {\bf 12}, 4180 (2021). 

\bibitem{Yazdani22} D. C\u{a}lug\u{a}ru, N. Regnault, M. Oh, K. P. Nuckolls, D. Wong, R. L. Lee, A. Yazdani, O. Vafek, and B. A. Bernevig, Spectroscopy of Twisted Bilayer Graphene Correlated Insulators, Phys. Rev. Lett. {\bf 129}, 117602 (2022).

\bibitem{Utama} M. I. B. Utama, R. J. Koch, K. Lee, N. Leconte, H. Li, S. Zhao, L. Jiang, J. Zhu, K. Watanabe, T. Taniguchi {\it et al.}, Visualization of the flat electronic band in twisted bilayer graphene near the magic angle twist, Nat. Phys. {\bf 17}, 184 (2021).

\bibitem{Efetov21} S. Lisi, X. Lu, T. Benschop, T. A. de Jong, P. Stepanov, J. R. Duran, F. Margot, I. Cucchi, E. Cappelli, A. Hunter, {\it et al.}, Observation of flat bands in twisted bilayer graphene, Nat. Phys. {\bf 17}, 189 (2021).

\bibitem{Sato} K. Sato, N. Hayashi, T. Ito, N. Masago, M. Takamura, M. Morimoto, T. Maekawa, D. Lee, K. Qiao, J. Kim, {\it et al.}, Observation of a flat band and bandgap in millimeter-scale twisted bilayer graphene, Commun Mater {\bf 2}, 117 (2021).

%%%%%%%%%%%%%%%% SP%%%%%%%%%%
\bibitem{Bauer1} Y. Tserkovnyak, A. Brataas, and G. E. W. Bauer, Enhanced Gilbert Damping in Thin Ferromagnetic Films, Phys. Rev. Lett. {\bf 88}, 117601 (2002).

\bibitem{Bauer2} Y. Tserkovnyak, A. Brataas, G. E. W. Bauer, and B. I. Halperin, Nonlocal magnetization dynamics in ferromagnetic heterostructures, Rev. Mod. Phys. {\bf 77}, 1375 (2005).

\bibitem{SP-book} Maekawa, Sadamichi and others (eds), Spin Current, 1st edn, Series on Semiconductor Science and Technology (Oxford, 2012; online edn, Oxford Academic, 17 Dec. 2013), https://doi.org/10.1093/acprof:oso/9780199600380.001.0001.

\bibitem{Hellman} F. Hellman, A. Hoffmann, Y. Tserkovnyak, G. S. D. Beach, E. E. Fullerton, C. Leighton, A. H. MacDonald, D. C. Ralph, D. A. Arena, H. A. D\"urr  {\it el al.}, Interface-induced phenomena in magnetism, Rev. Mod. Phys. {\bf 89}, 025006 (2017).

\bibitem{Qiu2016} Z. Qiu, J. Li, D. Hou, E. Arenholz, A. T. N'Diaye, A. Tan, K.-i. Uchida, K. Sato, S. Okamoto, Y. Tserkovnyak, Z. Q. Qiu, and E. Saitoh, Spin-current probe for phase transition in an insulator, Nat. Commun. {\bf 7}, 12670 (2016).

\bibitem{Yang2018} F. Yang and P. C. Hammel, FMR-driven spin pumping in Y$_3$
Fe$_5$O$_{12}$-based structures, J. Phys. D Appl. Phys. {\bf 51}, 253001 (2018).

\bibitem{Han2020} W. Han, S. Maekawa, and X. Xie, Spin current as a probe of quantum materials, Nat. Mater. {\bf 19}, 139 (2020).

\bibitem{Hait} S. Hait, S. Husain, H. Bangar, L. Pandey, V. Barwal, N. Kumar, N. K. Gupta, V. Mishra, N. Sharma, P. Gupta, {\it et al.}, Spin Pumping through Different Spin-Orbit Coupling Interfaces in $\beta$-W/Interlayer/Co$_2$FeAl Heterostructures, ACS Appl. Mater. Interfaces {\bf 14}, 37182 (2022).

%%%%%%%%%%%%%%%%%% Gr/TMD %%%%%%%%%%%%%%%%%%%
\bibitem{Castro} A. Avsar, J. Y. Tan, T. Taychatanapat, J. Balakrishnan, G. K. W.
Koon, Y. Yeo, J. Lahiri, A. Carvalho, A. S. Rodin, E. C. T.
O’Farrell, G. Eda, A. H. Castro Neto, and B. \"Ozyilmaz, Spin-orbit proximity effect in graphene, Nat.
Commun. {\bf 5}, 4875 (2014).

\bibitem{Morpu15} Z. Wang, D.-K. Ki, H. Chen, H. Berger, A. H. MacDonald, and A. F. Morpurgo, Strong interface-induced spin-orbit interaction in graphene on WS$_2$, Nat. Commun. {\bf 6}, 8339 (2015).

\bibitem{Morpu16} Z. Wang, D.-K. Ki, J. Y. Khoo, D. Mauro, H. Berger, L. S. Levitov, and A. F. Morpurgo, Origin and Magnitude of ‘Designer’ Spin-Orbit Interaction in Graphene on Semiconducting Transition Metal Dichalcogenides, Phys. Rev. X {\bf 6}, 041020 (2016).

\bibitem{Bock16} B. Yang, M.-F. Tu, J. Kim, Y. Wu, H. Wang, J. Alicea, R. Wu, M. Bockrath, and J. Shi, Tunable spin-orbit coupling and symmetry-protected edge states in graphene/WS$_2$, 2D Mater. {\bf 3}, 031012 (2016).

\bibitem{Casa} W. Yan, O. Txoperena, R. Llopis, H. Dery, L. E. Hueso, and F. Casanova, A two-dimensional spin field-effect switch, Nat. Commun. {\bf 7}, 13372 (2016).


\bibitem{Shi} B. Yang, M. Lohmann, D. Barroso, I. Liao, Z. Lin, Y. Liu, L. Bartels, K. Watanabe, T. Taniguchi, and J. Shi, Strong electron-hole symmetric Rashba spin-orbit coupling in graphene/monolayer transition metal dichalcogenide heterostructures, Phys. Rev. B {\bf 96}, 041409(R) (2017).

\bibitem{Wees} T. S. Ghiasi, J. Ingla-Ayn\'es, A. A. Kaverzin, and B. J. van Wees, Large Proximity-Induced Spin Lifetime Anisotropy in Transition-Metal Dichalcogenide/Graphene Heterostructures, Nano Lett. {\bf 17}, 7528 (2017).

\bibitem{Eroms} A. Dankert and S. P. Dash, Electrical gate control of spin current in van der Waals heterostructures at room temperature, Nat. Commun. {\bf 8}, 16093 (2017).

\bibitem{Eroms2} T. V\"{o}lkl, T. Rockinger, M. Drienovsky, K. Watanabe, T. Taniguchi, D. Weiss, and J. Eroms, Magnetotransport in heterostructures of transition metal dichalcogenides and graphene, Phys. Rev. B {\bf 96}, 125405 (2017).

\bibitem{Makk} S. Zihlmann, A. W. Cummings, J. H. Garcia, M.  Kedves, K. Watanabe, T. Taniguchi, C. Sch\"{o}nenberger, and P. Makk, Large spin relaxation anisotropy and valley-Zeeman spin-orbit coupling in 
WSe$_2$/graphene/$h$-BN heterostructures, Phys. Rev. B {\bf 97}, 075434 (2018).


\bibitem{BouchitaPRL} T. Wakamura, F. Reale, P. Palczynski, S. Gu\'eron, C. Mattevi, and H. Bouchiat, Strong Anisotropic Spin-Orbit Interaction Induced in Graphene by Monolayer WS$_2$, Phys. Rev. Lett. {\bf 120}, 106802 (2018).

\bibitem{BLG} J. C. Leutenantsmeyer, J. Ingla-Ayn\'es, J. Fabian, and B. J. van Wees, Observation of Spin-Valley-Coupling-Induced Large Spin-Lifetime Anisotropy in Bilayer Graphene, Phys. Rev. Lett. {\bf 121}, 127702 (2018).


\bibitem{Omar} S. Omar and B. J. van Wees, Spin transport in high-mobility graphene on WS$_2$ substrate with electric-field tunable proximity spin-orbit interaction, Phys. Rev. B {\bf 97}, 045414 (2018).


\bibitem{Valen} L. A. Ben\'{\i}tez, J. F. Sierra, W. S. Torres, A. Arrighi, F. Bonell, M. V. Costache, and S. O. Valenzuela, Strongly anisotropic spin relaxation in graphene–transition metal dichalcogenide heterostructures at room temperature, Nat. Phys. {\bf 14}, 303 (2018).

\bibitem{Roche} C. K. Safeer, J. Ingla-Ayn\'es, F. Herling, J. H. Garcia, M. Vila, N. Ontoso, M. R. Calvo, S. Roche, L. E. Hueso, and F. Casanova, Room-Temperature Spin Hall Effect in Graphene/MoS$_2$ van der Waals Heterostructures, Nano Lett. {\bf 19}, 1074 (2019).


\bibitem{Zaletel} J. O. Island, X. Cui, C. Lewandowski, J. Y. Khoo, E. M. Spanton, H. Zhou, D. Rhodes, J. C. Hone, T. Taniguchi, K. Watanabe, L. S. Levitov, M. P. Zaletel and A. F. Young, Spin-orbit-driven band inversion in bilayer graphene by the van der Waals proximity effect, Nature {\bf 571}, 85 (2019).

\bibitem{Wang} D. Wang, S. Che, G. Cao, R. Lyu, K. Watanabe, T. Taniguchi, C. Ning Lau, and M. Bockrath, Quantum Hall Effect Measurement of Spin–Orbit Coupling Strengths in Ultraclean Bilayer Graphene/WSe$_2$ Heterostructures, Nano Lett. {\bf 19}, 7028 (2019).

\bibitem{Bouchiat19} T. Wakamura, F. Reale, P. Palczynski, M. Q. Zhao, A. T. C. Johnson, S. Gu\'eron, C. Mattevi, A. Ouerghi, and H. Bouchiat, Spin-orbit interaction induced in graphene by transition metal dichalcogenides, Phys. Rev. B {\bf 99}, 245402 (2019).

\bibitem{David} A. David, P. Rakyta, A. Korm\'{a}nyos, and G. Burkard, Induced spin-orbit coupling in twisted graphene–transition metal dichalcogenide heterobilayers: Twistronics meets spintronics,
Phys. Rev. B {\bf 100}, 085412(2019).

\bibitem{Zaletel20} T. Wang, N. Bultinck, and M. P. Zaletel, Flat-band topology of magic angle graphene on a transition metal dichalcogenide,
Phys. Rev. B {\bf 102}, 235146 (2020).

\bibitem{Bouchiat21} For a review see, T. Wakamura, S. Gu\'eron and H. Bouchiat, Novel transport phenomena in graphene induced by strong spin-orbit interaction, Comptes Rendus. Physique, {\bf 22}, 145 (2021).


%%%%%%%%%%%%%%%%%%%%%%%%%%%%

\bibitem{Alex} H. S. Arora, R. Polski, Y. Zhang, A. Thomson,
Y. Choi, H. Kim, Z. Lin, I. Z. Wilson, X. Xu,
J. -H. Chu, K. Watanabe, T. Taniguchi, J. Alicea and S. Nadj-Perge, Superconductivity in metallic twisted bilayer graphene stabilized by WSe$_2$, Nature {\bf 583} 379 (2020).

\bibitem{Lin} J.-X. Lin, Y.-H. Zhang, E. Morissette, Z. Wang, S. Liu, D. Rhodes, K. Watanabe, T. Taniguchi, J. Hone, J. I. A. Li, Spin-orbit–driven ferromagnetism at half moir\'e filling in magic-angle twisted bilayer graphene, Science {\bf 375}, 437 (2022).

\bibitem{Bhowmick} S. Bhowmik, B. Ghawri, Y. Park, D. Lee, S. Datta, R. Soni, K. Watanabe, T. Taniguchi, A. Ghosh, J. Jung, and U. Chandni, Spin-orbit coupling-enhanced valley ordering of malleable bands in twisted bilayer graphene on WSe$_2$, Nature Communications {\bf 14}, 4055 (2023).

\bibitem{temp} We consider the relatively high temperature regime compared to the critical temperatures at which emerge the strongly correlated insulating states of TBG ($\sim 1\mathrm{K}$)~\cite{Herrero1}, or the ferromagnetic phase observed, at $20\, \mathrm{mK}$, in TBG aligned to $\mathrm{WSe}_2$~\cite{Lin}.

\bibitem{Alex2} Y. Zhang, R. Polski, A. Thomson, E. Lantagne-Hurtubise, C. Lewandowski, H. Zhou, K. Watanabe, T. Taniguchi, J. Alicea and S. Nadj-Perge, Enhanced superconductivity in spin-orbit proximitized bilayer graphene, Nature  {\bf 613}, 268 (2023).


\bibitem{Gmitra-BL} M. Gmitra, and J. Fabian, Proximity Effects in Bilayer Graphene on Monolayer WSe$_2$: Field-Effect Spin Valley Locking, Spin-Orbit Valve, and Spin Transistor, Phys. Rev. Lett. {\bf 119}, 146401 (2017).

\bibitem{SOC1} M. Gmitra, and J. Fabian, Graphene on transition-metal dichalcogenides: A platform for proximity spin-orbit physics and optospintronics, Phys. Rev. B {\bf 92}, 155403 (2015).

\bibitem{SOC2} M. Gmitra, D. Kochan, P. Hogl, and J. Fabian, Trivial and inverted Dirac bands and the emergence of quantum spin Hall states in graphene on transition-metal dichalcogenides, Phys. Rev. B {\bf 93}, 155104 (2016).


\bibitem{SOC7} A. W. Cummings, J. H. Garcia, J. Fabian, J. and S. Roche, Giant Spin Lifetime Anisotropy in Graphene Induced by Proximity Effects, Phys. Rev. Lett. {\bf 119}, 206601 (2017).


\bibitem{Mc11} R. Bistritzer and A. H. MacDonald, Moir\'e bands in twisted double-layer graphene, Proc. Natl. Acad. Sci.U.S.A., {\bf 108}, 12233 (2011).

\bibitem{supp} For details, see the Supplemental Material including detailed derivations of the continuum model of TBG/WSe$_2$ [Eq.~(\ref{H1})] and the correction to the Gilbert damping coefficient [Eq.~(\ref{alpha_R})], numerical calculations of the band structure of TBG/WSe$_2$ and the behavior of the Gilbert damping coefficient as a function of different model parameters.

\bibitem{Falko19} D. A. Ruiz-Tijerina and V. I. Fal'ko, Interlayer hybridization and moir\'e superlattice minibands for electrons and excitons in heterobilayers of transition-metal dichalcogenides, Phys. Rev. B {\bf 99}, 125424 (2019).

\bibitem{Bi} Z. Bi, N. F. Q. Yuan, and L. Fu, Designing flat bands by strain, Phys. Rev. B {\bf 100}, 035448 (2019).

\bibitem{Koshino18} M. Koshino, N. F. Q. Yuan, T. Koretsune, M. Ochi, K. Kuroki, and L. Fu, Maximally Localized Wannier Orbitals and the Extended Hubbard Model for Twisted Bilayer Graphene, Phys. Rev. X {\bf 8}, 031087 (2018).

\bibitem{Marwa} M. Manna\"{i} and S. Haddad, Twistronics versus straintronics in twisted bilayers of graphene and transition metal dichalcogenides, Phys. Rev. B {\bf 103}, L201112 (2021) and references therein.

\bibitem{w} G. Cantele, D.Alf{\`e}, F. Conte, V. Cataudella, D. Ninno, and P. Lucignano, Structural relaxation and low-energy properties of twisted bilayer graphene, Phys. Rev. Research {\bf 2}, 043127 (2020).


\bibitem{Ohnuma} Y. Ohnuma, H. Adachi, E. Saitoh, and S. Maekawa, Enhanced dc spin pumping into a fluctuating ferromagnet near 
$T_c$, Phys. Rev. B {\bf 89}, 174417 (2014).

\bibitem{Matsuo18} M. Matsuo, Y. Ohnuma, T. Kato, and S. Maekawa, Spin Current Noise of the Spin Seebeck Effect and Spin Pumping, Phys. Rev. Lett. {\bf 120}, 037201 (2018).

\bibitem{Kato19} T. Kato, Y. Ohnuma, M. Matsuo, J. Rech, T. Jonckheere, and T. Martin, Microscopic theory of spin transport at the interface between a superconductor and a ferromagnetic insulator, Phys. Rev. B {\bf 99}, 144411 (2019).

\bibitem{Kato20} T. Kato, Y. Ohnuma, and M. Matsuo, Microscopic theory of spin Hall magnetoresistance, Phys. Rev. B {\bf 102}, 094437 (2020).

\bibitem{Matsuo-JP} Y. Ominato and M. Matsuo, Quantum Oscillations of Gilbert Damping in Ferromagnetic/Graphene Bilayer Systems, J. Phys. Soc. Jpn. {\bf 89}, 053704 (2020).

\bibitem{Matsuo20} Y. Ominato, J. Fujimoto, and M. Matsuo,  Valley-Dependent Spin Transport in Monolayer Transition-Metal Dichalcogenides, Phys. Rev. Lett. {\bf 124}, 166803 (2020).

\bibitem{Yama} M. Yama, M. Tatsuno, T. Kato, and M. Matsuo, Spin pumping of two-dimensional electron gas with Rashba and Dresselhaus spin-orbit interactions,
Phys. Rev. B {\bf 104}, 054410 (2021).

\bibitem{Funato}T. Funato, T. Kato, and M. Matsuo, Spin pumping into anisotropic Dirac electrons, Phys. Rev. B {\bf 106}, 144418 (2022).

\bibitem{Guinea22} T. Cea, P. A. Pantale\'on, N. R. Walet and F. Guinea, Electrostatic interactions in twisted bilayer graphene, Nano Materials Science {\bf 4}, 27 (2022) and references therein.


%%%%%%%%%%%%%%%%%%%%%%%%%%%%%%%%%%%%%%

\bibitem{Guinea18} F. Guinea and N. R. Walet, Electrostatic effects, band distortions, and superconductivity in twisted graphene bilayers, PNAS {\bf 115}, 13174 (2018).

\bibitem{Guinea20} T. Cea and F. Guinea, Band structure and insulating states driven by Coulomb interaction in twisted bilayer graphene, Phys. Rev. B {\bf 102}, 045107 (2020).

\bibitem{Guinea21} T. Cea and F. Guinea, Coulomb interaction, phonons, and superconductivity in twisted bilayer graphene, PNAS {\bf 118}, e2107874118 (2021).
%%%%%%%%%%%%%%%%%%%%%

\bibitem{SP-Gr} M. Gurram, S. Omar, S. Zihlmann, P. Makk, C. Sch\"{o}nenberger, and B. J. van Wees, Spin transport in fully hexagonal boron nitride encapsulated graphene, Phys. Rev. B {\bf 93}, 115441 (2016).

\bibitem{SP-TMD} Z. Zhou, P. Marcon, X. Devaux, P. Pigeat, A. Bouch\'e, S. Migot, A. Jaafar, R. Arras, M. Vergnat, L. Ren, Large Perpendicular Magnetic Anisotropy in Ta/CoFeB/MgO on Full-Coverage Monolayer MoS$_2$ and First-Principles Study of Its Electronic Structure, ACS Appl. Mater. Interfaces, {\bf 13}, 32579 (2021).
 
\bibitem{Rebeca} R. Ribeiro-Palau, C. Zhang, K. Watanabe, T. Taniguchi, J. Hone, C. R. Dean, Twistable electronics with dynamically rotatable heterostructures, Science {\bf 361}, 690 (2018).

\bibitem{Hu} C. Hu, T. Wu, X. Huang, Y. Dong, J. Chen, Z. Zhang, B. Lyu, S. Ma, K. Watanabe, T. Taniguchi, {\it et al.}, In-situ twistable bilayer graphene, Scientific Reports {\bf 12}, 204 (2022).

\bibitem{Geim} Y. Yang, J. Li, J. Yin, S. Xu, C. Mullan, T. Taniguchi, K. Watanabe, A. K. Geim, K. S. Novoselov, A. Mishchenko, In situ manipulation of van der Waals heterostructures for twistronics, Sci. Adv. {\bf 6}, eabd3655 (2020).

\bibitem{Inbar} A. Inbar, J. Birkbeck, J. Xiao,  T. Taniguchi, K. Watanabe, B. Yan, Y. Oreg, A. Stern, E. Berg and S. Ilani, The quantum twisting microscope, Nature {\bf 614}, 682 (2023).
\end{thebibliography}
\end{document}